\documentclass[]{pasj02} 
\usepackage[switch,mathlines]{lineno} 

\newcommand{\actaa}{AcA.}

\jyear{2026}
\Received{}
\Accepted{}

\begin{document} 

\title{Investigating Potential Relationships Between Mass Transfer Indicators and $\delta$ Scuti Pulsations in oEA Systems}

\author{
 Eda \textsc{\c{C}elik},\altaffilmark{1}\altemailmark\orcid{0000-0003-3278-5140} \email{celikeda7575@gmail.com} 
 Filiz \textsc{Kahraman Ali\c{c}avu\c{s}}\altaffilmark{2,3}\orcid{0000-0002-9036-7476}\email{filizkahraman01@gmail.com}
}

\altaffiltext{1}{\c{C}anakkale Onsekiz Mart University, School of Graduate Studies, Physics Department, TR-17100, \c{C}anakkale, T\"{u}rkiye}
\altaffiltext{2}{\c{C}anakkale Onsekiz Mart University, Faculty of Science, Physics Department, TR-17100, \c{C}anakkale, T\"{u}rkiye}
\altaffiltext{3}{\c{C}anakkale Onsekiz Mart University, Astrophysics Research Center and Ulup{\i}nar Observatory, TR-17100, Çanakkale, T\"{u}rkiye}




\maketitle

\begin{abstract}
We explored potential associations between ongoing mass transfer and the pulsation properties of $\delta$ Scuti primaries in a sample of 17 oscillating eclipsing Algol (oEA) systems. For this purpose, we combined binary modelling, updated O-C analyses, and frequency characterization based on TESS photometry. To ensure homogeneity, the $T_{\rm eff}$ of the components were derived using a consistent SED-fitting approach. From the binary solutions and O--C diagrams, we determined the mass-transfer indicators ($Q$, $dP/dt$, and $dM/dt$) and examined their relations with the dominant (highest amplitude) pulsation period and amplitude. For the full sample, the results indicate weak to moderate positive trends, with the strongest relation found between $\log(dP/dt)$ and $\log P_{\rm puls}$, while the relations involving $dM/dt$ and pulsation amplitudes are weaker and statistically insignificant. After excluding the two prominent outliers, R CMa and BW Del, the correlations between $log P_{\rm puls}$ and the mass-transfer indicators become statistically more significant, including relations with $dP/dt$, $Q$, and $dM/dt$ under both conservative and non-conservative assumptions. These relationships may reflect the indirect effects of mass transfer on the structure of the accreting primary and on the orbital configuration, potentially reducing tidal constraints and allowing shifts toward longer pulsation periods. However, the large dispersion and limited sample size indicate that these relationships remain poorly constrained.  Our results suggest that pulsation properties are primarily governed by intrinsic stellar parameters and the well-established $P_{\mathrm{orb}}$--$P_{\mathrm{puls}}$ relation, while ongoing mass transfer may contribute as a secondary factor. Although the present sample does not allow firm conclusions regarding the role of mass transfer, this study provides a homogeneous observational framework for future investigations based on larger samples and improved constraints.  
\end{abstract}


\section{Introduction}\label{sec:Intro}

Eclipsing binaries are crucial astrophysical laboratories, particularly for the precise determination of fundamental stellar parameters such as mass ($M$) and radius ($R$) via combined light-curve and radial-velocity modeling \citep{2010AARv..18...67T,2021AARv..29....4S, 2024AA...691A.170H}. Within eclipsing binary systems Algol-type eclipsing binaries are significant objects for understanding the effects of mass transfer on stellar and orbital evolution, including changes in mass ratio, angular momentum, orbital period, surface rotation, and (in some systems) pulsation behavior.

Algols reach their current evolutionary status evolving from a detached binary configuration \citep{1971ARAA...9..183P}. During their evolution, the originally more massive component fills its Roche lobe and transfers mass to its companion, leading to a reversal of the mass ratio. As a result, in the current configuration, the less massive component acts as the Roche-lobe-filling donor, while the more massive component is the accretor. It was suggested that accretion onto the mass-gaining star could modify its fundamental parameters ($M$, $R$, and density $\rho$) \citep{2001icbs.book.....H, 2006epbm.book.....E,2006MNRAS.373..435I}.


A particularly interesting properties of Algol type system is that they may exhibit oscillations. The first pulsating Algols were found in the 1970s \citep{1971IBVS..596....1T, 1973IBVS..823....1B}. The definition of the Oscillating Eclipsing Algol (oEA) systems, in which the primary component is typically a A-, or F-type star exhibiting $\delta$ Scuti-type pulsations, was made by \citet{2002ASPC..259...96M}. These systems show clear evidence of mass transfer, making them especially valuable for studying pulsational behavior and mass-transfer effects simultaneously \citep{2002ASPC..259...96M}. The $\delta$ Scuti stars, observed as component star in oEA-type systems, are pulsating variables that exhibit oscillation behavior with relatively short pulsating periods ($P_{puls}$) \citep{2022Galax..10...97M}. Their spectral types range from A and early-F, with masses between approximately $1.5 \leq M \leq 2.5 M_\odot$ \citep{2010aste.book.....A}. $\delta$ Scuti stars are placed within the classical instability strip on the Hertzsprung-Russell Diagram (HRD). They exhibit low-order radial and non-radial pressure and gravity oscillation modes with their pulsation frequencies varying generally in the range of $\sim$\,4\,-\,80 d$^{-1}$. These oscillations are driven by $\kappa$ mechanism which operates in the Helium partial ionization zone \citep{1962ZA.....54..114B,1971AA....14...24C,2010aste.book.....A}.

The latest catalogs of binary member $\delta$ Scuti stars were presented by \cite{2017MNRAS.470..915K,2017MNRAS.465.1181L} where around a hundred of eclipsing binaries with $\delta$ Scuti component is listed. Fortunately, the known number of such systems has increased steadily thanks to advances in time-series photometry and large-scale sky surveys (e.g. \cite{2022ApJS..259...50S, 2022ApJS..263...34C, 2022RAA....22h5003K, 2023MNRAS.524..619K}). 

Because of the proximity of binary components in these kinds of systems, it is known that gravitational interactions change the oscillation behavior of pulsating stars. The correlation between the orbital period ($P_{orb}$) and $P_{puls}$ has been clearly demonstrated in various studies \citep{2006MNRAS.366.1289S, 2017MNRAS.470..915K, 2017MNRAS.465.1181L}. According to the study of \citet{2017MNRAS.470..915K}, $\delta$ Scuti stars in the eclipsing binary systems have shorter $P_{puls}$ and smaller pulsation amplitude ($Amp$) than single $\delta$ Scuti stars. This is because of the proximity effect of the evolved non-pulsating secondary component. Additionally, due to tidal locking, significant alterations can be occurred in the $Amp$ and $P_{puls}$. Similarly, it is believed that mass transfer may also affect pulsation properties. oEA systems provide a unique laboratory for studying the interplay between mass transfer and stellar pulsations. The influence of ongoing mass accretion on the pulsation properties of the primary component was first brought to attention by the seminal works of \citet{2002ASPC..259...96M,2004AA...419.1015M}, who suggested that the mass flow could perturb the stellar structure and affect the excitation of oscillation modes. Since then, further studies have explored how the accumulation of circumstellar material and the variation in orbital parameters shape the pulsational behavior of these systems \citep{2025arXiv250220626P,2024AA...692A.168L,2022MNRAS.514..622M,2022Galax..10...97M}.

The physical nature of the mass transfer process in Algol systems is primarily determined by orbital separation and the size of the accreting component. The mass can either form an accretion disk or, in more compact systems, directly impact the surface of the host star via a stream of gas \citep{2022Galax..10...97M}. This ongoing mass exchange imparts additional angular momentum and thermal energy to the accreting component's atmosphere. As discussed by \citet{2022Galax..10...97M}, this mass transfer process can disrupt pressure-driven oscillations and potentially trigger accretion-induced pulsation; making oEA systems unique laboratories for studying the internal structure of stars under external dynamic influence. Furthermore, since most oEA systems are observed in a slow mass transfer phase, they provide a stable configuration for long-term monitoring of these interactions. Well known oEA systems like RZ Cas, AB Cas, and AS Eri provide clear observational evidence for these effects. For instance, RZ Cas shows a direct link between mass-transfer events and pulsational changes \citep{2020AA...644A.121L}. In AB Cas, the combination of orbital period variations and a diverse spectrum of pulsation modes allows for a detailed study of how mass accumulation impacts the stellar interior \citep{2022MNRAS.514..622M}. Additionally, the precise parameters derived for AS Eri \citep{2022MNRAS.512..917L} highlight the importance of high-quality data in understanding the evolution of mass-gaining pulsators. In conclusion, these systems serve as essential benchmarks for exploring the relation between binary evolution and stellar oscillations.

The presence of mass transfer in oEA systems can be examined through orbital period variation analysis. In close binary systems, the orbital period may vary due to several physical mechanisms, including mass transfer between the binary components, apsidal motion, magnetic activity, or the presence of a third body. Among these, mass transfer and magnetic activity cause a real physical change in the orbital period \citep{1966AdAA...4..233K,1992ApJ...385..621A}, while apsidal motion and third-body mechanisms create only an observational change \citep{1993AA...277..487C,1959AJ.....64..149I}. Such variations are typically investigated by constructing and analyzing the orbital period changing; O–C diagram, which represents the difference between the ''Observed`` and the ''Calculated`` times of minimum light as a function of epoch (or orbital phase). Since several studies have suggested that mass transfer may affect the $\delta$ Scuti-type pulsations in Algol-type binaries \citep{2002ASPC..259...96M, 2004AA...419.1015M, 2020AA...644A.121L, 2022Galax..10...97M}, the present study aims to investigate a possible relationship between the mass transfer rate, inferred from the O–C analysis, and the pulsation period and amplitude. In this context, for Algol-type systems, a positive parabolic trend in the O–C diagram is interpreted as mass transfer from the evolved secondary component to the more massive pulsating primary component. The main objective of this work is to determine the amount of mass transfer using O–C analysis, based on theoretical modeling of the parabolic variation in the O–C diagram. To achieve this, we selected systems reported in the literature to exhibit a positive parabolic trend and to contain a $\delta$ Scuti-type pulsating component and aim to find possible potential effects of mass transfer on the pulsation properties.

The paper is organized as follows. The target selection criteria and the observational data are described in Sect.\,2. Sect.\,3 outlines the methodology used to determine the effective temperature ($T_{\rm eff}$) values. The binary modeling of the selected system and the corresponding results are presented in Sect.\,4. Sect.\,5 details the derivation of new times of minima and the subsequent O–C analysis. Sect.\,6 focuses on the frequency analysis. The discussion and conclusions are provided in Sect.\,7 and Sect.\,8, respectively.

\section{Target Selection and Observational Data} \label{sec:targetanddata}

In this study, our aim is to determine the characteristics of mass transfer in some oEA systems using the O-C diagram and revealing possible effects of mass transfer on the oscillations. Mass transfer is observed as an upward parabolic variation in the O-C diagram if it occurs from the less massive secondary component to the more massive primary component \citep{2001icbs.book.....H, 2005ApSS.296...91R}. The reverse scenario is also possible. To investigate the effect of mass transfer from the evolved component to the more massive pulsating component in oEA systems, we examined the most recent catalog of $\delta$ Scuti stars in eclipsing binaries presented by \citet{2017MNRAS.470..915K}, along with the more recent discoveries reported by \citet{2022RAA....22h5003K} and \citet{2023MNRAS.524..619K}. We then searched for oEA systems exhibiting an upward parabolic trend in their orbital period variation by using the VarAstro (O–C Gateway)\footnote{https://var.astro.cz/en}. Based on this search, we identified 17 $\delta$ Scuti components in Algol-type eclipsing binaries. The number of targets is limited not by choice but by observational constraints. The reliable identification of a genuine parabolic trend in an O–C diagram requires long-term, well-sampled eclipse timing data spanning several decades. However, only a subset of known oEA systems currently possess sufficiently extended and dense timing records to allow this trend to be distinguished confidently from other possible effects, such as magnetic activity cycles or third-body perturbations. Information on the selected stars is provided in Table\,\ref{table1}.

\begin{table*}
    \centering 
      \caption{Information about the selected targets. The time of primary minimum ($T_0$) and $P_{orb}$ were taken from \citet{2004AcA....54..207K}. In column 6, only the counts of TESS sectors at 120s, 200s or 600s cadence are listed. * and ** represent the 200s and 600s data respectively, while others are 120s cadence data.}\label{table1}
    \begin{tabular}{lllllll}
    \\
    \hline
    Star    &\textit{V} (mag)    &Spectral        &$T_0$                     &$P_{orb}$        & TESS      &References\footnotemark[$*$] \\
    Name  &$\pm\,0.01$         & Type            &(HJD+2450000)        &(day)               &  sectors   &\\
    \hline
    XZ Aql  &10.18      &A2/3\,II/III   &2501.0900 (10)         &2.1392120 (10)   & 81* & 6,15 \\
    Y Cam   &10.60      &A9IV\,+\,K1IV  &2502.2020 (20)         &3.3057400 (20)   & 40, 47, 53 &6,8\\
    AB Cas  &10.31      &A3V\,+\,K1V   &2501.3480 (10)         &1.3668887 (7)    & 18, 19, 25, 52 & 1,8\\
    RZ Cas  &6.26       &A3V\,+\,K0IV   &2500.5672 (7)          &1.1952578 (4)    & 18, 19, 25, 52 & 6,8   \\
    XX Cep  &9.18       &A6V\,+\,K3IV   &2501.5283 (9)          &2.3373080 (10)   & 17, 18, 24 &6,12    \\
    WY Cet  &9.28       &F0\,V          &2500.8148 (20)         &1.9397504 (20)   & 3 &6,14 \\
    R CMa   &5.70       &F0V\,+\,K1V    &2501.1430 (8)          &1.1359495 (2)    & 7, 33, 34 &11,8   \\
    UW Cyg  &10.86      &A7V            &2503.3570 (10)         &3.4507750 (10)   & 14, 15, 41, 55 &6,13     \\
    BW Del  &11.28      &F2             &2500.8164 (10)         &2.4231633 (8)    & 81* & 6,16 \\
    SX Dra  &10.40      &A7V\,+\,K7IV   &2501.1581 (10)         &5.1695831 (10)   & 14-26, 41, 47-51 & 6,9 \\
    TW Dra  &7.46       &A5V\,+\,K0III  &2500.7340 (10)         &2.8068690 (30)   & 14-16, 21-24, 41, 48-50 & 4,3  \\
    TZ Eri    &9.61       &A5/6V\,+\,K0/1III   &2502.5560 (20)         &2.6061530 (30)   & 5, 32 & 6,10   \\
    CT Her  &11.32      &A3\,V          &2501.6230 (30)         &1.7863750 (30)   & 78, 79* &6,7  \\
    RX Hya  &9.56       &A8\,+\,K2IV    &2501.3170 (30)         &2.2816590 (10)   & 8, 61 & 6,5    \\
    Y Leo    &10.07      &A3\,+\,K6IV    &2501.2353 (20)         &1.6860850 (20)   & 45, 46, 48**& 6,5   \\
    FR Ori   &10.64      &A7             &2501.0900 (30)         &0.8831655 (10)   & 33** & 6,2 \\
    AC Tau  &11.09      &A7V            &2500.2610 (60)         &2.0433190 (70)   & 5 & 6,13   \\
    \hline
    \end{tabular}
\begin{tabnote}
\footnotemark[$*$] (1)\,\citet{2015AAS...22533616H}, (2)\,\citet{1993yCat.3135....0C}, (3)\,\citet{2010AJ....139.1327T}, (4)\,\citet{2002AA...384..180F}, (5)\,\citet{1984BICDS..27...91B}, (6)\,\citet{2000AA...355L..27H}, (7)\,\citet{1984IBVS.2549....1H}, (8)\,\citet{2001AA...366..178R}, (9)\,\citet{2013AJ....145...87S}, (10)\,\citet{1998AAS..132..367B}, (11)\,\citet{2002yCat.2237....0D},  (12)\,\citet{2016AJ....151...77K}, (13)\,\citet{2023ApJS..264...17Z}, (14)\,\citet{1988mcts.book.....H}, (15)\,\citet{1999MSS...C05....0H}, (16)\,\citet{2006AA...446..785M}
\end{tabnote}
\end{table*}

The O–C diagrams, along with the corresponding times of minimums used to generate them were obtained from the VarAstro database. VarAstro is a comprehensive portal where photometric data is published and shared \citep{2025epsc.conf.1685W}.

The light curves of the targets were taken from the Transiting Exoplanet Survey Satellite (TESS) and used to perform binary modeling, pulsation analysis and measure new minimum times. TESS is a space telescope designed to detect planets around the stars with a brightness lower than $13^m.0$, with a broad infrared filter corresponding to the central wavelength I band \citep{2014SPIE.9143E..20R}. TESS collects data in different exposure times such as \textbf{20-}, 120-, 200-, 600-, and 1800s. 
For the pulsation analysis, binary modeling, and to measure new minimum time, we compiled 120-, 200-, or 600-s cadence TESS data available at the time of our analyses, because these cadences have Nyquist frequencies \textbf{above} the typical $\delta$ Scuti pulsation range. 

Because of the relatively large TESS pixel scale, particular care was taken to minimize possible contamination from neighbouring sources during the light-curve extraction. For each target, Gaia DR3 sources in the vicinity of the target were projected onto the corresponding TESS target-pixel image, and the adopted photometric pixels were inspected with respect to the positions and brightnesses of neighbouring sources. In most systems, no comparably bright source was found sufficiently close to the target to produce a significant contribution. However, a few systems contain fainter Gaia sources located within, or very close to, the same TESS pixel/point-spread function as the target. This is particularly evident for CT Her and TW Dra. For CT Her, nearby Gaia sources approximately 2.0–2.4 mag fainter than the target are projected very close to the target position, corresponding individually to approximate flux ratios of about 0.16 and 0.11 relative to the target if their full flux were included. For TW Dra, the nearest source is about 2.2 mag fainter, corresponding to an approximate flux ratio of about 0.13. These estimates are based on Gaia $G$ magnitudes and therefore provide only an indication of the possible contamination in the broader TESS passband; the actual contribution also depends on the TESS point-spread function and the adopted extraction pixels. The photometric extraction was therefore performed by inspecting the target-pixel files together with these Gaia source positions and selecting the target-dominated pixels whenever possible. The list of the number of the used TESS sectors are listed in Table\,\ref{table1}.

\section{Determination of $T_{\rm eff}$ values}\label{sec:three}

The aim of this study is to explore possible associations between mass-transfer indicators and the pulsation properties of $\delta$ Scuti stars in oEA systems. To derive the mass-transfer indicators consistently across the sample, estimates of the component masses are required. These masses were obtained through binary light-curve modelling.

Binary light-curve modelling relies on several input parameters, including the effective temperature ($T_{\rm eff}$) of the primary component, which affects the relative luminosities of the stellar components. Since the $T_{\rm eff}$ values reported in the literature for our targets were derived using different techniques and assumptions, adopting them directly could introduce systematic differences within the sample. Therefore, rather than relying on heterogeneous literature values, we derived $T_{\rm eff}$ estimates for all systems using a uniform spectral energy distribution (SED) fitting procedure. This approach was adopted primarily to establish a homogeneous temperature scale for the subsequent binary modelling and comparative analyses.


The SED fitting procedure compares the observed multi-band photometric fluxes with synthetic stellar atmosphere models. The observed SEDs were constructed using photometric measurements covering ultraviolet and optical wavelengths retrieved through the VOSA service \citep{2014ASPC..485..321R}. Each photometric measurement is characterized by its filter identifier, effective wavelength, observed flux, and uncertainty.


The theoretical fluxes were adopted from the ATLAS9 stellar atmosphere model grids \citep{2003IAUS..210P.A20C}. These models provide monochromatic flux distributions ($F_\lambda$) as a function of wavelength and are characterized by their $T_{\rm eff}$ and $\log g$. The grid used in this study spans a range of $T_{\rm eff}$ between 3000 and 12000\,K in steps of 100\,K, and $\log g$ between 3.0 and 4.5 in steps of 0.1\,dex. During the fitting procedure the synthetic fluxes are corrected for interstellar extinction using the extinction law of \citet{1989ApJ...345..245C}. The extinction correction is parameterized by the colour excess $E(B-V)$ and the ratio of total-to-selective extinction $R_V$, and is applied to each wavelength element of the synthetic flux before computing synthetic photometric fluxes. The $E(B-V)$ values adopted in this work were determined using the three-dimensional Bayestar dust map \citep{2019ApJ...887...93G}, based on the Gaia DR3 parallaxes \citep{2023AA...674A...1G} and the Galactic coordinates of the targets. The derived $E(B-V)$ values are listed in Table\,\ref{table2}.

In order to account for the relative flux contributions of both components in the composite SED, the luminosity ratios in the $V$ band were adopted whenever available in the literature. For systems lacking published $V$-band luminosity ratios, preliminary binary light-curve modelling as described in Sect.\,4, was performed using available $V$-band photometric data, primarily from the ASAS survey \citep{2019MNRAS.486.1907J}. These preliminary models were used to estimate the relative flux contributions of the components in the $V$ band (see Appendix Table~\ref{tableA1}), which were subsequently used as constraints in the SED fitting procedure.

The optimal stellar parameters are determined by performing a grid search over all available combinations of $T_{\rm eff_1}$, $\log g_1$, $T_{\rm eff_2}$, and $\log g_2$ within the adopted model grid, where indices 1 and 2 refer to the primary and secondary components, respectively. For each parameter combination the predicted composite flux is compared with the observed flux distribution using a $\chi^2$ minimization approach. To minimize the influence of potentially discrepant photometric measurements, a two-pass fitting strategy with robust outlier rejection is employed. In the first pass, the full dataset is fitted to obtain an initial solution. Residuals between observed and model fluxes are then evaluated and a robust scatter estimate is computed using the median absolute deviation. Photometric measurements deviating by more than three times this scatter are rejected, and the fitting procedure is repeated using the remaining data points.

\begin{figure*}
    \centering
    \includegraphics[alt={ALT text: Spectral energy distribution plot for RZ Cas showing multiwavelength photometric data points with error bars and the best-fitting composite model curve alongside individual stellar component contributions.},width=0.8\linewidth]{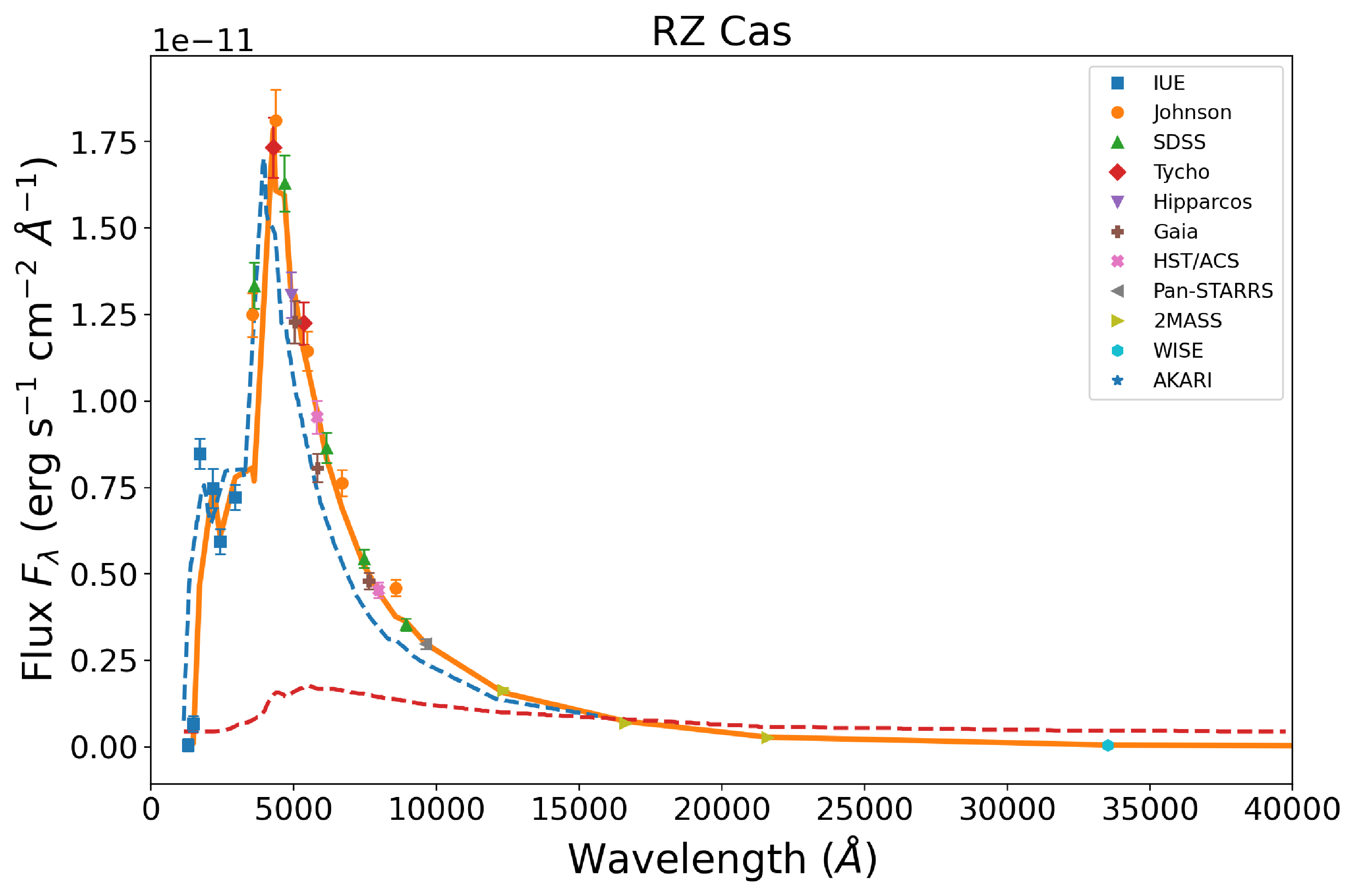}
    \caption{Example SED fit for RZ\,Cas. The observed photometric fluxes from different surveys are shown with different symbols as indicated in the legend. The solid orange curve represents the best-fitting composite SED model, while the blue and red dashed curves indicate the flux contributions of the hot primary and the cool secondary components, respectively. The observed SEDs were constructed using multiwavelength photometric data, including ultraviolet measurements from IUE \citep{1978Natur.275..372B}, optical photometry from the Johnson system \citep{1953ApJ...117..313J}, SDSS \citep{2000AJ....120.1579Y}, Tycho and Hipparcos \citep{1997AA...323L..57H, 1997AA...323L..49P}, as well as Gaia \citep{2016AA...595A...1G}. Additional data from HST/ACS \citep{2005PASP..117.1049S}, Pan-STARRS \citep{2016arXiv161205560C}, and infrared surveys such as 2MASS \citep{2006AJ....131.1163S}, WISE \citep{2010AJ....140.1868W}, and AKARI \citep{2007PASJ...59S.369M} were also included.}
    \label{fig:RZCas_SED}
\end{figure*}

Broadband SED fitting primarily constrains $T_{\rm eff}$, whereas $\log g$ is only weakly constrained because variations in surface gravity have a limited impact on the overall shape of the SED. Therefore, the derived $\log g$ values should be interpreted as approximate estimates. Since the secondary component contributes only a small fraction of the total flux, its surface gravity was fixed at $\log g_2 = 3.5$ to reduce parameter degeneracy. The uncertainties in $T_{\rm eff,1}$, $T_{\rm eff,2}$, and $\log g_1$ were estimated from the $\chi^2$ distribution of the grid search using the $\Delta\chi^2 = 1$ criterion. For simplicity and clarity, Table\,\ref{table2} reports typical average uncertainties of $\pm300$\,K for $T_{\rm eff_1}$, $\pm400$\,K for $T_{\rm eff_2}$, and $\pm0.25$\,dex for $\log g_1$, which represent the characteristic uncertainties of the grid-based SED analysis. The resulting SED parameters are listed in Table\,\ref{table2}, while an example of the SED fitting for one of the best known targets is presented in Fig.\,\ref{fig:RZCas_SED}.

\begin{table}
    \centering 
      \caption{The $T_{\rm eff}$ of the components and the $\log g$ of the primary star were derived from the SED fitting procedure. The subscripts 1 and 2 refer to the primary and secondary components, respectively. Because the flux contribution of the secondary component to the total SED is significantly smaller than that of the primary, the $\log g$ of the secondary star was fixed at 3.5 in all models. The uncertainties reported in the table represent typical average errors obtained from the SED analysis.}\label{table2}
    \begin{tabular}{llllc}
    \\
    \hline
    Star     &$E(B-V)$ &$T_{\rm eff_1}$ & $\log g_1$ &$T_{\rm eff_2}$ \\
             & (mag\,$\pm$\,0.02) & (K\,$\pm$\,300)  & (\,$\pm$\,0.25)  & (K\,$\pm$\,400) \\
    \hline
    XZ Aql   & 0.04 &7800  & 3.8 &4400 \\
    Y Cam    & 0.08 &7700  & 3.7 &4700 \\
    AB Cas   & 0.04 &7700  & 4.3 &4200 \\
    RZ Cas   & 0.00 &8250  & 4.3 &4800 \\
    XX Cep   & 0.17 &8300  & 4.2 &4800 \\
    WY Cet   & 0.00 &7000  & 3.8 &4200 \\
    R CMa    & 0.00 &6900  & 4.2 &4800 \\
    UW Cyg   & 0.13 &7700  & 4.1 &4600 \\
    BW Del   & 0.05 &7500  & 4.0 &4500 \\
    SX Dra   & 0.04 &7100  & 3.6 &4600 \\
    TW Dra   & 0.00 &7700  & 4.2 &4400 \\
    TZ Eri   & 0.00 &7600  & 4.1 &4500 \\
    CT Her   & 0.05 &7800  & 4.3 &4400 \\
    RX Hya   & 0.06 &7600  & 4.1 &4500 \\
    Y Leo    & 0.00 &7600  & 4.0 &4100 \\
    FR Ori   & 0.18 &8000  & 4.0 &4600 \\
    AC Tau   & 0.17 &7700  & 3.9 &4300 \\
    \hline
    \end{tabular}
\end{table}

\begin{figure}
\begin{center}
\includegraphics[alt={Scatter plot of SED-based versus literature spectroscopic $T_{\rm eff}$ values for 10 oEA systems. The dashed line indicates equality and error bars show uncertainties.},width=85mm]{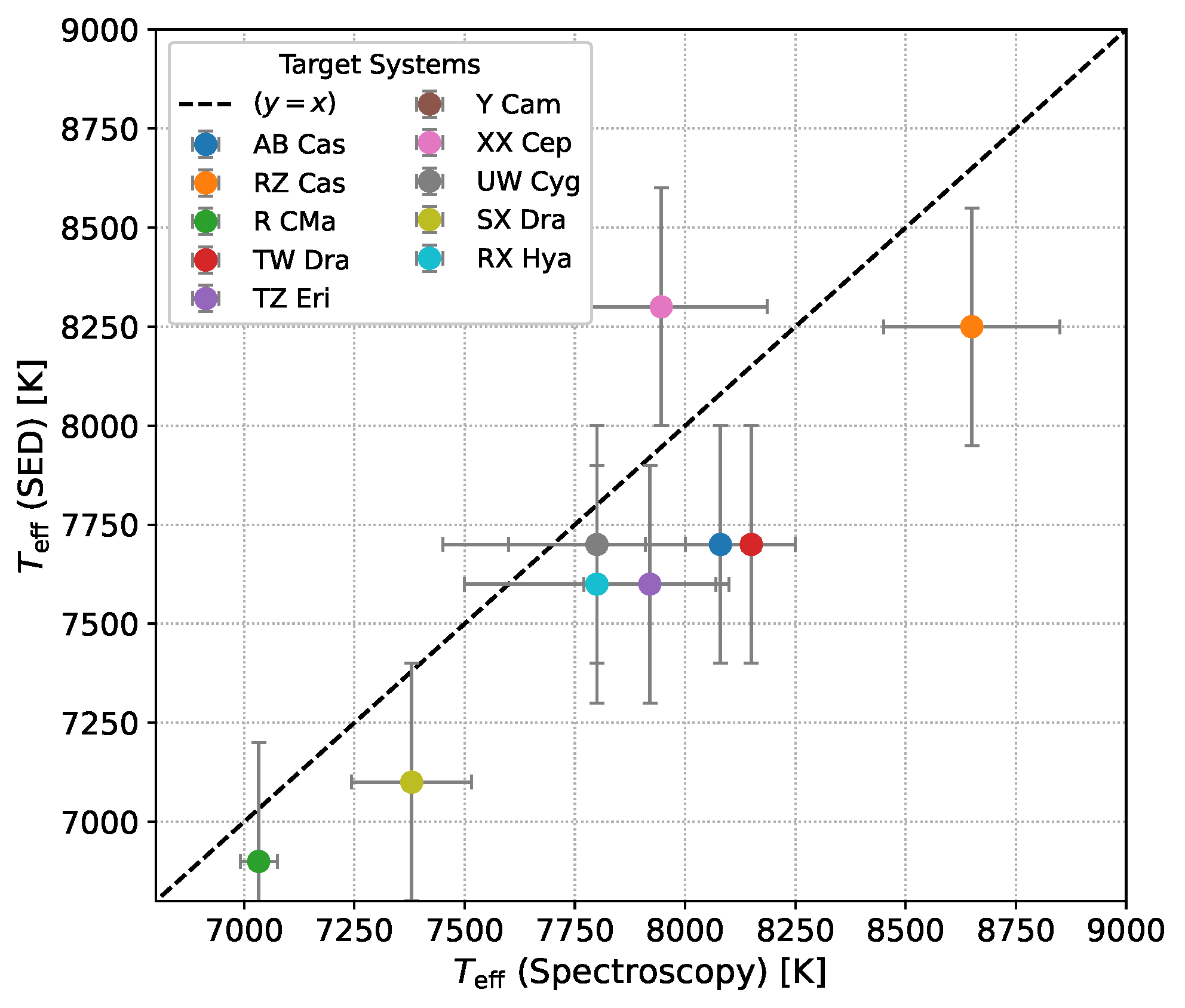}
\end{center}
\caption{Comparison of the $T_{\rm eff}$ values determined from SED analysis in this study with those derived from spectroscopic analyses in the literature for 10 oEA systems.}\label{fig:sed_spec_teff_comp}
\end{figure}

The $T_{\rm eff}$ values derived from the SED analysis were compared with previously published values obtained from spectroscopic analyses (see, Fig.\,\ref{fig:sed_spec_teff_comp}). For several systems, $T_{\rm eff}$ values were determined using spectroscopic techniques such as spectral disentangling or composite spectrum modelling. For example, $T_{\rm eff}=8080\pm170$\,K was reported for AB\,Cas from spectral disentangling analysis by \citet{2017AJ....153..247H}, while \citet{2020AA...644A.121L} derived $T_{\rm eff}=8650\pm200$\,K for RZ\,Cas using a similar method. For R\,CMa, \citet{2018AA...615A.131L} obtained $T_{\rm eff}=7033\pm42$\,K from spectral disentangling, whereas $T_{\rm eff}=8150\pm20$\,K was reported for TW\,Dra by \citet{2010AJ....139.1327T}. In addition, \citet{2025arXiv250220626P} derived $T_{\rm eff}=7920\pm150$\,K for TZ\,Eri using spectroscopic analysis. Composite spectrum fitting techniques yielded $T_{\rm eff}=7800\pm200$\,K for Y\,Cam \citep{2024PASJ...76..787C} and $T_{\rm eff}=7946\pm240$\,K for XX\,Cep \citep{2016AJ....151...77K}. For several other systems, the $T_{\rm eff}$ values were estimated assuming that the observed spectra are dominated by the primary component, leading to $T_{\rm eff}=7800\pm350$\,K for UW\,Cyg \citep{2017MNRAS.470..915K}, $T_{\rm eff}=7380\pm136$\,K for SX\,Dra \citep{2018ApJS..235....5Q}, and $T_{\rm eff}=7800\pm300$\,K for RX\,Hya (Kahraman Aliçavuş, in preparation). The $T_{\rm eff}$ values obtained from our SED modelling show a generally good agreement with these literature values. In most cases the differences between the two determinations remain within $\sim$100–400\,K, which is comparable to the typical uncertainties associated with both SED fitting and spectroscopic temperature determinations. 
Overall, the comparison indicates a reasonable agreement between the $(T_{\rm eff}$ values derived from the SED analysis and previously published spectroscopic estimates as shown in Fig.\ref{fig:sed_spec_teff_comp}. However, some systems show differences of several hundred Kelvin, with the SED-based temperatures tending to be slightly lower than the spectroscopic values in several cases. 

\section{Binary Modeling}


In this section, binary light-curve modelling was performed to derive the fundamental parameters of the 17 target oEA systems. As described in Sect.~3, homogeneous $T_{\rm eff}$ estimates were adopted for all targets. Another important input parameter for the modelling is the mass ratio, $q=M_2/M_1$.

Whenever available, we adopted $q$ values derived from published double-lined spectroscopic (SB2) analyses, as these provide direct observational constraints on the component masses. The corresponding literature values for XZ~Aql, Y~Cam, AB~Cas, RZ~Cas, R~CMa, XX~Cep, TW~Dra, and TZ~Eri are listed in Table\,\ref{tab:tableA2}.

\begin{figure}
  \centering
  \begin{minipage}[t]{0.5\textwidth}
    \centering
    \includegraphics[alt={Light curve of WY Cet with TESS data, Wilson-Devinney model curves, and residuals.}, width=\linewidth]{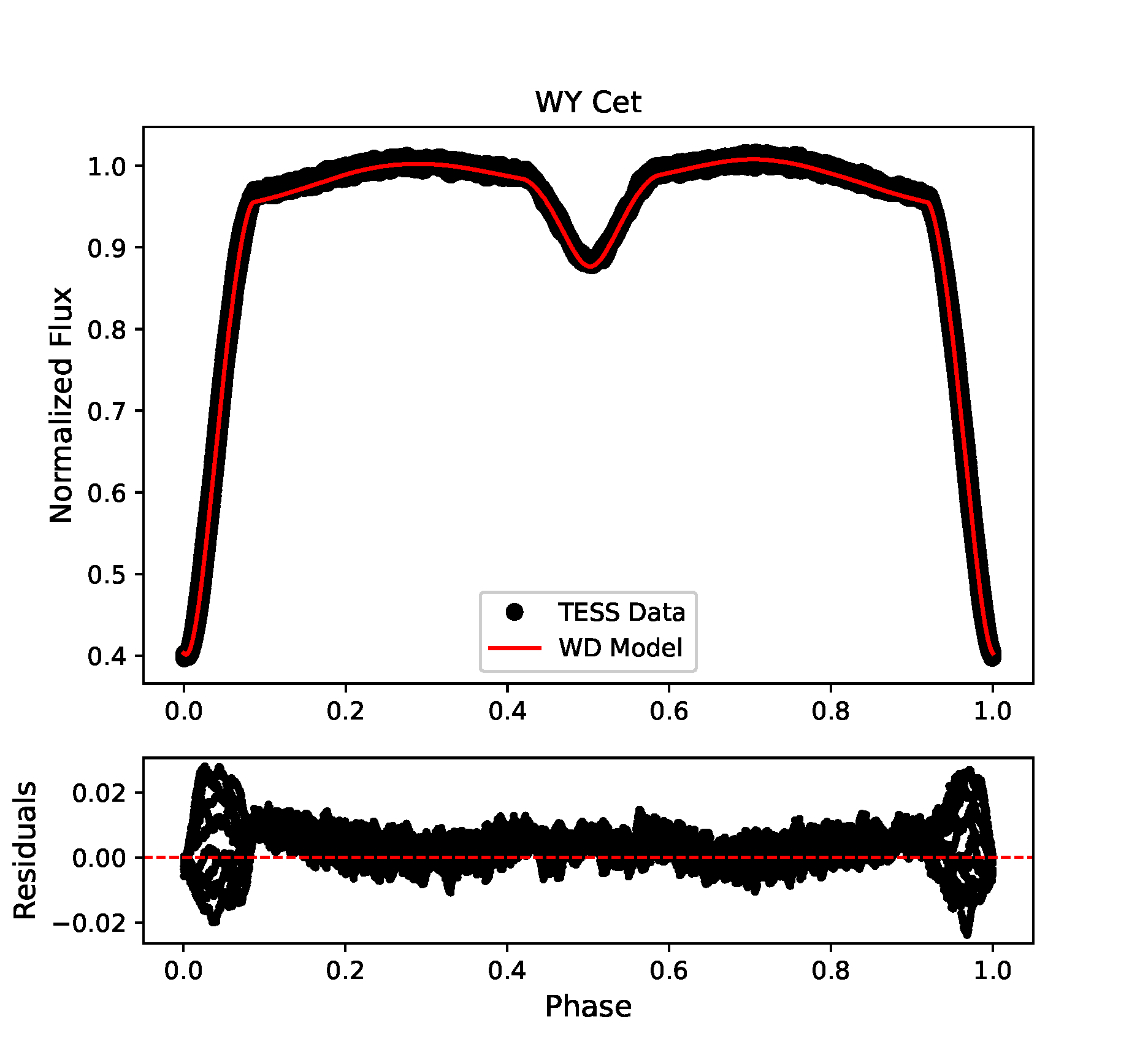}
  \end{minipage}\hfill
  \begin{minipage}[t]{0.5\textwidth}
    \centering
    \includegraphics[alt={Light curve of XX Cep with TESS data, Wilson-Devinney model curves, and residuals.},width=\linewidth]{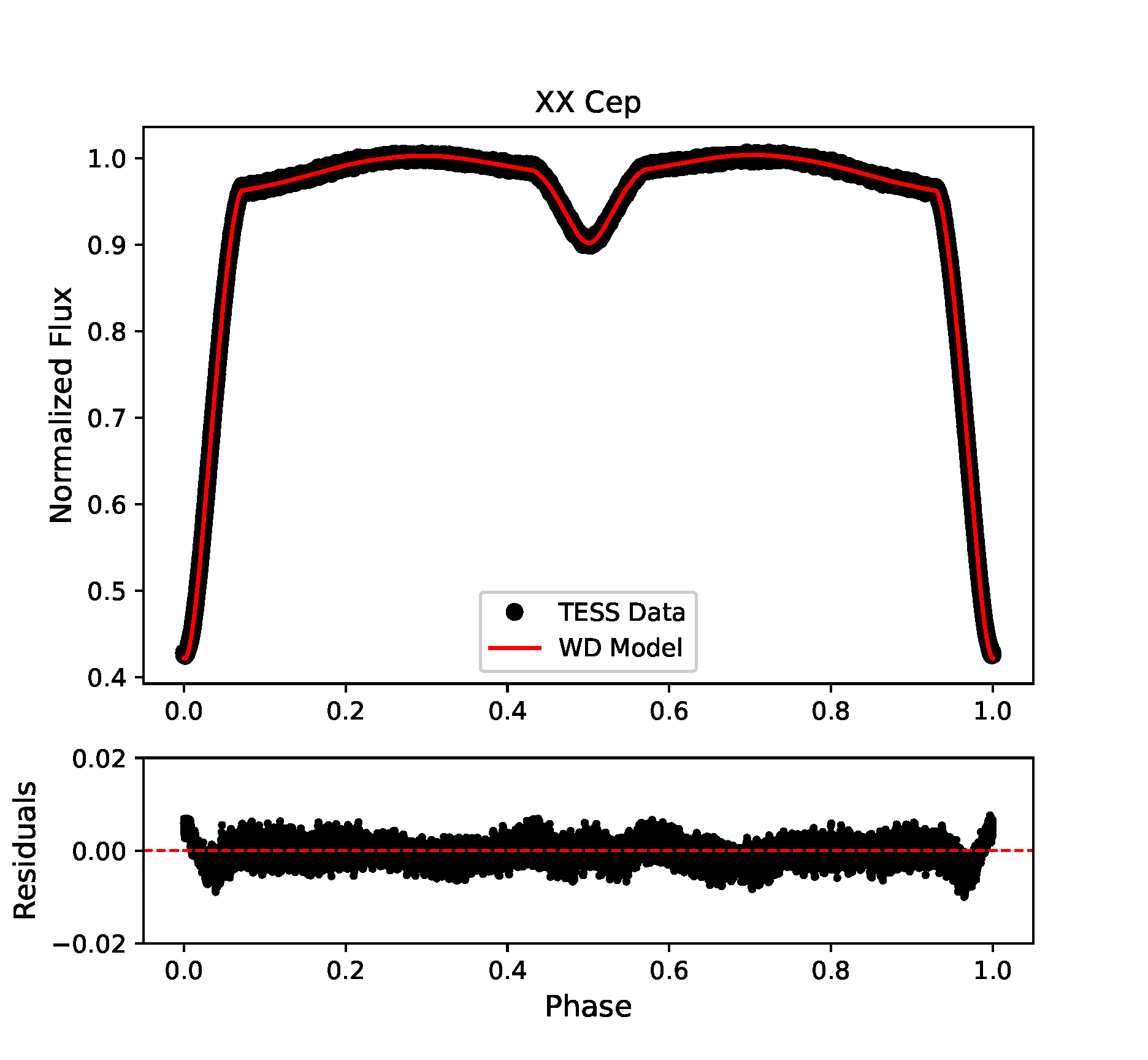}
  \end{minipage}
  \caption{Binary modeling of two targets. Red line represents theoretical Wilson-Devinney (WD) model. Bottom panels in each figures show the residuals from the data.}
  \label{fig:figure1}
\end{figure}

{Two-panel light curves of WY Cet and XX Cep with TESS data, Wilson-Devinney model curves, and residuals.}


For systems lacking published SB2 solutions, $q$ values were estimated using the $q$-search method based on photometric light curve modeling of high-quality TESS data.Although photometric mass ratios are generally less robust than spectroscopic determinations, previous studies have shown that photometric $q$ values derived from photometric light-curve modelling can provide physically meaningful estimates in semi-detached systems, particularly for systems with high orbital inclination ($i$) (e.g.\cite{2005ApSS.296..221T,2022Galax..10....8T}). According to this approach, by varying the $q$ values in 0.01 step and following the procedure applied in F. \citet{2024MNRAS.527.4076K} using the Wilson-Devinney (WD) code, the optimal solution was determined from the minimum of the residuals between the observed and synthetic light curves. The $q$ values obtained for all systems are listed in Table \,\ref{tab:tableA2}.

After this step, we analyzed the \textit{TESS} binary light curves of all selected systems by fixing the $q$ values of the system and SED $T_{\rm eff}$ values of the primary components. The aim of this analysis was to determine the binary parameters of the systems and calculate the corresponding physical parameters using the Kepler equation. To achieve this, the WD code along with Monte-Carlo uncertainty estimation simulation \citep{2004AcA....54..299Z,2010MNRAS.408..464Z} was used. 

Before the analysis, the TESS fluxes of each system were converted into magnitude values using the method outlined by \citet{2022RAA....22h5003K}, and the binary light curve which are affected by the pulsations, was examined. Due to oscillations, the eclipse shapes and depths can be significantly distorted in some systems. Therefore, we removed the pulsations from the binary light curves by theoretically modeling them with the Period04 program \citep{2005CoAst.146...53L} as outlined by \citet{2022RAA....22h5003K}. After this step, the binary light curves were phase-folded and normalized at 1.

Since TESS light curves were used in this study, the binary modeling was carried out under the presumption of the Cousins $I_c$ band, corresponding to the central wavelength of the TESS photometric band. During the binary modeling, some parameters were fixed, while some others were treated as free. The SED $T_{\rm eff}$ values of the primary components were adopted as fixed parameters. Based on the $T_{\rm eff}$ of the each components, the logarithmic limb darkening coefficients were derived from \citet{1993AJ....106.2096V}. The gravity-darkening exponents adopted in the binary modelling are bolometric quantities rather than passband-specific coefficients. We adopted the classical values for the radiative envelopes of the hotter primary components \citep{1924MNRAS..84..665V} and for the convective envelopes of the cooler secondary components \citep{1967ZA.....65...89L}. The bolometric albedos were correspondingly set to $A_1 = 1.0$ for primary and $A_2 = 0.5$ for secondary star \citep{1969AcA....19..245R}. If the systems are not mentioned in the literature to have an eccentric orbit or if the second minimum does not shift according to 0.5 phase, the eccentricity ($e$) parameter is taken as 0. Finally, $q$ values were determined using the previously mentioned methods and were taken as fixed parameters during the analysis. On the other hand, phase shift ($\phi$), $i$, $T_{\rm eff}$ of secondary component, surface potentials ($\Omega_1, \Omega_2$), relative luminosities ($l_1, l_2$), and third light ($l_3$) were set as free parameters.  Binary modeling for two example systems is presented in Fig.\,\ref{fig:figure1} (for other stars, see Fig.\,\ref{fig:all_LC}), and the parameters obtained from the modeling are listed in Table\,\ref{table3}. During this analysis we also found variation caused by the spot on the surface of one of the binary components in eight system. These spot variations were also examined during the binary modeling and the determined parameters for the spots' are also given in Table\,\ref{tableA4}.

It is noteworthy that the third-light solutions obtained for some systems are consistent with the independent contamination assessment described in Sect.\,\ref{sec:targetanddata}. In particular, CT Her and TW Dra yield third-light fractions of $l_3=0.112$ and 0.124, respectively. These values are of the same order as the possible flux contribution inferred from the nearby Gaia DR3 sources projected close to the targets on the TESS pixel images. This agreement suggests that a substantial fraction of the fitted third light in these systems may originate from unresolved field-star contamination rather than from a physically bound third component. Since $l_3$ was included as a free parameter in the light-curve solutions, such approximately constant additional flux is accounted for directly in the binary modelling. Nevertheless, the fitted $l_3$ values should primarily be interpreted as the total unresolved additional flux within the TESS aperture rather than as direct evidence for a tertiary component. As third light was explicitly included in the light-curve modelling, its contribution is incorporated into the adopted binary solutions. Therefore, the possible field-star contamination does not invalidate the binary modelling, although the usual covariance between $l_3$ and parameters affecting the eclipse depths should be kept in mind.

\begin{table*}
\scriptsize
\centering
\caption{Parameters obtained from the binary light curve modeling. Subscripts 1 and 2 refer to the primary and secondary components, respectively. The values of $T_{\rm eff1}$ and $q$ were adopted as fixed parameters.}
\label{table3}
\begin{tabular}{l@{\hskip 3pt}c@{\hskip 3pt}r@{\hskip 3pt}r@{\hskip 3pt}r@{\hskip 3pt}r@{\hskip 3pt}r@{\hskip 3pt}c@{\hskip 3pt}c@{\hskip 3pt}c@{\hskip 3pt}r@{\hskip 3pt}r@{\hskip 3pt}r}
\hline
 Star & $i$     & $T_{\rm eff_1}$  & $T_{\rm eff_2}$  & $\Omega_1$ & $\Omega_2$ & Phase &        $q$ & $r_{1,mean}$ & $r_{2,mean}$ & $l_1/(l_1+l_2)$ & $l_2/(l_1+l_2)$ &     $l_3$ \\
  & ($^\circ$)  &  (K)         &  (K)         &              &          & shift &            &              &              &  &  &     \\
 \hline
XZ Aql & 85.00 (1) & 7800 (300) & 4343 (70) &  4.324 (9) &  2.193 (4) & -0.0052 (4) &  0.184 (4) &   0.2433 (6) &  0.2442 (15) & 0.874 (1) & 0.127 (1) & -   \\
 Y Cam & 85.83 (3) & 7700 (300) & 4488 (95) &  4.302 (3) &  2.324 (4) & -0.0029 (4) & 0.238 (10) &   0.2481 (2) &  0.2621 (15) & 0.837 (2) & 0.163 (2) & -   \\
AB Cas & 89.81 (13) &7700 (300) & 4509 (110) &  3.950 (4) &  2.188 (4) & 0.0010 (2) &  0.182 (2) &   0.2680 (3) &  0.2434 (15)& 0.872 (2) & 0.128 (2) & -   \\
RZ Cas & 82.39 (2) & 8250 (300) & 4334 (80) & 4.620 (17) &  2.576 (4) & -0.0016 (4) &  0.351 (2) &   0.2362 (9) &  0.2912 (15) & 0.844 (3) & 0.156 (6) & -   \\
XX Cep & 83.32 (2) & 8300 (300) & 4514 (70) &  4.712 (6) &  2.106 (4) &  0.0007 (4) &  0.151 (2) &   0.2204 (3) &  0.2309 (15) & 0.863 (2) & 0.137 (2) & 0.060 (1) \\
WY Cet & 82.36 (2) & 7000 (300) & 4275 (85) &  4.220 (3) &  2.365 (4) &  0.0025 (4) & 0.250 (20) &   0.2542 (2) &  0.2621 (15) & 0.841 (3) & 0.159 (3) & -   \\
 R Cma & 84.35 (1) & 6900 (300) & 4600 (120) &  3.400 (3) &  2.043 (4) & -0.0005 (4) &  0.128 (3) &   0.3100 (3) &  0.2205 (15)& 0.883 (1) & 0.117 (1) & 0.128 (2) \\
UW Cyg & 88.66 (5) & 7700 (300) & 4306 (90) &  5.647 (9) &  2.056 (4) & -0.0003 (4) &  0.133 (12) & 0.1667 (15) &  0.2228  (15) & 0.824 (1) & 0.176 (1) & 0.040 (1) \\
BW Del & 78.32 (1) & 7500 (300) & 4199 (80) &  4.152 (9) &  2.058 (4) & -0.0046 (4) & 0.135 (20) &   0.2507 (6) &  0.2230 (15) & 0.903 (2) & 0.097 (2) & -   \\
SX Dra & 84.83 (3) & 7100 (300) & 4309 (105) & 7.026 (16) &  2.643 (4) &  0.0052 (4) &  0.383 (8) &   0.1509 (4) &  0.2980 (15)& 0.591 (1) & 0.409 (1) & 0.031 (1) \\
TW Dra & 86.89 (6) & 7700 (300) & 4240 (75) & 5.084 (15) &  2.733 (4) &  0.0013 (4) & 0.427 (11) &   0.2162 (7) &  0.2300 (15) & 0.790 (1) & 0.210 (1) & 0.124 (1) \\
TZ Eri & 87.18 (6) & 7600 (300) & 4353 (85) & 6.246 (16) &  2.185 (4) & -0.0002 (4) &  0.181 (8) &   0.1653 (4) &  0.2430 (15) & 0.749 (1) & 0.252 (5) &  -   \\
CT Her & 83.21 (27) &7800 (300) & 4096 (85) & 4.288 (50) &  2.019 (4) & -0.0031 (4) & 0.120 (20) &  0.2415 (29) &  0.2163 (15) & 0.920 (2) & 0.080 (2) & 0.112 (7) \\
RX Hya & 82.70 (28) &7600 (300) & 4066 (50) & 5.437 (60) &  2.318 (4) &  0.0005 (4) &  0.235 (8) &  0.1930 (42) &  0.2613 (15) & 0.829 (6) & 0.172 (6) &   -  \\
Y Leo & 85.90 (10) & 7600 (300) & 3665 (60) & 4.804 (29) &  2.532 (4) & -0.0136 (4) &  0.330 (9) &  0.2250 (15) &  0.2865 (15) & 0.887 (4) & 0.113 (4) &   - \\
FR Ori & 81.11 (2) & 8000 (300) & 4435 (95) &  3.811 (8) &  2.950 (4) &  0.0055 (4) &  0.539 (9) &   0.3131 (8) &  0.3255 (15) & 0.860 (1) & 0.140 (1) & 0.139 (2) \\
AC Tau & 83.60 (3) & 7700 (300) & 4260 (95) &  4.745 (4) &  2.606 (4) &  0.0032 (4) &  0.365 (8) &   0.2300 (2) &  0.2942 (15) & 0.818 (1) & 0.182 (1) &   - \\
\hline
\end{tabular}
\end{table*}

The fundamental physical parameters of the component stars, such as $M$, $R$, luminosity ($L$), bolometric ($M_{\rm bol}$) and absolute magnitudes ($M_{\rm V}$) were calculated from the quantities obtained from the binary modeling. In order to calculate these parameters, the widely used Kepler and Pogson equations were applied, together with the bolometric correction (BC) from the study by \citet{2020MNRAS.496.3887E}. For the double-lined spectroscopic systems, the $M$ values were directly calculated from the $M\sin^3 i$ values obtained from the radial-velocity solutions. In contrast, for the other system the $M$ values of the primary components were estimated using the empirical $T_{\rm eff}$\,–\,$M$ relation presented by \citet{2018MNRAS.479.5491E}. Therefore, particular emphasis was placed on the accurate determination of the $T_{\rm eff}$ in Sect.\,3. The calculated physical parameters of the component stars are presented in Table\,\ref{Table3}. The uncertainties in the fundamental parameters were estimated by propagating the errors of the derived quantities listed in Table\,\ref{table3} together with the uncertainties of the BC values, and were computed as their quadratic sum.


\begin{table*}
\scriptsize
\centering
\caption{Fundamental parameters of targets derived from the light curve modeling. Subscripts 1 and 2 represent the primary and secondary components, respectively.}\label{Table3}
\begin{tabular}{lrrrrrrrrrrrr p{3.3cm}}
\hline
Star  & $M_1$ & $M_2$  & $R_1$    & $R_2$  & log($L_1/L_\odot$) & log($L_2/L_\odot$) & log$g_1$  & log$g_2$ & $M_{bol1}$ & $M_{bol2}$ & $M_{V1}$ & $M_{V2}$\\ 
      & $(M_\odot)$ & $(M_\odot)$ & $(R_\odot)$ & $(R_\odot)$  &    & & &  &  (mag) & (mag) & (mag) &(mag) \\
\hline
XZ Aql & 2.44(1) & 0.45(1) &  2.42(1) & 2.43(1) &  1.29(7) &  0.28(3) & 4.06(1) & 3.32(1) & 1.51(1) &  4.05(1) & 1.47(3) &  4.71(3) \\
 Y Cam & 2.04(2) & 0.49(2) &  3.15(3) & 3.33(2) &  1.50(7) &  0.61(4) & 3.75(1) & 3.08(2) & 1.00(2) &  3.22(1) & 0.95(4) &  3.78(3) \\
AB Cas & 2.01(4) & 0.37(3) &  1.85(2) & 1.68(4) &  1.04(7) &  0.02(4) & 4.21(1) & 3.55(4) & 2.15(2) & 4.68(5) & 2.10(4) & 5.23(6) \\
RZ Cas & 1.96(1) & 0.69(2) &  1.55(1) & 1.91(1) &  0.10(6) &  0.06(3) & 4.35(1) & 3.71(1) & 2.24(1) &  4.58(1) & 2.25(3) &  5.26(3) \\
XX Cep & 2.21(1) & 0.33(2) &  2.23(3) & 2.34(2) &  1.33(6) &  0.31(3) & 4.09(1) & 3.22(3) & 1.42(3) &  3.97(2) & 1.43(4) &  4.51(3) \\
WY Cet & 1.60(8) & 0.40(4) & 2.21(0) & 2.28(1) & 1.02(7) & 0.19(4) & 3.95(2) & 3.32(4) & 2.18(2) & 4.25(9) & 2.10(2) & 4.99(9) \\
 R Cma & 1.60(1) & 0.20(2) &  1.73(1) & 1.23(1) &  0.78(8) & -0.22(5) & 4.17(1) & 3.57(4) & 2.78(1) & 5.28(2) & 2.70(3) & 5.76(3) \\
UW Cyg & 1.80(9) & 0.24(1) & 2.04(18) & 2.73(2) & 1.12(10) & 0.36(4) & 4.07(8) & 2.94(3) & 1.94(3) & 3.83(9) & 1.89(3) & 4.53(9) \\
BW Del & 1.74(9) & 0.23(4) & 2.35(1) & 2.09(1) & 1.20(7) & 0.09(3) & 3.94(2) & 3.17(7) & 1.75(2) & 4.52(8) & 1.69(2) & 5.34(8) \\
SX Dra & 1.63(9) & 0.62(3) & 2.48(1) & 4.89(2) & 1.15(7) & 0.87(4) & 3.86(2) & 2.85(2) & 1.87(2) & 2.56(11) & 1.79(2) & 3.26(11) \\
TW Dra & 2.17(2) & 0.93(2) &  2.64(1) & 2.81(2) &  1.34(7) &  0.36(3) & 3.93(1) & 3.51(1) & 1.38(1) &  3.84(2) & 1.33(3) &  4.56(3) \\
TZ Eri & 1.77(1) & 0.32(1) &  1.68(1) & 2.47(2) &  0.93(7) &  0.30(3) & 4.23(1) & 3.16(2) & 2.42(1) &  4.00(2) & 2.36(3) &  4.65(4) \\
CT Her & 1.83(12) & 0.22(4) & 1.98(2) & 1.77(1) & 1.12(9) & -0.10(4) & 4.11(3) & 3.28(8) & 1.95(2) & 4.99(9) & 1.91(2) & 5.91(9) \\
RX Hya & 1.77(9) & 0.42(2) & 1.84(4) & 2.49(1) & 1.01(7) & 0.19(2) & 4.15(3) & 3.26(2) & 2.22(2) & 4.28(5) & 2.16(2) & 5.23(5) \\
Y Leo & 1.77(9) & 0.58(3) & 1.94(1) & 2.47(1) & 1.05(7) & -0.003(3) & 4.11(2) & 3.42(2) & 2.11(2) & 4.75(7) & 2.05(2) & 6.16(7) \\
FR Ori & 1.89(9) & 1.02(5) & 1.68(0) & 1.74(1) & 1.02(7) & 0.03(4) & 4.26(2) & 3.96(2) & 2.20(2) & 4.68(9) & 2.18(2) & 5.27(9) \\
AC Tau & 1.80(9) & 0.66(4) & 2.03(0) & 2.60(1) & 1.12(7) & 0.30(4) & 4.08(2) & 3.42(2) & 1.95(2) & 3.99(10) & 1.90(2) & 4.74(10) \\
\hline
\end{tabular}
\end{table*}

\section{Orbital Period Variation Analysis}

\subsection{Determination of Times of Minimum}

Before the orbital-period analysis we read the new minimum times from the TESS data to include them in the current O-C diagrams of our targets. Based on all available TESS sector data listed in Table\,\ref{table1}, one to three minima were measured from each sector light curve. To determine the times of minimum, we first normalized each light curve and interactively selected a time window around the eclipses. Within this window, the local dip was modeled with a Gaussian profile using a non-linear least-squares fitting procedure. The center of the fitted Gaussian was adopted as the minimum time. All fits were visually inspected, and the resulting minima together with their uncertainties from the fit covariance were recorded for further O–C analysis.

Since the TESS light curves are provided in Barycentric Julian Date (BJD),and most historical minimum times reported in the literature spanning several decades are given in Heliocentric Julian Date (HJD), we converted the TESS minimum times to HJD to maintain epoch-long homogeneity in our $O-C$ analysis. Considering the analyzed orbital period variations and collective $O-C$ shifts are on the order of hours, the slight difference between barycentric and heliocentric time scales (at most a few seconds) is physically negligible and does not affect the derived $O-C$ trends or system parameters. The transformation corrects for the difference between the barycentric and heliocentric reference frames by subtracting the light-travel time between the Sun and the Solar System barycenter projected along the line of sight to the target. For this purpose, we applied the standard barycentric-to-heliocentric correction following the procedure described by \citet{2010PASP..122..935E}. The measured and HJD converted minimum times are provided in Table\,\ref{tab:tab_ap_min}.

\subsection{O-C Analysis}
Four main factors contribute to orbital period variation in binary star systems are mass transfer between components, apsidal motion, magnetic activity, and the presence of a third body. Of these, mass transfer and magnetic activity creates a real physical change in the orbital period. In semi-detached Algol-type systems, long-term monotonic orbital period variations are commonly interpreted as signatures of mass transfer between the components. In addition, systemic mass loss from the system may be observed (e.g. \cite{1980AA....83..217M,2001ApJ...552..664N,2011AA...528A..16V}). These mass transfer processes and mass losses result in genuine orbital period variations, which appear as a parabolic trend in the O-C diagram. The orbital period variation and the annual mass transfer rate can be determined by modeling the O-C curve, and the mass transfer rate can be calculated using the derived parameters.

\begin{figure}
  \centering
  \begin{minipage}[t]{0.85\linewidth}
    \centering\includegraphics[alt={O–C diagram for WY Cet Del showing observed period variations, model curves, and residuals.},width=\linewidth]{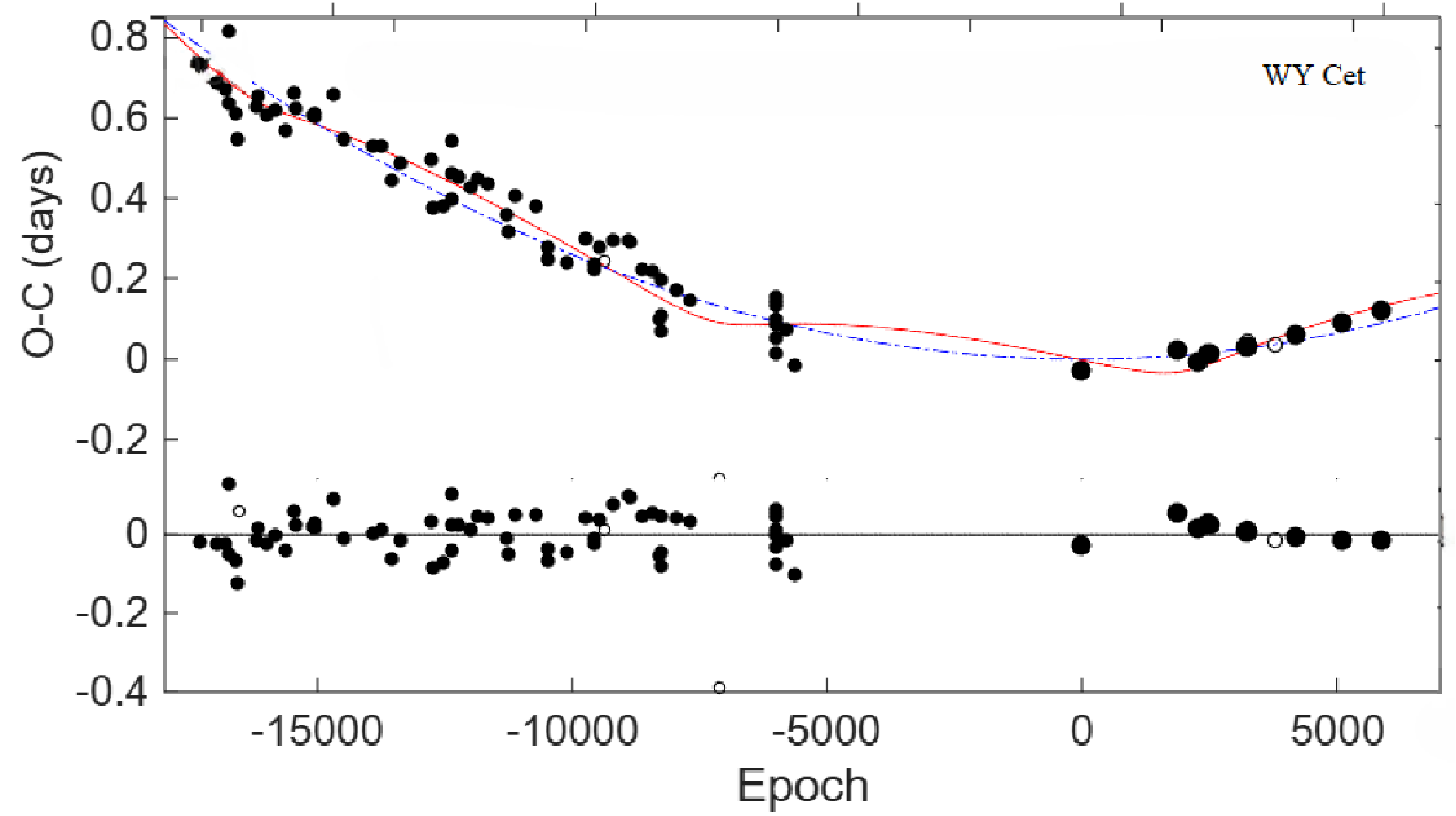}
  \end{minipage}\hfill
  \begin{minipage}[t]{0.85\linewidth}
    \centering\includegraphics[alt={O–C diagram for BW Del showing observed period variations, model curves, and residuals.},width=\linewidth]{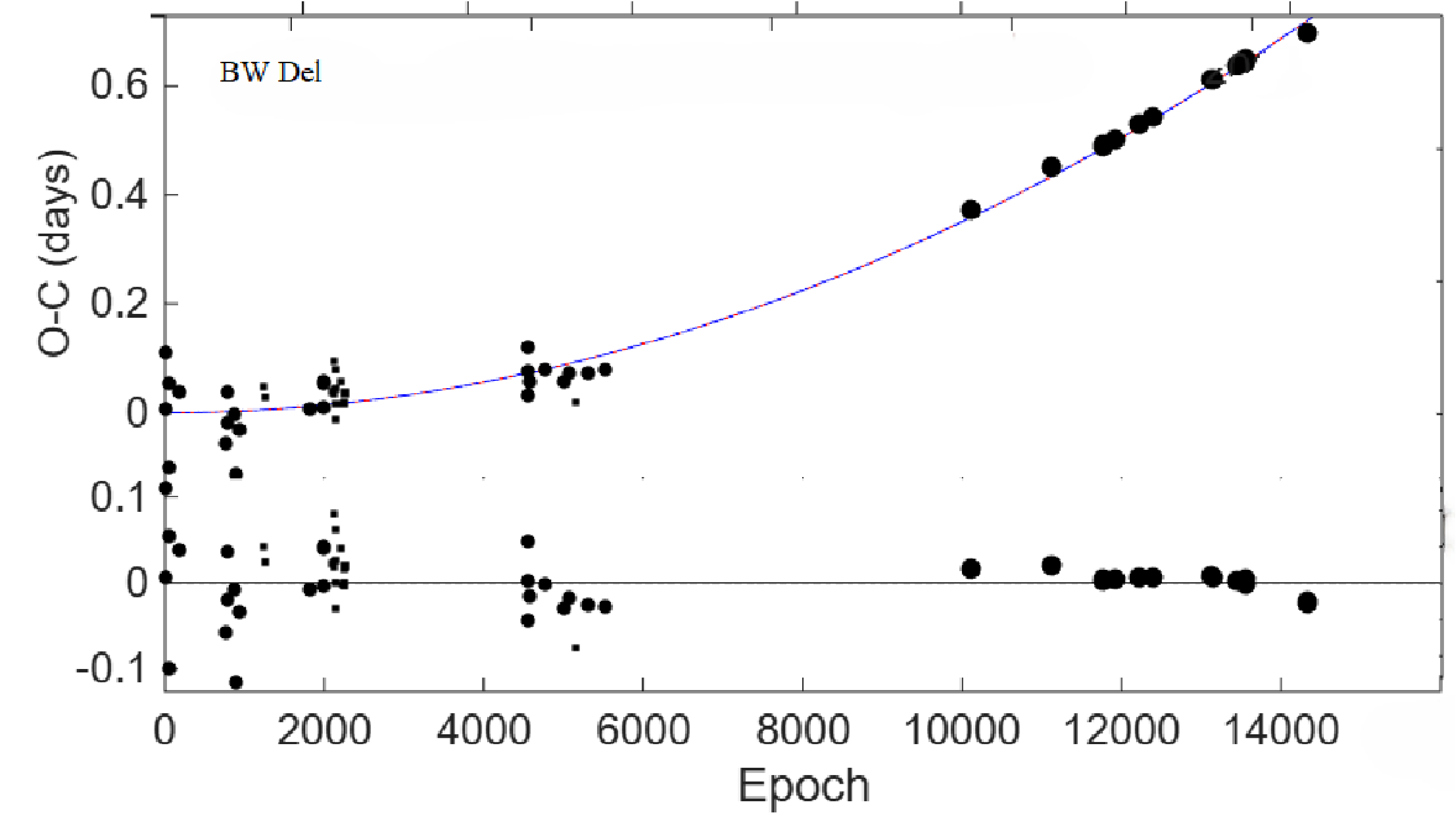}
  \end{minipage}
  \caption{O–C diagram showing cyclic period variations for two of our targets. The blue dashed lines represent the upward parabolic variation, while the red lines represent the combined cyclic variation models.}
  \label{fig:fig_oc1}
\end{figure}

In this study, we selected 17 Algol-type eclipsing binaries whose O-C curves exhibit a clear upward parabolic trend indicating ongoing mass transfer from the evolved secondary to the pulsating primary component.Our primary aim is to estimate mass-transfer indicators by modelling the long-term parabolic variations observed in the O-C diagrams and to characterize any additional cyclic variations present in the residuals.

As a first step, we applied a second-order polynomial fit to characterize the main variation in the O-C curve for all selected systems using the following equation.

\vspace{0.5em}
$(O-C) = Q\,\times\,E^2 + \Delta P\,\times\,E + \Delta T_0$
\vspace{0.5em}

In this equation, Q is the coefficient that represents the orbital period change rate due to mass transfer, $\Delta P$ denotes the orbital period change, and $\Delta T_0$ accounts for the correction of the ephemeris time $T_0$. The orbital period variation rate is calculated using the Q parameter in the following equation \citep{2006MNRAS.373..435I}:

\vspace{0.5em}
$\mathrm{d}P/\mathrm{d}t = 2\,\times\,Q\,\times\,365.25/P^2$
\vspace{0.5em}


The rate of orbital-period variation allows the annual mass–transfer rate to be inferred. We therefore computed $dM/dt$ under both conservative and non-conservative assumptions, following the procedure of \citet{2014MNRAS.441.1166E}, using the parameters listed in Table~\ref{Table3}; the resulting values are given in Table~\ref{table4}, and the modeled O-C variations for all targets are shown in Fig.\,\ref{fig:fig_oc1} and \ref{fig:ap}. As a guide, we also adopt a representative mass-loss rate of $\mathrm{d}M/\mathrm{d}t\,=\,10^{-8}\,M_\odot\,\mathrm{yr}^{-1}$ for our Algol-type systems with A-F ($\sim$7000--8000\,K) $\delta$~Scuti accretors. This value lies within the typical range observed in classical Algols and, for representative component masses $M_1\simeq 2\,M_\odot$ and $M_2\simeq 0.8\,M_\odot$, implies $\mathrm{d}P/\mathrm{d}t\, \approx 2\times 10^{-8}\,\mathrm{yr}^{-1}$ (i.e., a few $\mathrm{ms}\,\mathrm{yr}^{-1}$ for $P_{\rm orb}\sim 1\text{--}3\,\mathrm{d}$), consistent with the upward parabolic curvature of our O-C diagrams \citep{2006MNRAS.373..435I}. 

The aim of this study regarding O-C analysis is primarily to determine the fundamental parameters of mass transfer. However, the O-C diagrams clearly show that, in addition to the parabolic variation caused by mass transfer, most systems exhibit a cyclic variation. To enhance the reliability of the analysis, this cyclic change was taken into account. Except for UW Cyg, BW Del, FR Ori, and AC Tau, this cyclic variation was included in the analysis for 13 systems (for these four systems, only the parabolic variation was examined).

Orbital period's cyclical change can be attributed either to magnetic activity (Applegate hypothesis, \cite{1992ApJ...385..621A}) or the presence of a third body (Light-Time Effect, LITE \cite{1959AJ.....64..149I}). The analyses were carried out using the method provided by \citet{2009NewA...14..121Z}. From these analyses, we obtained the amplitude ($A_{LITE}$) and period ($P_{LITE}$) of the cyclic variation, as well as the eccentricity ($e_3$) and periastron longitude ($\omega_3$) of the third body. All of these parameters are given in the Appendix (Table\,\ref{tab:a6}). Using these results, the magnetic field strength and the mass function of the third body were calculated following the methods given by \citet{1992ApJ...385..621A} and \citet{1990BAICz..41..231M}.

According to the results listed in Table\,\ref{tab:a6}, although R CMa ($l_3 = 0.128, f(m_3) = 0.054$) has a low mass function, its third-light contribution is around 13\%. On the other hand, SX Dra ($l_3 = 0.031, f(m_3) = 0.268$) has a relatively low third-light contribution but a quite high mass function. Therefore, the presence of a third body may be considered for these two systems. For the other systems with high mass functions\,$-$\,XZ Aql  ($f(m_3) = 0.134$), Y Cam ($f(m_3) = 0.316$), WY Cet ($f(m_3) = 0.168$), TZ Eri ($f(m_3) = 0.216$), and Y Leo ($f(m_3) = 0.146$)\,$-$\,no third-light contribution was found in the light curve analysis. Considering the relationship between magnetic activity and mass transfer (RZ Cas is most important example for this relation, \cite{2020AA...644A.121L}) in oEA-type systems, as well as the spectral types of the secondary components (mostly K-type), the cyclic variations observed in these four systems and the remaining ones can be explained with the Applegate hypothesis. Additionaly, the uncertainty in the parameters obtained from analysis for the TW Dra system is quiet high. Therefore, the calculated magnetic field strength and third body mass function values are unreliable.




\begin{table*}
\begin{center} 
    \caption{Results of the orbital period analysis. The subscripts ‘con’ and ‘ncon’ indicate the results for the conservative and non-conservative assumptions.}\label{table4}
    \begin{tabular}{lrrrrrr}
    \\
    \hline
    Star & $T_0$       & $P_{orb}$  &  Q               & dP/dt   & $dM/dt_{con}$ & $dM/dt_{ncon}$\\
         &  (HJD)      & (day)           & ($10^{-10} day$) & (s/yr)  & ($M_{sun}$/yr) & ($M_{sun}$/yr) \\
    \hline
XZ Aql & $2441903.4382\,(38)$ & $2.139186\,(1)$ & $26.00\,(2)$ & $0.0767\,(1)$ & $7.611\,(45)\times10^{-8}$ & $7.668\,(45)\times10^{-8}$ \\
Y Cam  & $2445259.3947\,(36)$ & $3.305651\,(2)$ & $192.11\,(16)$ & $0.3668\,(3)$ & $2.726\,(22)\times10^{-7}$ & $2.733\,(22)\times10^{-7}$ \\
AB Cas & $2446849.2773\,(3)$  & $1.366883\,(0)$ & $6.03\,(1)$    & $0.0278\,(1)$ & $1.851\,(46)\times10^{-6}$ & $1.861\,(31)\times10^{-6}$ \\
RZ Cas & $2452500.5579\,(8)$  & $1.195253\,(0)$ & $2.34\,(1)$    & $0.0124\,(1)$ & $4.218\,(29)\times10^{-8}$ & $4.333\,(30)\times10^{-8}$ \\
R Cma  & $2445391.2393\,(23)$ & $1.135945\,(0)$ & $1.65\,(1)$    & $0.0092\,(1)$ & $1.226\,(9)\times10^{-8}$  & $1.383\,(11)\times10^{-8}$ \\
XX Cep & $2444839.7823\,(26)$ & $2.337320\,(1)$ & $27.05\,(1)$   & $0.0731\,(1)$ & $7.574\,(45)\times10^{-8}$ & $7.629\,(45)\times10^{-8}$ \\
WY Cet & $2448501.0938\,(141)$& $1.939730\,(3)$ & $26.02\,(1)$   & $0.0847\,(1)$ & $8.967\,(11)\times10^{-8}$ & $9.043\,(11)\times10^{-8}$ \\
UW Cyg & $2452503.3573\,(95)$ & $3.450779\,(2)$ & $14.53\,(152)$ & $0.0266\,(28)$& $8.190\,(9)\times10^{-9}$  & $8.613\,(9)\times10^{-9}$ \\
BW Del & $2425795.3891\,(362)$& $2.423101\,(5)$ & $35.05\,(370)$ & $0.0913\,(96)$& $3.945\,(4)\times10^{-8}$ & $3.988\,(4)\times10^{-8}$ \\
SX Dra & $2444705.5111\,(103)$& $5.169436\,(3)$ & $382.50\,(22)$ & $0.4670\,(3)$ & $3.517\,(20)\times10^{-7}$ & $3.530\,(20)\times10^{-7}$ \\
TW Dra & $2433310.3996\,(60)$ & $2.806757\,(1)$ & $88.43\,(1)$   & $0.1989\,(1)$ & $4.428\,(15)\times10^{-7}$ & $4.437\,(15)\times10^{-7}$ \\
TZ Eri & $2452502.5354\,(24)$ & $2.606124\,(1)$ & $20.95\,(1)$   & $0.0507\,(1)$ & $2.939\,(17)\times10^{-8}$ & $2.994\,(17)\times10^{-8}$ \\
CT Her & $2452501.6102\,(18)$ & $1.786379\,(1)$ & $7.87\,(1)$    & $0.0278\,(1)$ & $1.496\,(16)\times10^{-8}$ & $1.535\,(17)\times10^{-8}$ \\
RX Hya & $2446530.1960\,(27)$ & $2.281657\,(1)$ & $31.39\,(4)$   & $0.0868\,(1)$ & $7.974\,(9)\times10^{-8}$  & $8.046\,(93)\times10^{-8}$ \\
Y Leo  & $2437368.5037\,(15)$ & $1.686086\,(0)$ & $6.78\,(1)$    & $0.0254\,(1)$ & $5.060\,(22)\times10^{-8}$ & $5.166\,(22)\times10^{-8}$ \\
FR Ori & $2427862.1720\,(189)$& $0.883159\,(1)$ & $1.10\,(18)$   & $0.0079\,(13)$& $7.570\,(12)\times10^{-8}$ & $7.800\,(13)\times10^{-8}$ \\
AC Tau & $2425889.6512\,(361)$& $2.043330\,(4)$ & $20.20\,(123)$ & $0.0624\,(38)$& $1.217\,(7)\times10^{-7}$  & $1.229\,(7)\times10^{-7}$ \\
    \hline
    \end{tabular}
\end{center}
\end{table*}

\section{Frequency Analysis}

Since our aim was to investigate possible relationship between the calculated mass transfer indicators ($Q$, $dP/dt$, and $dM/dt$) and the pulsation properties ($P_{\rm puls}$ and $Amp$), the pulsation analysis was performed using \textit{TESS} data to ensure a homogeneous set of parameters across the sample.

Our goal was to determine pulsation frequencies ($f$) and corresponding $Amp$ values. For this purpose, we analyzed the TESS light curves using the Period04 program \citep{2005CoAst.146...53L}, which applies the Fourier transform method to determine individual pulsation frequencies. In the first step of the analysis, the binary contribution was removed from the light curves following the procedure described by \citet{2022MNRAS.510.1413K}. Subsequently, frequency analysis was performed on the residual light curves. Since our aim was to investigate the impact of mass-transfer on the pulsation, rather than performing an exhaustive multi-sector frequency analysis, using a single \textit{TESS} sector was sufficient for reliably determining the highest-amplitude frequency. A more detailed frequency study lies beyond the scope of the present work. 

\begin{figure}
\begin{center}
\includegraphics[alt={Two-panel frequency spectra plot with 4.5 sigma significance level for BW Del and FR Ori },width=85mm]{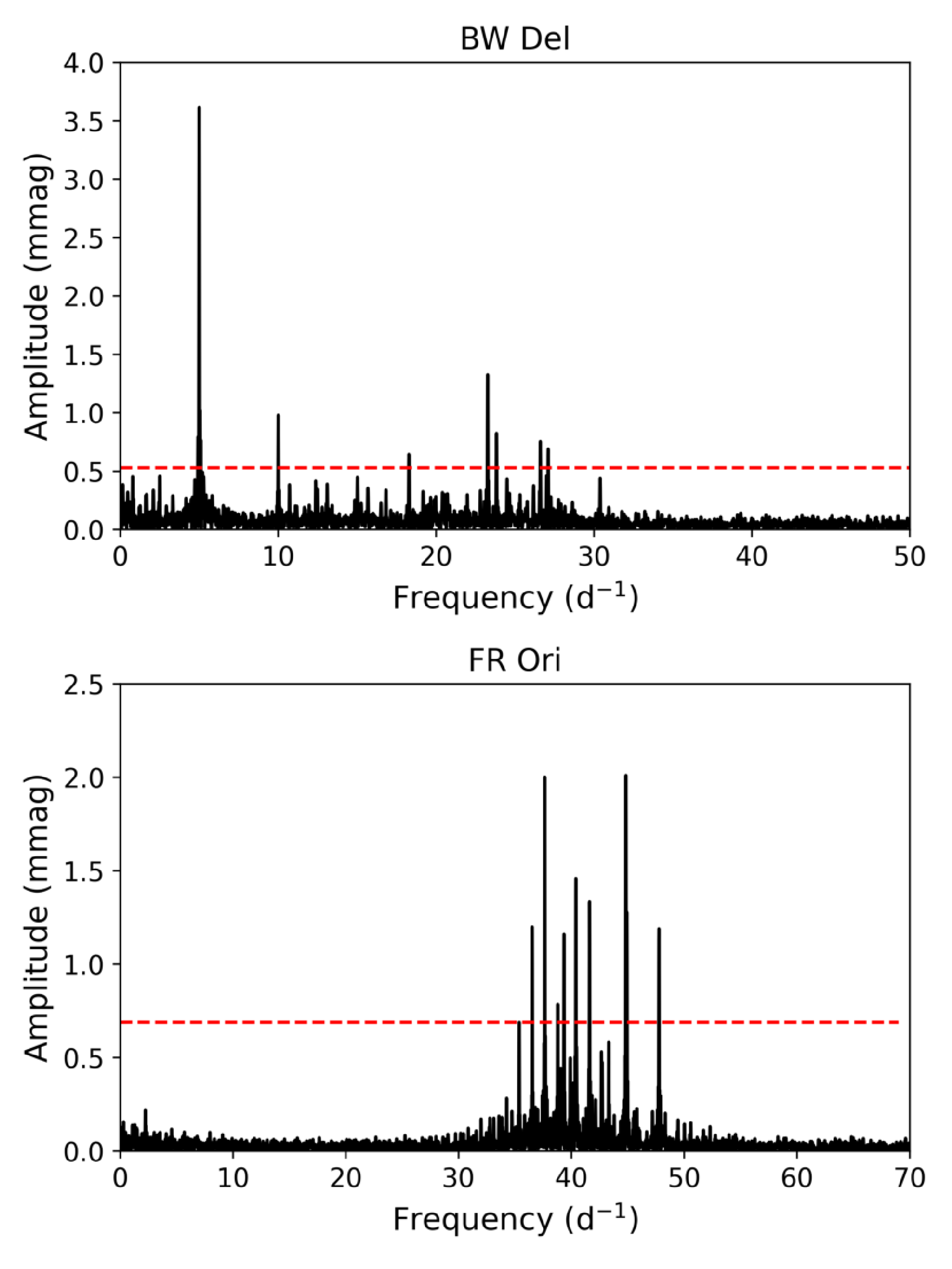}
\end{center}
\caption{Frequency spectra of two of the targets. Red dashed lines represents 4.5 $\sigma$ level.}\label{fig:frekans}
\end{figure}

To detect $\delta$ Scuti-type pulsations, frequency scanning was conducted in the range of 4–80\,d$^{-1}$, and a signal-to-noise ratio (SNR) threshold of 4.5$\sigma$ was adopted as the significance criterion \citep{2021AcA....71..113B}.The five strongest pulsation frequencies and their amplitudes for each target are summarized in Table\,\ref{tab:all_frequencies}, and example frequency spectra for two representative systems are displayed in Fig\,\ref{fig:frekans}.

For the frequencies listed in Table\,\ref{tab:all_frequencies}, the pulsation constant ($Q{_p}$) values were calculated following the method of \cite{1990DSSN....2...13B}. Based on these $Q{_p}$ values, the pulsation modes were classified using the criteria outlined by \citet{2024AA...690A.104D} and \citet{1981ApJ...249..218F}. While for some stars (e.g., AB Cas, TZ Eri, WY Cet) the dominant frequency (highest-amplitude frequency) corresponds to the fundamental mode, and for some it corresponds to the overtones. The full set of identified frequencies across the sample does not allow secure mode identification for every target. This is primarily due to the limited mode resolution in some light curves and the intrinsic complexity of pulsation spectra in mass-transferring systems. Interestingly, the distribution of the derived $Q{_p}$ values shows an apparent dichotomy. While some dominant modes have $Q_p$ values close to the fundamental radial-mode value of $\sim0.03$ d, many cluster around $Q{_p}\sim0.015$ d. The latter is approximately half the fundamental-mode pulsation constant and resembles the “second ridge” identified in the period–luminosity and period–density distributions of $\delta$ Scuti stars by \citet{2022MNRAS.516.2080B}, which occurs at periods approximately half those of the fundamental-mode ridge. \citet{2022MNRAS.516.2080B} noted that these periods are more consistent with third- or fourth-overtone pulsations than with the first overtone. Although the similarity is intriguing, a direct identification with the second-ridge population would require a dedicated period–luminosity analysis and is beyond the scope of the present study.

Among all detected modes, the frequency with the highest amplitude is the most likely to respond measurably to such global structural changes, as it carries the dominant contribution to the observable light variation. Therefore, to ensure consistency and avoid speculative mode assignments, our mass-transfer correlation analysis was performed exclusively using the highest-amplitude (dominant) frequency for each system, rather than attempting to interpret the full mode spectrum. This approach provides a physically motivated and observationally robust basis for examining the possible link between pulsation behavior and mass-transfer processes.

\section{Discussion}

In order to investigate the possible effects of mass transfer on the pulsation properties, namely on the $P_{\rm puls}$ and $Amp$, we examined potential relationships between the mass-transfer indicators ($Q$, $dP/dt$, $dM/dt$) inferred from the O$-$C analysis and $P_{\rm puls}$ and $Amp$ of the dominant (highest amplitude) frequency. As a first step, we focused on systems for which the $q$ is reliably determined from double-lined spectroscopic radial velocity solutions, as listed in Table\,\ref{tab:tableA2}. For these systems, possible relations between mass transfer indicators and pulsation properties were explored separately. However, due to the limited number of these systems with spectroscopically determined $q$ value in our sample, the resulting correlations do not reveal statistically significant or well-defined trends. These results are presented in Appendix (see Fig.\,\ref{fig:relationship_app}). Subsequently, we extended the analysis to the full sample of systems considered in this study. The correlations obtained for all systems are presented and discussed in the following subsections.

\subsection{Relationship between $Q$, $dP/dt$ and pulsation properties}\label{sec:7.1}

\begin{figure}
  \centering
  \begin{minipage}[t]{0.46\textwidth}
    \centering
    \includegraphics[alt={Scatter plot showing the correlation between $\log Q$ and $\log P_{\text{puls}}$ with a linear fit and $1\sigma$ error band.},width=\linewidth,height=5.4cm]{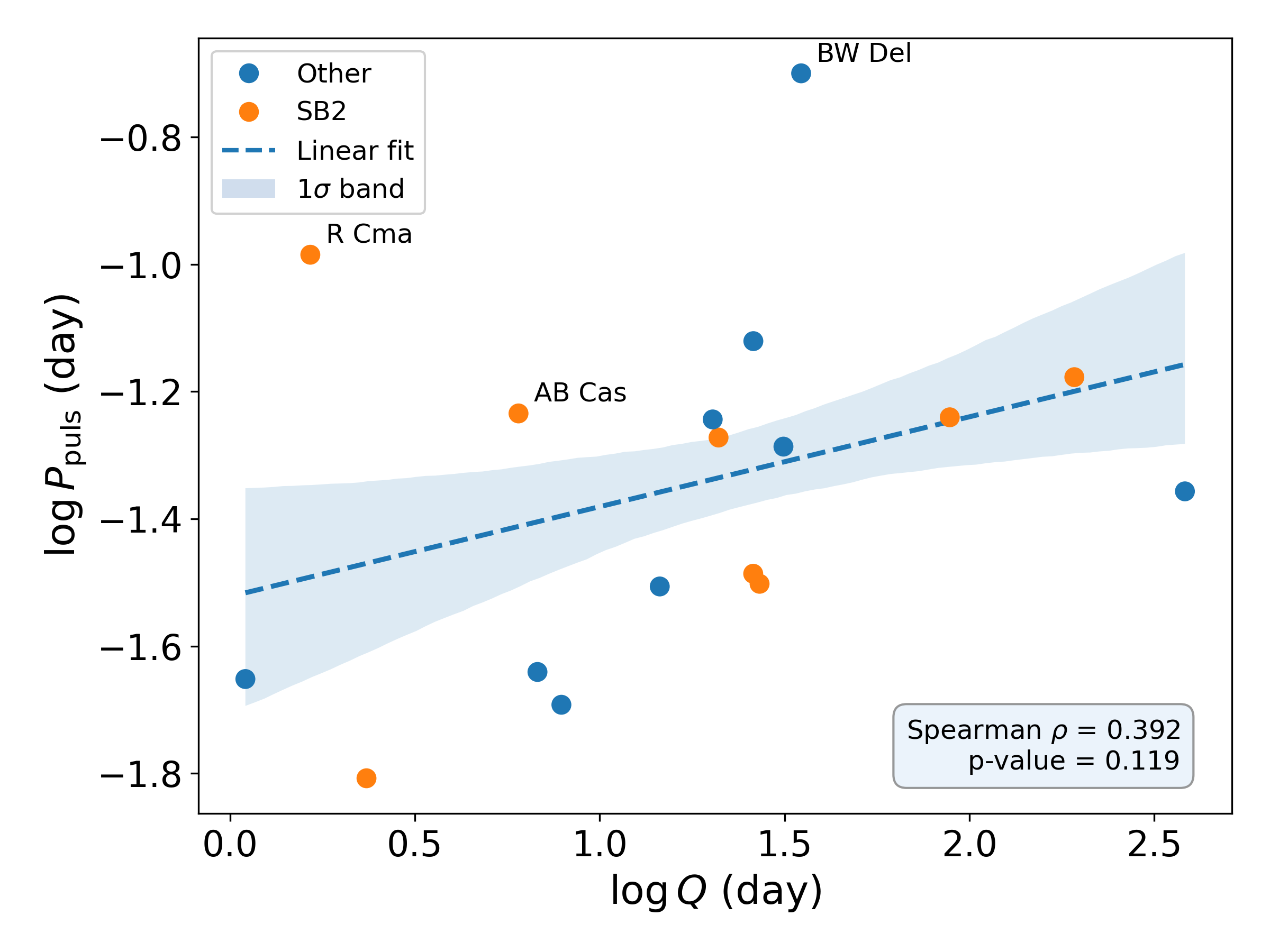}
  \end{minipage}\hfill
  \begin{minipage}[t]{0.46\textwidth}
    \centering
    \includegraphics[alt={Scatter plot showing the correlation between $\log Q$ and $\log \text{Amp}$ with a linear fit and $1\sigma$ error band.},width=\linewidth,height=5.4cm]{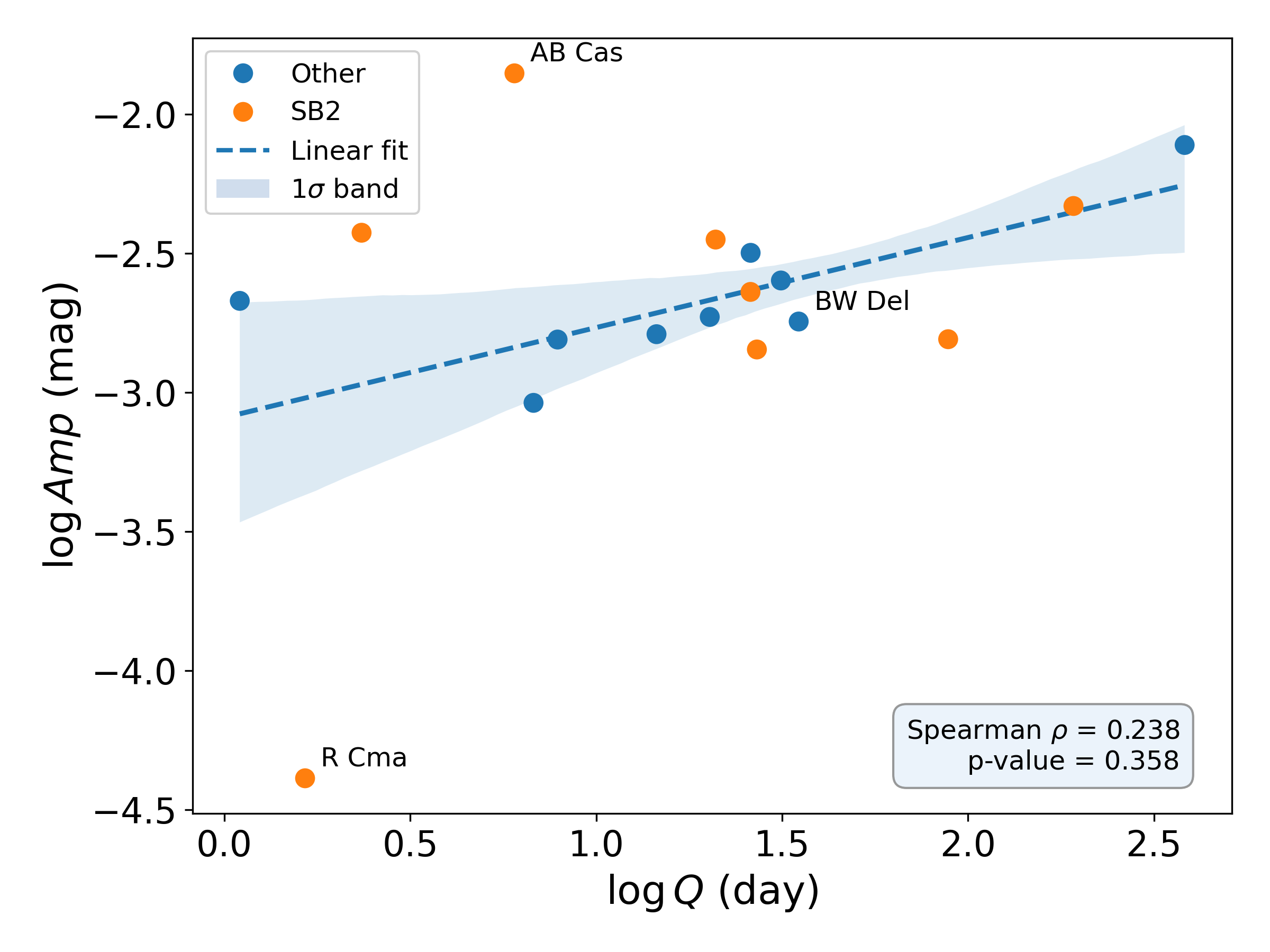}
  \end{minipage}
  \caption{Relations between $\log Q$ and the logarithm of the pulsation period, $\log P_{\mathrm{puls}}$ (left panel), and between $\log Q$ and the logarithmic amplitude, $\log Amp$ (right panel). Systems with double-lined spectroscopic radial velocity solutions (SB2) are shown in orange dots. The dashed lines represent the best-fit linear relation, while the shaded region corresponds to the $\pm1\sigma$ residual standard deviation about the fit. The inset reports the Spearman rank correlation coefficient, $\rho$, together with the corresponding p-value.}
  \label{fig:fig_Q}
\end{figure}

\begin{figure}
  \centering
  \begin{minipage}[t]{0.46\textwidth}
    \centering
    \includegraphics[alt={Scatter plot showing the correlation between $\log dP/dt$ and $\log P_{\text{puls}}$ with a linear fit and $1\sigma$ error band.},width=\linewidth,height=5.4cm]{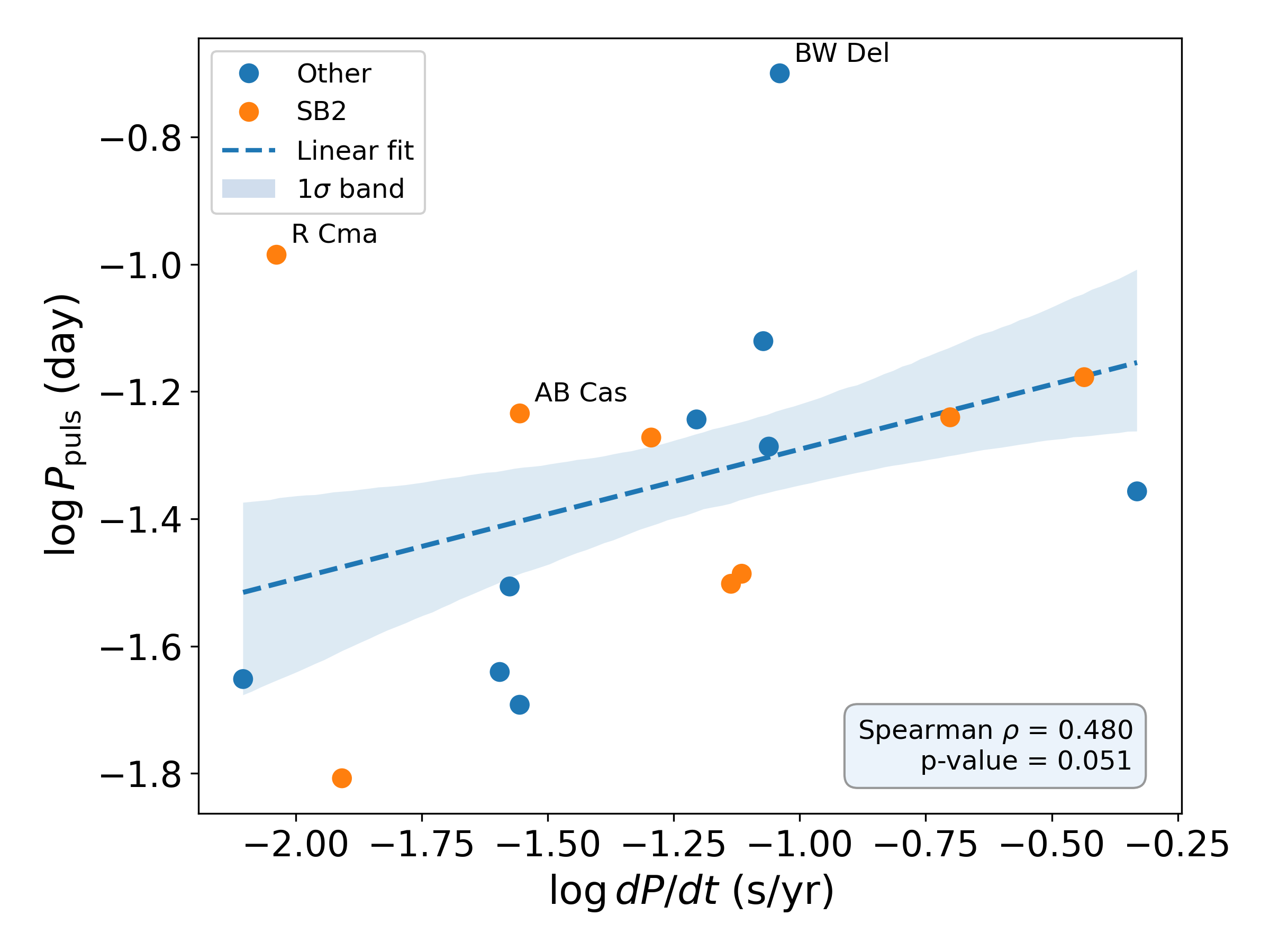}
  \end{minipage}\hfill
  \begin{minipage}[t]{0.46\textwidth}
    \centering
    \includegraphics[alt={Scatter plot showing the correlation between $\log dP/dt$ and $\log \text{Amp}$ with a linear fit and $1\sigma$ error band.},width=\linewidth,height=5.4cm]{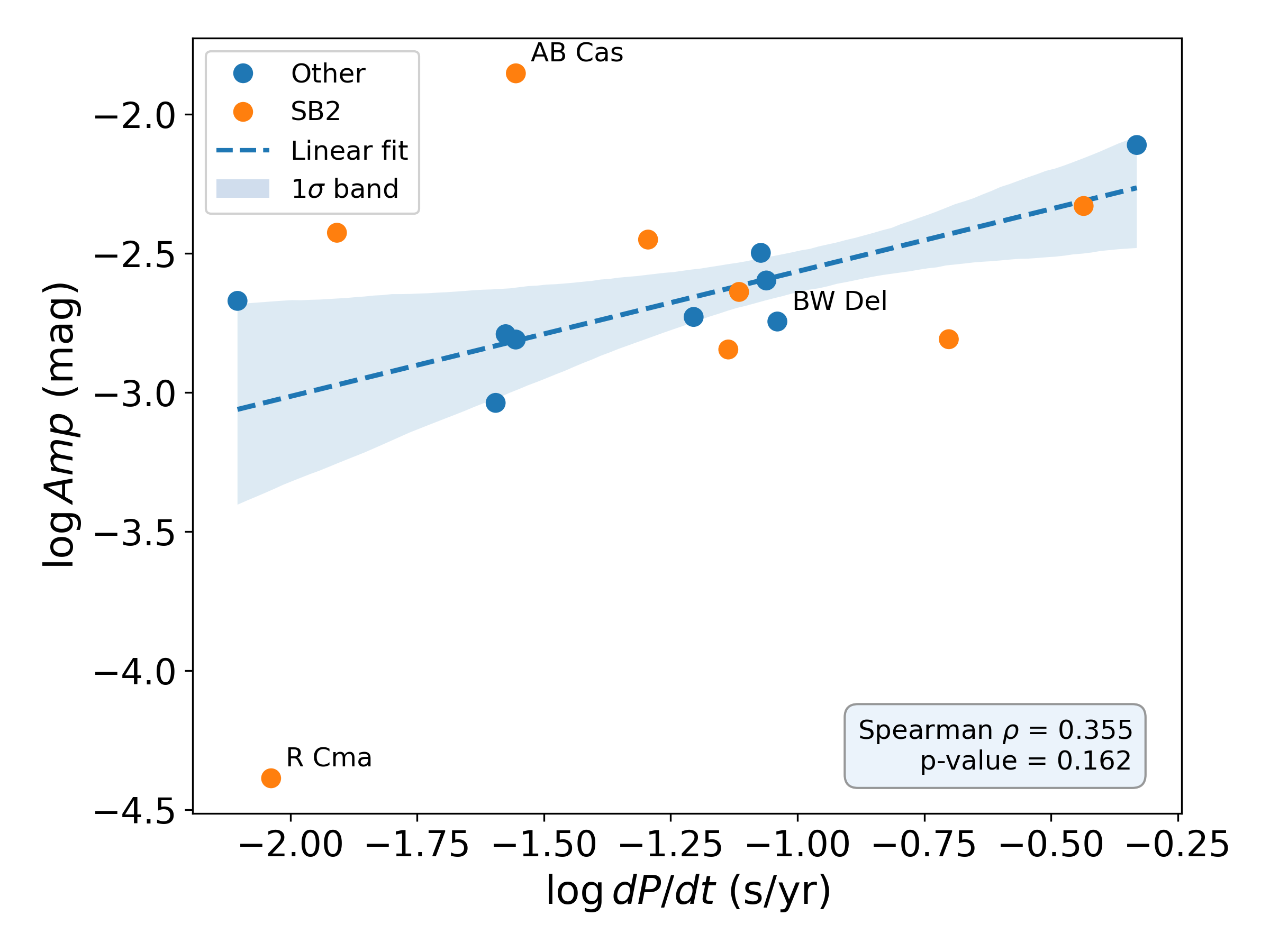}
  \end{minipage}
  \caption{Relations between $dP/dt$ and the logarithm of the pulsation period, $\log P_{\mathrm{puls}}$ (left panel), and between $dP/dt$ and the logarithmic amplitude, $\log Amp$ (right panel). Systems with double-lined spectroscopic radial velocity solutions (SB2) are shown in orange dots. The dashed lines represent the best-fit linear relation, while the shaded region corresponds to the $\pm1\sigma$ residual standard deviation about the fit. The inset reports the Spearman rank correlation coefficient, $\rho$, together with the corresponding p-value. }
  \label{fig:fig_dpdt}
\end{figure}

In Figs.~\ref{fig:fig_Q} and \ref{fig:fig_dpdt}, we examine the relationships between the mass-transfer indicators, $Q$ and $dP/dt$, and the pulsation properties of the primary components ($P_{\rm puls}$ and $Amp$). Since $dP/dt$ is directly linked to $Q$, both parameters are expected to trace similar evolutionary effects; therefore, the corresponding diagrams are considered together.

The $\log Q$–$\log P_{\rm puls}$ and $\log dP/dt$–$\log P_{\rm puls}$ planes show weak positive trends, in the sense that systems with higher mass-transfer indicators tend to exhibit longer dominant pulsation periods. This tendency is reflected in the Spearman coefficients, which remain in the low-to-moderate range (e.g., $\rho \sim 0.39$ for $\log Q$–$\log P_{\rm puls}$ and $\rho \sim 0.48$ for $\log dP/dt$–$\log P_{\rm puls}$), with p-values indicating that the correlations are not statistically significant but it should be noted that $\log dP/dt$–$\log P_{\text{puls}}$ is closer to the statistically significant limit (p < $ 0.05$). A similar, but even weaker, behavior is observed in the amplitude diagrams, where both $\log Q$–$\log Amp$ and $\log dP/dt$–$\log Amp$ relations have lower correlation coefficients. In all panels, the dispersion is considerable, and no tight or well-defined relation is present. Furthermore, restricting the sample to SB2 systems, which provide more reliable mass ratios, does not significantly strengthen the correlations. These results indicate that, although a monotonic tendency may be present, it remains weak and strongly dependent on the sample composition. The lack of tight correlations, especially when considering amplitudes, is not surprising given that the pulsating behavior of $\delta$ Scuti stars is still not fully understood and pulsating amplitudes have not been clearly delimited.


From a physical perspective, the observed trends may reflect the combined effects of ongoing mass transfer and the resulting changes in the binary configuration. Increasing values of $Q$ or $\mathrm{d}P/\mathrm{d}t$ are associated with secular orbital period growth, which may be accompanied by changes in the orbital separation and tidal interactions. However, the large scatter indicates that pulsation properties are likely governed primarily by intrinsic stellar parameters, such as mass, evolutionary stage, rotation, and mode selection, together with the well-established $P_{\rm orb}$--$P_{\rm puls}$ relation. Any contribution from mass transfer should therefore be regarded as tentative and secondary.

A notable exception in nearly all diagrams shown in Fig.\,\ref{fig:fig_Q} and \ref{fig:fig_dpdt} is R\,CMa, which consistently deviates from the general distribution. This behaviour is expected, as R\,CMa is the prototype of a distinct subgroup of Algol-type binaries characterized by very low mass ratios and a peculiar evolutionary history. Previous studies have shown that R\,CMa-type systems have undergone extensive mass transfer and follow a different evolutionary pathway compared to classical Algols (e.g. \cite{2016AJ....151...25L}). In particular, the extremely low mass ratio and semi-detached configuration can significantly affect both orbital evolution and pulsation properties. In addition, observational effects such as eclipse-related amplitude modulation and geometric visibility may further influence the measured pulsation amplitudes. For these reasons, the position of R\,CMa in our diagrams most likely reflects its intrinsically different nature rather than statistical scatter. Other systems also show noticeable departures from the overall distributions. In particular, AB\,Cas tends to deviate mainly in the amplitude-based relations. This behaviour may be related to its complex evolutionary history, since recent studies indicate that AB\,Cas is a product of rapid non-conservative mass transfer and may represent an evolutionary channel similar to that of EL\,CVn-type systems \citep{2022MNRAS.514..622M}. Moreover, its dominant pulsation has been identified as a radial fundamental mode, and both binary evolution and rotational properties have been shown to play an important role in shaping its pulsational behaviour. Since the primary eclipse in AB\,Cas is also very deep, geometric effects may further influence the observed pulsation amplitudes \citep{2023ApJ...948...16K, 2022MNRAS.514..622M}. BW\,Del, on the other hand, appears to deviate more clearly in the relations involving the pulsation period. This system has been reported as a classical Algol/oEA star with a secular orbital-period increase attributed to mass transfer from a highly evolved secondary component, while its pulsating primary is located close to, or slightly beyond, the terminal-age main sequence. Its more advanced evolutionary status, together with the ongoing mass-transfer process, may therefore explain why it does not fully follow the general trends defined by the rest of the sample \citep{2013ApSS.343..123L}. Overall, these systems should be regarded as physically informative special cases when interpreting the global relations.

\subsection{Relationship between $dM/dt$ and pulsation parameters}\label{sec:7.2}

\begin{figure}
  \centering
  \begin{minipage}[t]{0.46\textwidth}
    \centering
    \includegraphics[alt={Scatter plot showing the correlation between conservative $\log dM/dt$ and $\log P_{\text{puls}}$ with a linear fit and $1\sigma$ error band.},width=\linewidth,height=5.4cm]{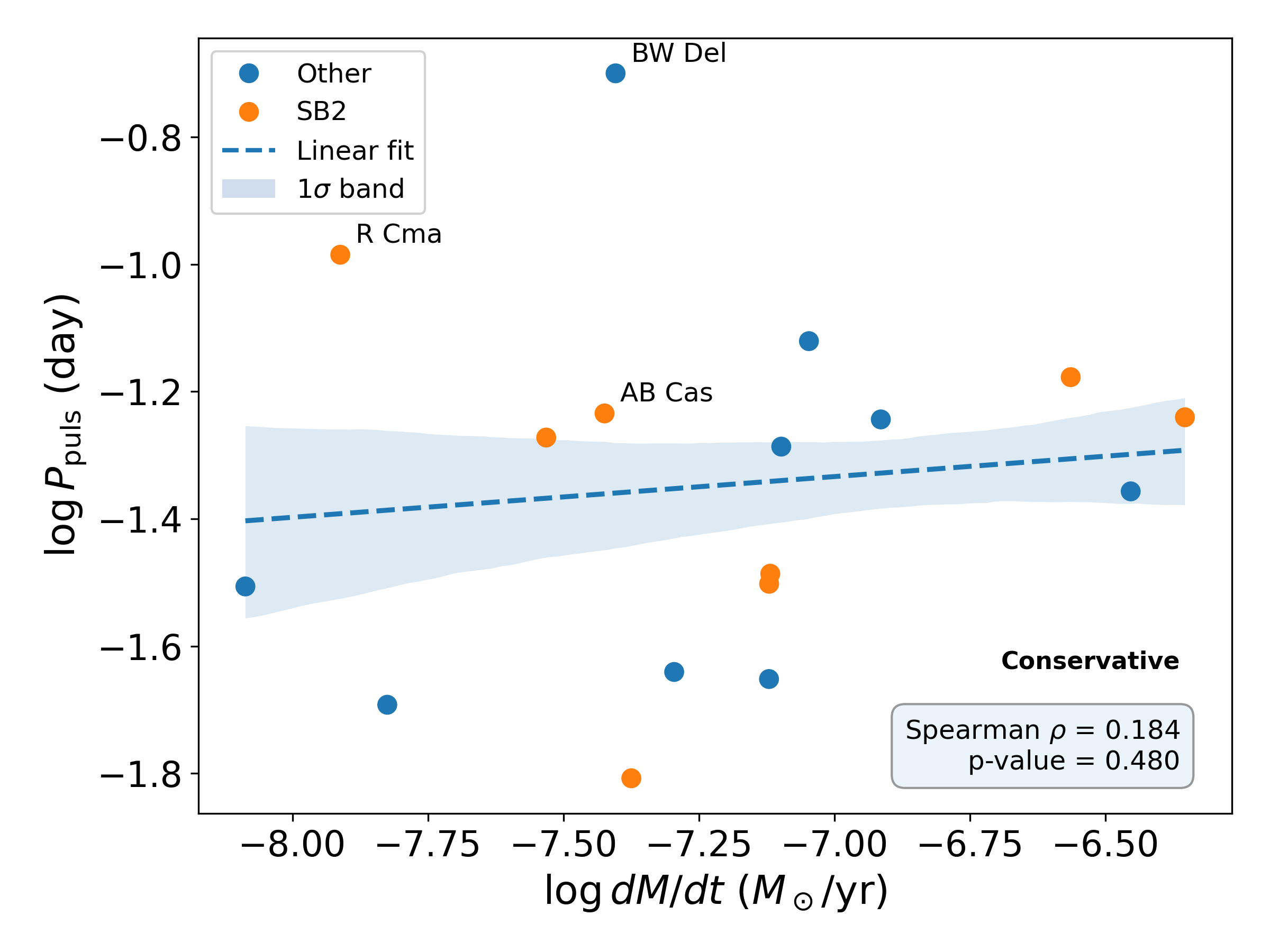}
  \end{minipage}\hfill
  \begin{minipage}[t]{0.46\textwidth}
    \centering
    \includegraphics[alt={Scatter plot showing the correlation between conservative $\log dM/dt$ and $\log \text{Amp}$ with a linear fit and $1\sigma$ error band.}width=\linewidth,height=5.4cm]{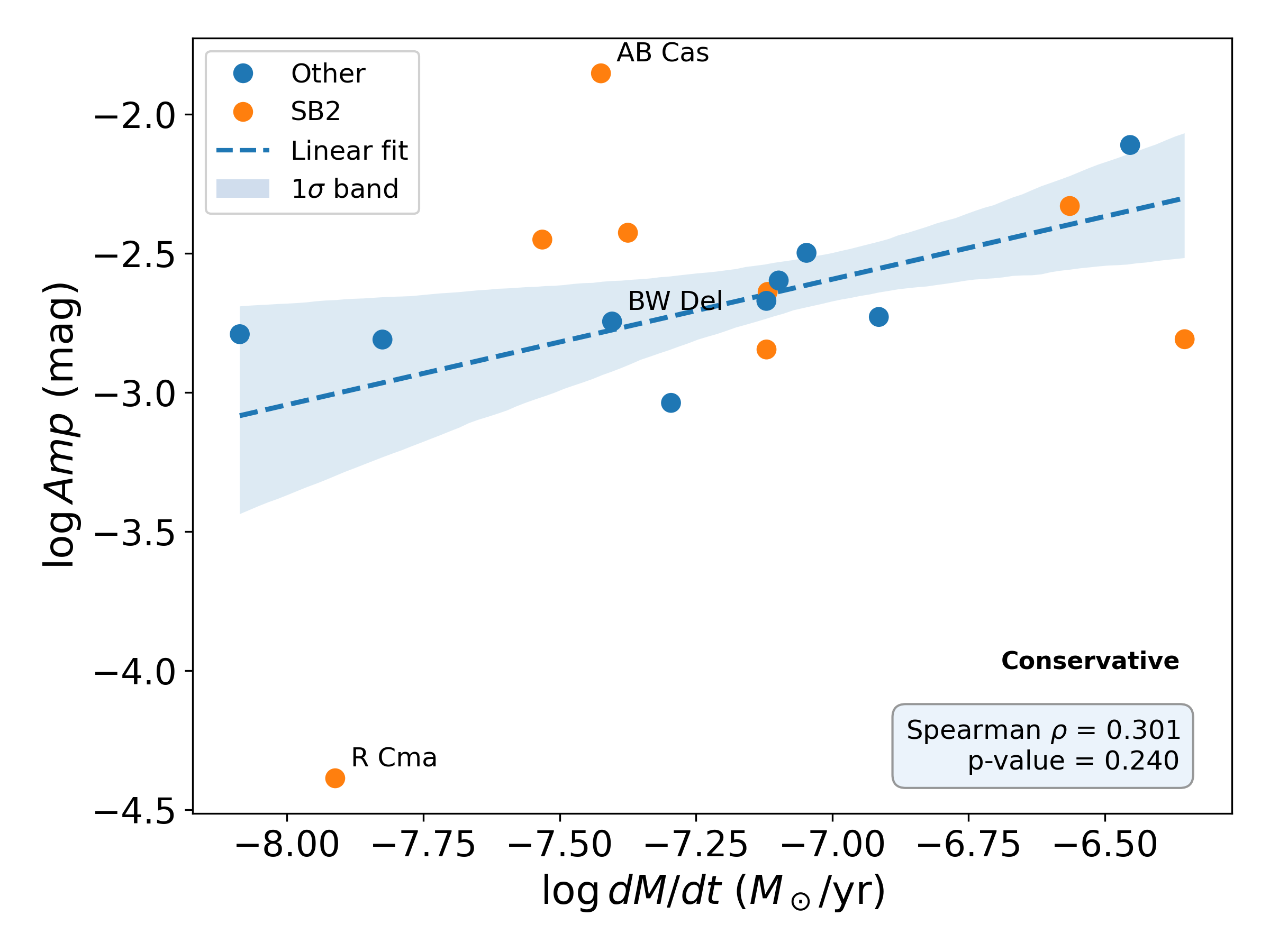}
  \end{minipage}
  \caption{Relations between the conservative mass transfer rate, $dM/dt$, and the logarithm of the pulsation period, $\log P_{\mathrm{puls}}$ (left panel), and between conservative $dM/dt$ and the logarithmic amplitude, $\log Amp$ (right panel). Systems with double-lined spectroscopic radial velocity solutions (SB2) are shown as orange dots. The dashed lines represent the best-fit linear relations, while the shaded regions correspond to the $\pm1\sigma$ residual standard deviation about the fits. The inset reports the Spearman rank correlation coefficient, $\rho$, together with the corresponding p-value.}
  \label{fig:fig_dmdt_con}
\end{figure}

\begin{figure}
 \centering
 \begin{minipage}[t]{0.46\textwidth}
  \centering
  \includegraphics [alt={Scatter plot showing the correlation between non-conservative $\log dM/dt$ and $\log P_{\text{puls}}$ with a linear fit and $1\sigma$ error band.},width=\linewidth,height=5.4cm]{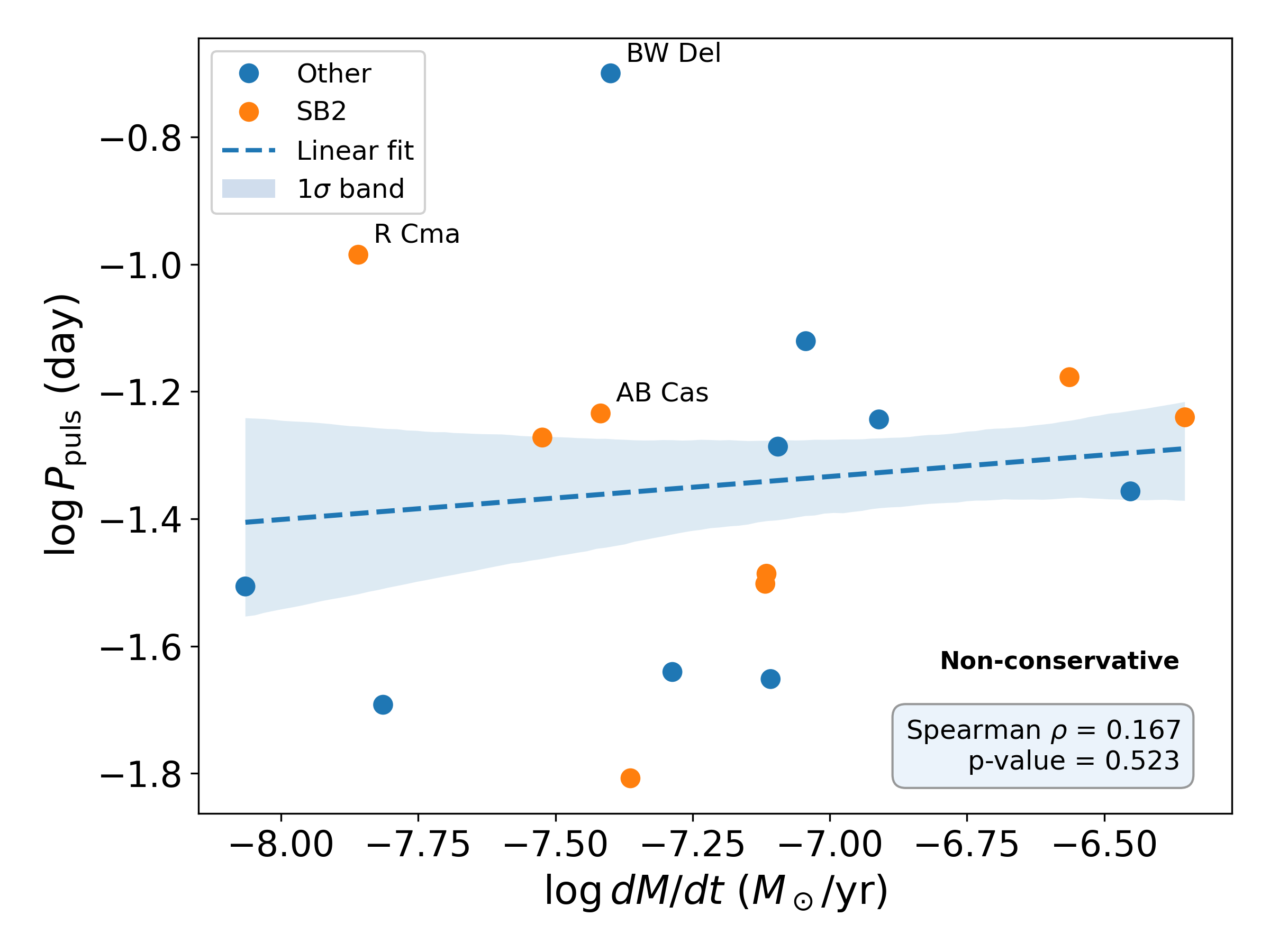}
 \end{minipage}\hfill
\begin{minipage}[t]{0.46\textwidth}
   \centering
  \includegraphics[alt={Scatter plot showing the correlation between non-conservative $\log dM/dt$ and $\log \text{Amp}$ with a linear fit and $1\sigma$ error band.},width=\linewidth,height=5.4cm]{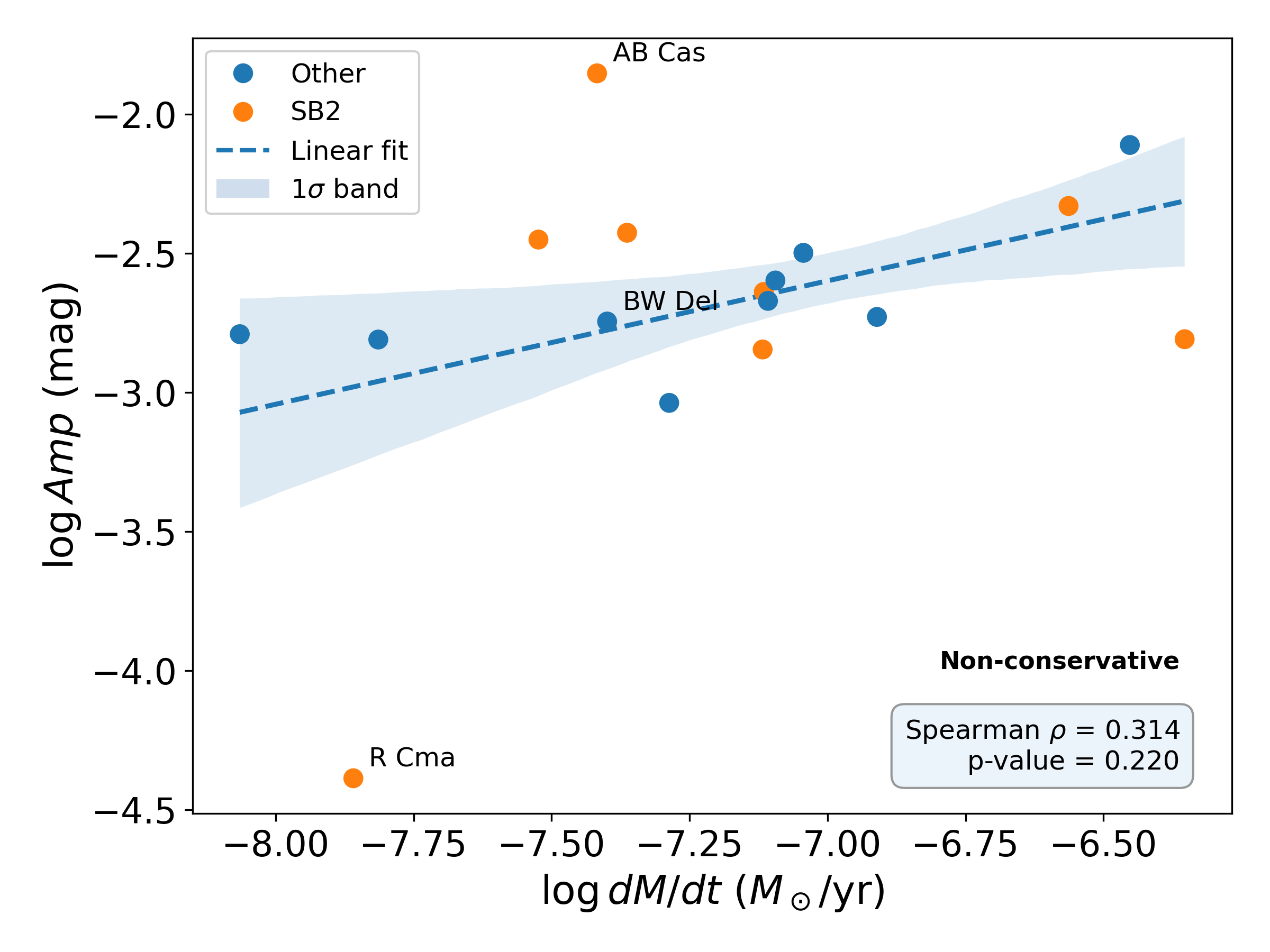}
 \end{minipage}
 \caption{Relations between the non-conservative mass transfer rate, $dM/dt$, and the logarithm of the pulsation period, $\log P_{\mathrm{puls}}$ (left panel), and between non-conservative $dM/dt$ and the logarithmic amplitude, $\log Amp$ (right panel). Systems with double-lined spectroscopic radial velocity solutions (SB2) are shown as orange dots. The dashed lines represent the best-fit linear relations, while the shaded regions correspond to the $\pm1\sigma$ residual standard deviation about the fits. The inset reports the Spearman rank correlation coefficient, $\rho$, together with the corresponding p-value.}
 \label{fig:fig_dmdt_nc}
\end{figure}

We calculated the mass-transfer rate $dM/dt$ under both conservative and non-conservative assumptions and examined its relation with $P_{\rm puls}$ and $Amp$ (Figs.~\ref{fig:fig_dmdt_con} and \ref{fig:fig_dmdt_nc}). In both cases, the distributions show at most very weak positive tendencies, but with large scatter. The corresponding Spearman coefficients remain low (in the conservative case, $\rho \sim 0.18$ for $P_{\text{puls}}$ and $\rho \sim 0.30$ for $Amp$, and in the non-conservative case, $\rho \sim 0.17$ for $P_{\text{puls}}$ and $\rho \sim 0.31$ for $Amp$., with p-values indicating that the correlations are not statistically significant. 

In particular, R CMa appears as an outlier in several diagrams, likely reflecting its distinct evolutionary history and extreme mass ratio. Although the relations with both pulsation properties are not statistically significant, the correlation with $Amp$ appears relatively stronger compared to that with $P_{\rm puls}$. The amplitude-based relations show considerable scatter and are not statistically significant. Since the observed amplitudes of non-radial modes can also be affected by geometrical cancellation, these relations should be interpreted with particular caution. The similarity of the results obtained under conservative and non-conservative assumptions indicates that the adopted mass-transfer scenario does not significantly affect the overall trends. These results suggest that any influence of mass transfer on pulsation properties is not clearly detectable within the current sample and may be secondary to other factors, such as stellar structure, evolutionary state, rotation, and mode selection.

\subsection{Correlation analysis after excluding outliers}\label{sec:7.3}
As discussed in Sects.\,\ref{sec:7.1} and \ref{sec:7.2}, R\,CMa and BW\,Del show clear deviations from the general distributions, particularly in the relations involving the dominant pulsation period. R\,CMa belongs to a distinct evolutionary group of Algol-type systems characterized by an extreme mass ratio and a peculiar mass-transfer history. BW\,Del also differs from the general distribution, possibly because of the more advanced evolutionary state of its pulsating primary. Therefore, to examine the influence of these two systems on the inferred correlations, we repeated the correlation analysis after excluding R\,CMa and BW\,Del.

\begin{figure}
 \centering
 \begin{minipage}[t]{0.46\textwidth}
  \centering
  \includegraphics[alt={Scatter plot showing the correlation between $\log Q$ and $\log P_{\text{puls}}$ excluding R CMa and BW Del, with a linear fit and $1\sigma$ error band.},width=\linewidth,height=5.4cm]{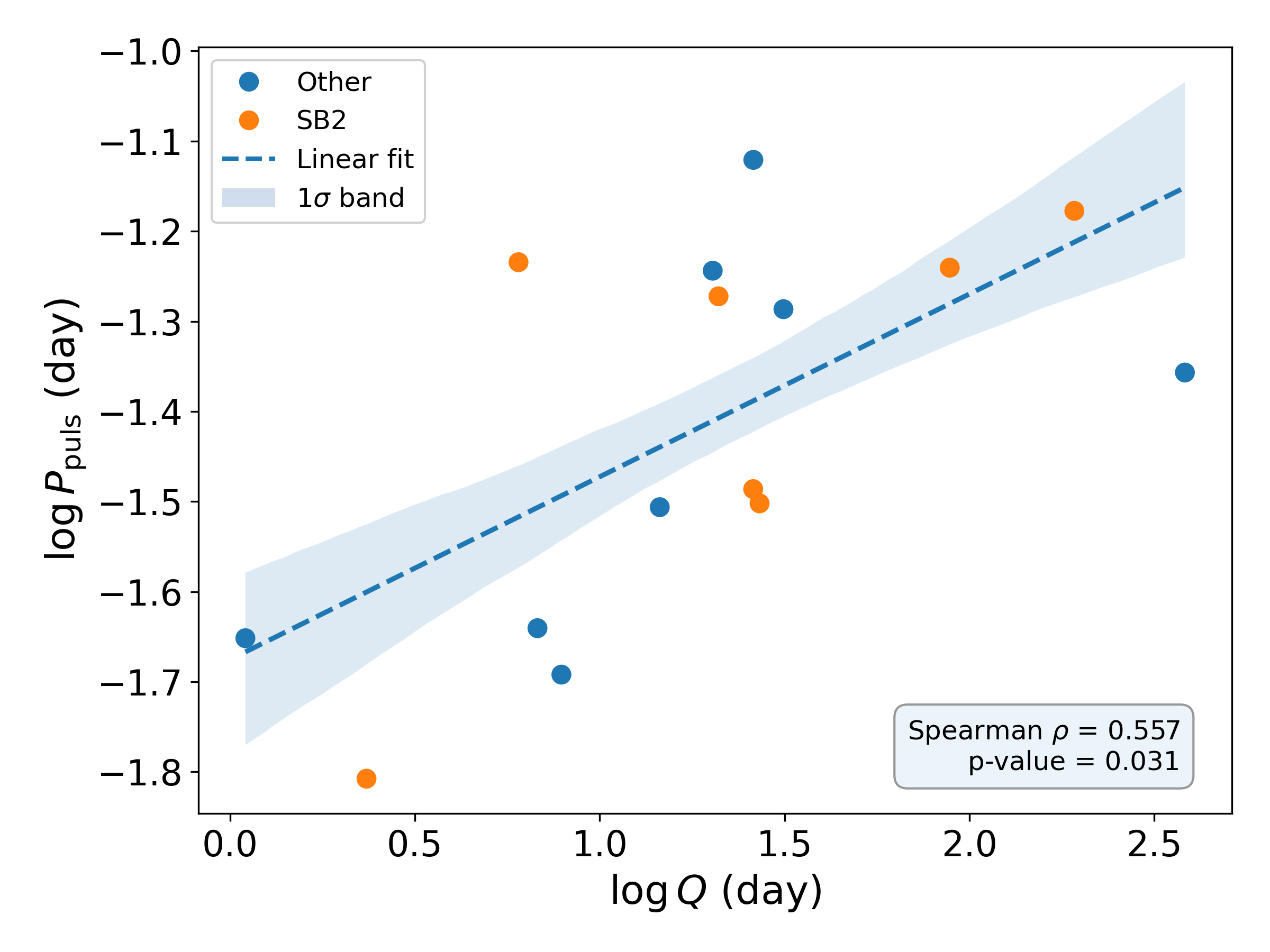}
 \end{minipage}\hfill
\begin{minipage}[t]{0.46\textwidth}
   \centering
  \includegraphics[alt={Scatter plot showing the correlation between $\log dP/dt$ and $\log P_{\text{puls}}$ excluding R CMa and BW Del, with a linear fit and $1\sigma$ error band.},width=\linewidth,height=5.4cm]{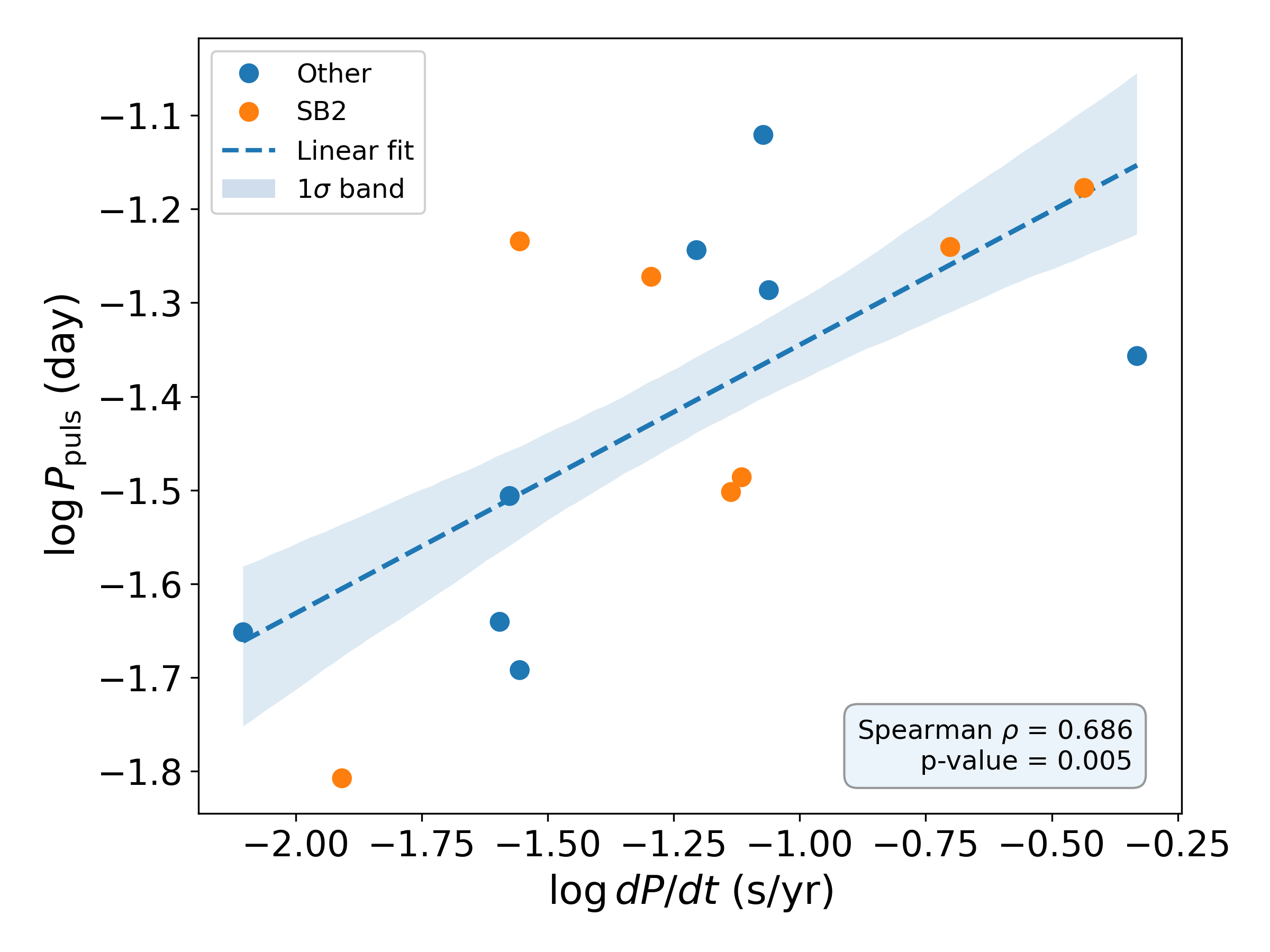}
 \end{minipage}
 \caption{Correlation analysis after excluding the R CMa and BW Del systems. The upper panel shows the relationship between $\log Q$ and $\log P_{\text{puls}}$, while the lower panel shows the relationship between $\log dP/dt$ and $\log P_{\text{puls}}$.}
 \label{fig:fig_Ppuls_removed_outlier}
\end{figure}

After excluding these two systems, the correlations involving the dominant $P_{\rm puls}$ become considerably stronger. As shown in Fig.\,\ref{fig:fig_Ppuls_removed_outlier}, a statistically significant positive correlation is found between $\log Q$ and $\log P_{\rm puls}$, with a Spearman correlation coefficient of $\rho = 0.557$ and $p = 0.031$. An even stronger positive correlation is obtained between $\log(dP/dt)$ and $\log P_{\rm puls}$, with $\rho = 0.686$ and $p = 0.005$. The latter represents the strongest correlation identified in the present sample. These results suggest that systems showing larger secular orbital-period changes tend to have longer dominant pulsation periods.

\begin{figure}
 \centering
 \begin{minipage}[t]{0.46\textwidth}
  \centering
  \includegraphics[alt={Scatter plot showing the correlation between conservative $\log dM/dt$ and $\log P_{\text{puls}}$ excluding R CMa and BW Del, with a linear fit and $1\sigma$ error band.},width=\linewidth,height=5.4cm]{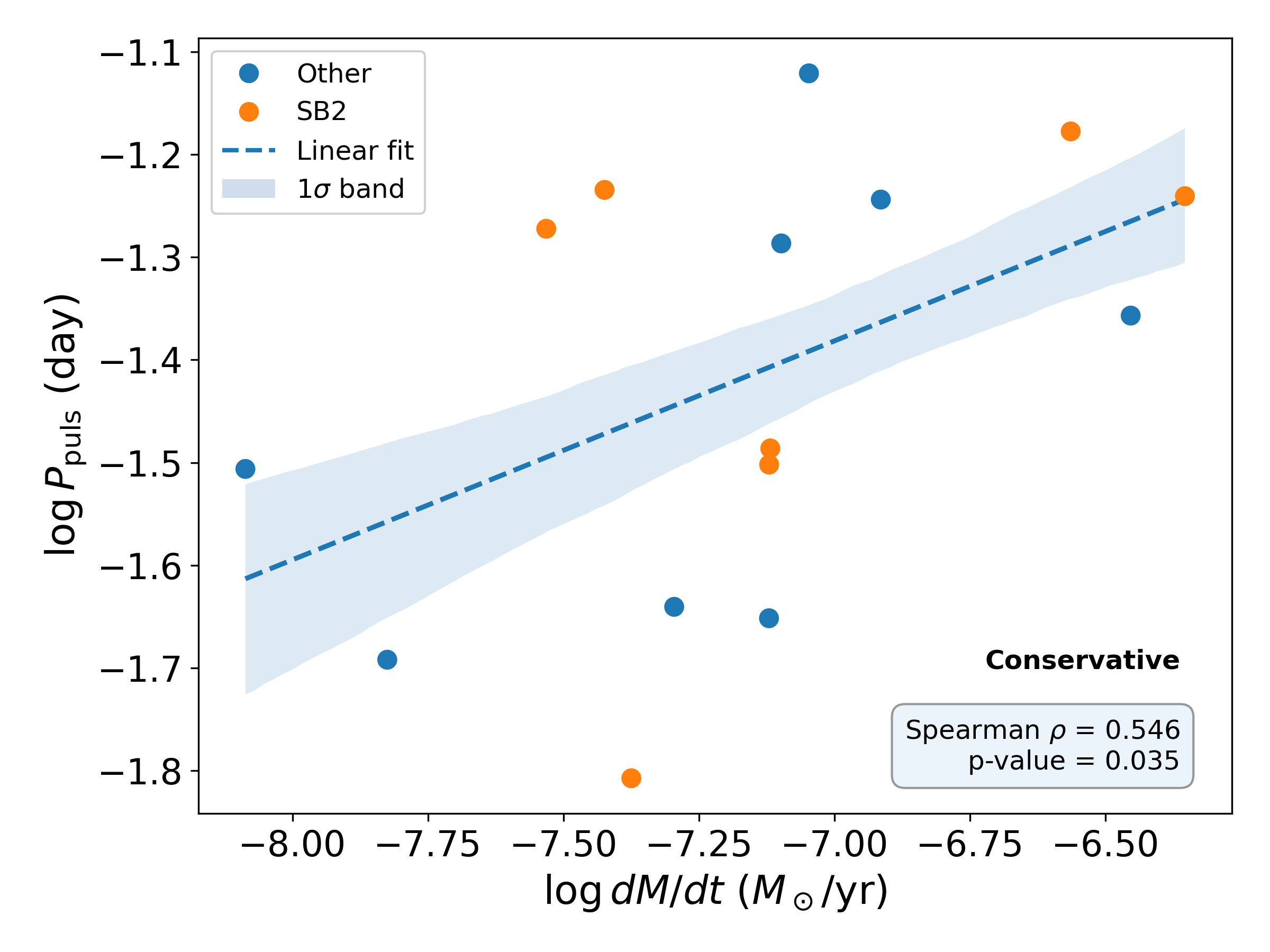}
 \end{minipage}\hfill
\begin{minipage}[t]{0.46\textwidth}
   \centering
  \includegraphics[alt={Scatter plot showing the correlation between non-conservative $\log dM/dt$ and $\log P_{\text{puls}}$ excluding R CMa and BW Del, with a linear fit and $1\sigma$ error band.},width=\linewidth,height=5.4cm]{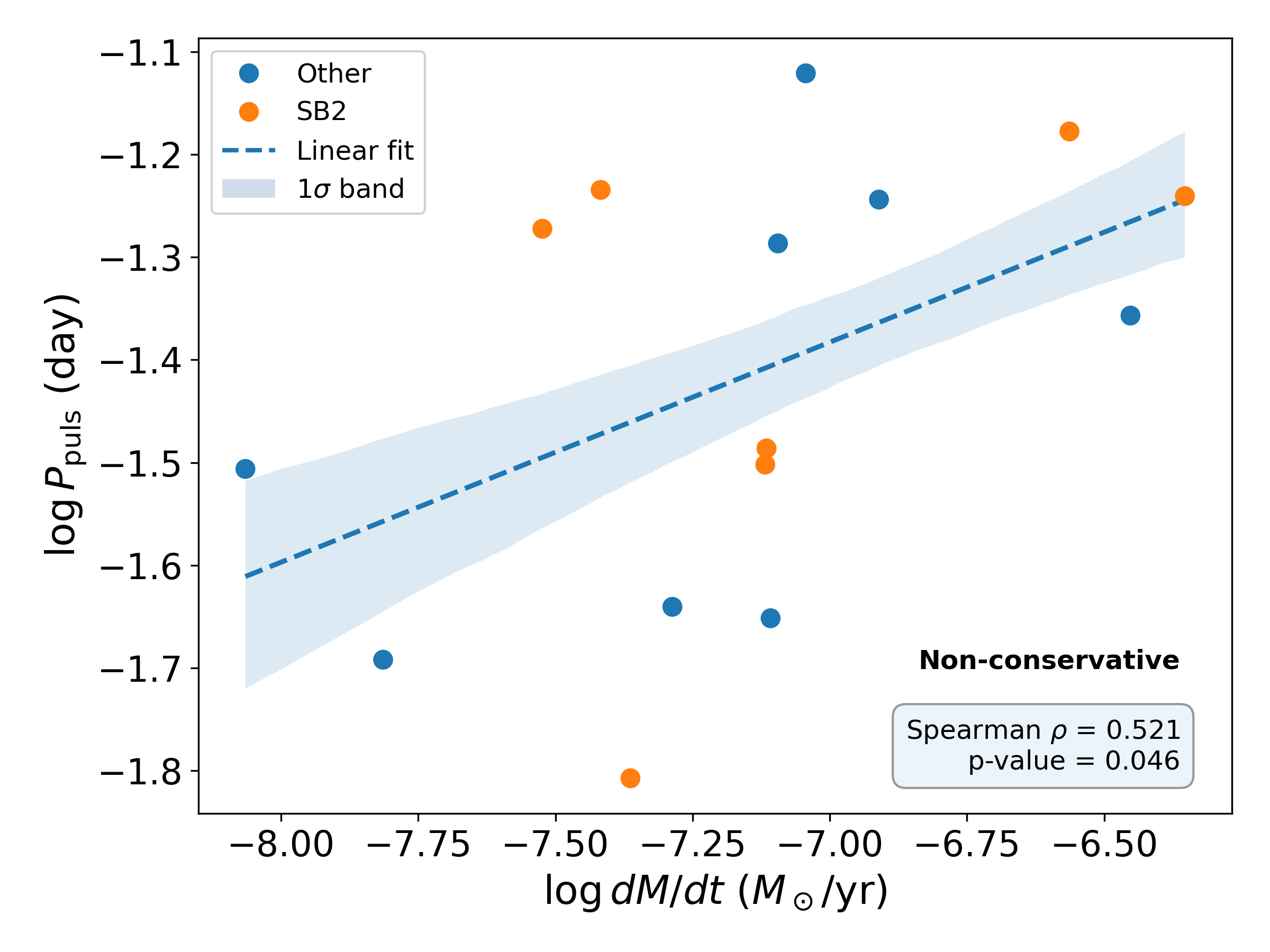}
 \end{minipage}
 \caption{Correlation analysis showing the relationships between the mass-transfer rate $dM/dt$  and $\log P_{\text{puls}}$ after excluding the R CMa and BW Del systems. The upper panel shows the relationship for conservative mass transfer, while the lower panel shows the relationship for non-conservative mass transfer.}
 \label{fig:fig_dM_removed_outliers}
\end{figure}

A similar behaviour is found for the mass-transfer rate. As shown in Fig.\,\ref{fig:fig_dM_removed_outliers}, $\log(dM/dt)$ is positively correlated with $\log P_{\rm puls}$ under both conservative and non-conservative mass-transfer assumptions. For the conservative case, the Spearman coefficient is $\rho = 0.546$ with $p = 0.035$, while for the non-conservative case we obtain $\rho = 0.521$ with $p = 0.046$. The similar behaviour in the two cases suggests that the presence of the observed trend is not strongly dependent on the adopted mass-transfer assumption.

It should be noted, however, that $Q$, $dP/dt$, and $dM/dt$ are not independent quantities. The orbital-period change rate $dP/dt$ is derived directly from the quadratic coefficient $Q$, while $dM/dt$ is subsequently estimated using the orbital-period variation together with the binary parameters. Therefore, the correlations shown in Figs.\,\ref{fig:fig_Ppuls_removed_outlier} and \ref{fig:fig_dM_removed_outliers} should be regarded as different manifestations of a common relation between secular orbital evolution and the dominant $\log P_{\rm puls}$ rather than as completely independent correlations.

The strengthening of the period-based correlations after excluding R\,CMa and BW\,Del suggests that these two systems have a considerable influence on the relations obtained from the full sample. Nevertheless, the remaining scatter in Figs.\,\ref{fig:fig_Ppuls_removed_outlier} and \ref{fig:fig_dM_removed_outliers} shows that the dominant pulsation period cannot be explained by mass-transfer-related parameters alone. Other properties of the pulsating primary, including its mass, evolutionary stage, rotation, and pulsation-mode selection, are also expected to contribute to the observed distribution. Although statistically significant correlations are obtained after excluding R~CMa and BW~Del, the reduced sample contains only 15 systems. Moreover, the significance of the correlations is sensitive to the composition of the sample. Therefore, the relations identified here should be interpreted cautiously. They suggest a possible connection between mass-transfer-related binary evolution and the dominant pulsation period, but they do not by themselves demonstrate a direct causal effect of mass transfer on the pulsations.

\section{Conclusions}
In this study, we investigated the possible connection between ongoing mass transfer and the pulsation properties of $\delta$ Scuti primaries in 17 oEA systems. For this purpose, we carried out binary modelling, long-term O--C analysis, and frequency characterization based on TESS photometry for these 17 systems, and derived the fundamental stellar parameters. Additionally, the mass-transfer indicators ($Q$, $dP/dt$, $dM/dt$) were determined.

We first derived the $T_{\rm eff}$ values of the binary components. To reduce systematic differences and place all systems on a more homogeneous scale, we determined the $T_{\rm eff}$ values through a uniform SED-fitting procedure. The resulting $T_{\rm eff}$ values are generally consistent with previously published spectroscopic values and provide a more homogeneous basis for the binary modelling and subsequent statistical analysis.

To determine the mass-transfer properties, we modelled the binary light curves of the selected systems and derived their fundamental stellar parameters, such as $M$ and $R$. We then constructed updated O--C diagrams by combining literature minima with new minima times derived from TESS data. These diagrams were used to estimate the mass-transfer indicators. In addition, the binary-variation--removed TESS light curves were analysed to extract the dominant pulsation frequencies and amplitudes.

In order to investigate the possible relationships between the mass-transfer indicators and the pulsation properties, we analysed both the full sample and a subsample of systems with spectroscopically determined mass ratios (SB2 systems), which provide more reliable binary parameters. However, owing to the limited number of such systems, no statistically meaningful correlations could be established within this subsample, indicating that the trends seen in the full sample should be regarded as tentative.

For the full sample of 17 systems, the relations between the mass-transfer indicators and the pulsation properties generally show considerable scatter and weak-to-moderate positive tendencies. The strongest relation is found between $\log(dP/dt)$ and $\log P_{\rm puls}$, with a Spearman correlation coefficient of $\rho=0.480$ and $p=0.051$. The $\log Q$--$\log P_{\rm puls}$ relation is weaker, while the relations involving $\log(dM/dt)$ are weak under both conservative and non-conservative mass-transfer assumptions. No statistically significant correlations are found between the pulsation amplitude and the mass-transfer indicators.

As discussed in Sect.\,\ref{sec:7.3}, R\,CMa and BW\,Del show clear deviations from the general distributions, particularly in the relations involving the dominant pulsation period. We therefore repeated the correlation analysis after excluding these two systems in order to examine their influence on the inferred trends. The period-based correlations become considerably stronger in the reduced sample. The strongest relation is obtained between $\log(dP/dt)$ and $\log P_{\rm puls}$, with $\rho=0.686$ and $p=0.005$. Positive correlations are also found between $\log Q$ and $\log P_{\rm puls}$ ($\rho=0.557$, $p=0.031$), and between $\log(dM/dt)$ and $\log P_{\rm puls}$ under both conservative ($\rho=0.546$, $p=0.035$) and non-conservative ($\rho=0.521$, $p=0.046$) assumptions. In contrast, the amplitude-based relations remain statistically insignificant.

These results suggest that the dominant pulsation period may be more closely associated with secular binary evolution than the observed pulsation amplitude. However, $Q$, $dP/dt$, and $dM/dt$ are not independent quantities, since $dP/dt$ is derived from $Q$, while $dM/dt$ is subsequently inferred from the orbital-period variation and the binary parameters. Therefore, the strengthened correlations should be regarded as different manifestations of a common underlying trend rather than as independent evidence for several separate effects.

We also examined the relation between the pulsation constant, $Q_{\rm p}$, and the mass-transfer coefficient, $Q$, but found no statistically significant correlation either for the full sample or after excluding R\,CMa and BW\,Del. Together with the absence of significant amplitude correlations, this indicates that the pulsation properties of the primary components cannot be explained by mass-transfer-related parameters alone.

Overall, our results suggest that mass-transfer-related binary evolution may have a secondary influence on the dominant pulsation period of $\delta$ Scuti primaries in oEA systems. Nevertheless, the correlations are sensitive to the composition of the sample, and the reduced sample contains only 15 systems. Intrinsic stellar properties such as mass, evolutionary stage, rotation, and mode selection, together with the well-established $P_{\rm orb}$--$P_{\rm puls}$ relation, are therefore likely to remain the primary factors governing the observed pulsation behaviour. The correlations identified here should consequently be interpreted cautiously and should not be taken as evidence for a direct causal effect of mass transfer on the pulsations.

Larger and more homogeneous samples, particularly systems with well-constrained spectroscopic mass ratios and long-term eclipse-timing data, will be needed to determine whether the trends identified here represent a general property of oEA systems.

\section*{Supplementary data} 

No supplementary data are associated with this article.

\begin{ack}
This study has been supported by the Scientific and Technological Research Council (TUBITAK) project through 120F330. This study is partly based on the MSc thesis of E.Ç., which was carried out under the supervision of F.K.A. We are grateful to Dr. Fahri Aliçavuş for his insightful comments and suggestions. The TESS data presented in this paper were obtained from the Mikulski Archive for Space Telescopes (MAST).  Funding for the TESS mission is provided by the NASA Explorer Program. This work has made use of data from the European Space Agency (ESA) mission Gaia (http://www.cosmos.esa.int/gaia), processed by the Gaia Data Processing and Analysis Consortium (DPAC, http://www.cosmos.esa.int/web/gaia/dpac/consortium). Funding for the DPAC has been provided by national institutions, in particular, the institutions participating in the Gaia Multilateral Agreement. This research has made use of the SIMBAD database, operated at CDS, Strasbourg, France
\end{ack}

\section*{Funding}
This study has been supported by the Scientific and Technological Research Council (TUBITAK) project through 120F330.


\appendix 

\section{Additional Tables and Figures}
In this section, we present the complete parameters obtained from the comprehensive analyses of the 17 oEA stars presented in the main text in Tables \ref{tableA1}, \ref{tab:tableA2}, \ref{tableA4}, \ref{tab:tab_ap_min}, \ref{tab:a6} and \ref{tab:all_frequencies}. We also provide the light curve models and O-C plots of the orbital period variation analyses for the 17 oEA stars in Fig.\,\ref{fig:all_LC} and \ref{fig:ap}, respectively. Fig. \ref{fig:relationship_app} is also included to support the possible relationships we investigated.
\clearpage
\setcounter{table}{0}
\renewcommand{\thetable}{A\arabic{table}}
\begin{table}
    \centering 
    \caption{Relative luminosity contributions in $V$ band and their references for the systems with temperatures (B-V) determined from the color scale.}\label{tableA1}
    \begin{tabular}{lll}
    \hline
    Star    & $l_1/l_2$ & Reference \\
    \hline 
    XZ Aql  & 16.85 & \citet{2016AJ....152...33Z} \\
    Y Cam   & 12.33 & \citet{2015AJ....150..131H} \\
    AB Cas  & 23.39 & \citet{2007ApSS.307..409A} \\
    RZ Cas  &13.71  &\citet{2006AN....327..905S} \\
    XX Cep  &11.35  & \citet{2016AJ....151...77K} \\
    WY Cet  &30.25 & this study \\
    R CMa   &19.41  &\citet{2011MNRAS.418.1764B} \\
    UW Cyg  &25.35  &\citet{2012MNRAS.422.1250L} \\
    BW Del  & 22.84 & this study \\
    SX Dra  &3.00   &\citet{2013AJ....145...87S} \\
    TW Dra  &8.25   & \citet{2013IBVS.6086....1B}  \\
    TZ Eri  &8.98   &\citet{2008CoAst.157..336L} \\
    CT Her  & 13.71 & \citet{2011AA...534A.111L} \\
    RX Hya  & 8.76 & this study \\
    Y Leo   &19.85 & \citet{2011ApSS.331..105T} \\
    FR Ori  & 20.28 & \citet{2014AJ....147...35Y}\\
    AC Tau  &10.24 & this study \\
    \hline       
    \end{tabular}
\end{table}

\setcounter{table}{1}
\renewcommand{\thetable}{A\arabic{table}}
\begin{table}
    \centering 
      \caption{The $q$ values and references used in the binary modeling.} \label{tab:tableA2}
    \begin{tabular}{lrrr}
    \\
    \hline
    Star      &$q (M_2/M_1)$   & Methods    		        &Ref. \footnotemark[$*$]\\
    \hline
    XZ Aql    &0.184\,(4)       &radial velocity analysis &8 \\
    Y Cam     &0.238\,(10)      &radial velocity analysis &1 \\
    AB Cas    &0.182\,(2)       &radial velocity analysis &2 \\
    RZ Cas    &0.351\,(2)       &radial velocity analysis &5 \\
    XX Cep   &0.151\,(2)       &radial velocity analysis &3 \\
    WY Cet    &0.250\,(20)       &$q$ search     	        &9 \\
    R CMa     &0.128\,(3)       &radial velocity analysis &4 \\
    UW Cyg    &0.133\,(12)       &$q$ search           	&9 \\
    BW Del    &0.135\,(20)       &$q$ search     &9 \\
    SX Dra    &0.383\,(8)       &$q$ search       		&9 \\
    TW Dra    &0.427\,(11)      &radial velocity analysis &7 \\
    TZ Eri    &0.181\,(8)       &radial velocity analysis &6 \\
    CT Her    &0.120\,(20)       &$q$ search               &9 \\
    RX Hya    &0.235\,(8)      &$q$ search     &9 \\
    Y Leo     &0.330\,(9)       &$q$ search        &9 \\
    FR Ori    &0.539\,(9)       &$q$ search    	&9 \\
    AC Tau    &0.365\,(8)       &$q$ search     &9 \\
    \hline
    \end{tabular}
\begin{tabnote}
\footnotemark[$*$] (1)\,\citet{2024PASJ...76..787C}, (2)\,\citet{2017AJ....153..247H}, (3)\citet{2016AJ....151...77K}, (4)\,\citet{2018AA...615A.131L}, (5)\,\citet{2020AA...644A.121L}, (6)\,\citet{2025arXiv250220626P}, (7)\,\citet{2010AJ....139.1327T}, (8)\,\citet{2016AJ....152...33Z}, (9)\,This study.
\end{tabnote}
\end{table}

\setcounter{table}{2}
\renewcommand{\thetable}{A\arabic{table}}
\begin{table}
\caption{Spot parameters found in the binary modeling. TF stands for temperature factor which is the ratio of the star’s surface temperature to the spot temperature. $\varphi$ is fixed parameters and means the latitude. $\lambda$ and $\theta$ are the longitude and spot angular radius, respectively. All hot spots were assumed on the surface of primary components.}\label{tableA4}
\centering
\begin{tabular}{lcccc}
\hline
 Stars   & $\varphi$ ($^{\circ}$) & $\lambda$ ($^{\circ}$) & $\theta$  ($^{\circ}$) & TF \\
\hline
 Y Cam   & 90.00 & 173.25 (14)& 14.62 (5)& 1.102 (2)\\
 AB Cas  & 90.00 & 174.00 (14)& 13.67 (6)& 1.059 (3)\\
 RZ Cas  & 90.00 & 194.00 (14)& 21.30 (6)& 1.035 (6)\\
 XX Cep  & 90.00 & 178.15 (18)& 10.56 (5)& 1.210 (4)\\
 WY Cet  & 90.00 & 168.27 (16)& 14.49 (6)& 1.081 (6)\\
 R CMa   & 90.00 & 169.42 (15)& 16.64 (10)& 1.060 (2)\\
 TZ Eri  & 90.00 & 176.75 (14)& 19.29 (5) & 1.051 (4) \\
 Y Leo   & 90.00 & 74.32 (13)& 10.96 (6)& 0.881 (4)\\
 AC Tau  & 90.00 & 78.04 (15)& 11.11 (10)& 1.090 (7)\\
\hline
\end{tabular}
\end{table}

\setcounter{table}{3}
\renewcommand{\thetable}{A\arabic{table}}
\begin{table*} 
    \scriptsize
    \centering
    \caption{Calculated minimum times in HJD (-2457000) from the TESS data listed in Table\,\ref{table1}.}\label{tab:tab_ap_min}
    \begin{tabular}{lllllllll p{3cm}}
    \hline
XZ Aql         & Y Cam         & AB Cas        & RZ Cas        & R CMa        & XX Cep         & WY Cet        & UW Cyg & BW Del  \\ \hline
3508.0966 (3)   & 2391.3956 (3) & 1791.7773 (3) & 1792.3840 (1) & 1492.1557 (1) & 1765.6233 (1) & 1386.0490 (2) & 1714.7702 (3) & 3511.8575 (5) \\ 
3531.6274 (2)   & 2417.8413 (3) & 1813.6476 (3) & 1800.7496 (1) & 1502.3793 (1) & 1786.6595 (1) & 1393.8080 (2) & 1732.0243 (3) & \\ 
               & 2583.1270 (4) & 1816.3812 (3) & 1813.8975 (1) & 1516.0109 (1) & 1793.6716 (1) & 1405.4465 (2) & 2422.1818 (3) &   \\ 
               & 2606.2667 (3) & 1827.3161 (3) & 1816.2880 (1) & 2202.1267 (1) & 1812.3705 (1) &               & 2446.3373 (3) &  \\ 
               & 2745.1067 (4) & 1840.9851 (3) & 1827.0451 (1) & 2213.4859 (1) & 1959.6237 (1) & ~ & ~ & ~  \\ 
               & 2768.2466 (3) & 1987.2115 (2) & 1840.1929 (1) & 2227.1173 (1) & 1980.6599 (1) & ~ & ~ & ~  \\ 
               &               & 2009.1116 (2) & 1984.8168 (1) & 2229.3892 (1) & ~ & ~ & ~ & ~  \\ 
               &               & 2719.8931 (3) & 1995.5735 (1) & 2240.7488 (1) & ~ & ~ & ~ & ~  \\ 
               &               & 2729.4612 (3) & 2008.7214 (1) & 2253.2444 (1) & ~ & ~ & ~ & ~  \\ 
               &               & 2741.7632 (2) & 2718.6918 (1) & ~ & ~ & ~ & ~ & ~ \\ 
               &               & 2882.5529 (3) & 2729.4489 (1) & ~ & ~ & ~ & ~ & ~  \\ \hline
SX Dra         & TW Dra        & TZ Eri        & CT Her        & RX Hya        & Y Leo        & FR Ori        & AC Tau &  ~ \\ \hline
1683.9843 (7)  & 1684.1314 (1) & 1439.3090 (1) & 3434.9248 (1) & 1518.1047 (1) & 2527.1178 (6) & 2202.1002 (2) & 1438.3962 (4) & ~  \\ 
1709.5118 (7)  & 1706.5856 (2) & 1449.7359 (2) & 3449.2164 (1) & 1527.2315 (1) & 2537.2343 (7) & 2213.5814 (2) & 1448.6132 (4) & ~ \\ 
1740.8517 (9)  & 1712.1991 (3) & 1462.7591 (2) & 3452.7895 (1) & 1540.9217 (1) & 2549.0367 (6) & 2226.8288 (2) & 1462.9169 (3) & ~  \\ 
1766.6998 (7)  & 1734.6536 (2) & 2181.0644 (3) & 3465.2942 (1) & 2964.6957 (1) & 2552.4089 (7) &               &  & ~  \\ 
1864.9236 (7)  & 1740.2690 (3) &               & 3479.5854 (1) & 2973.8225 (1) & 2564.2111 (6) &               &  & ~  \\ 
1895.9421 (8)  & 1759.9143 (3) &               &               & 2987.5127 (1) & 2577.6997 (6) &               &  & ~ \\ 
1901.1109 (7)  & 1872.1846 (5) &               &               &               & 2609.7354 (6) &               &  & ~ \\ 
1937.3021 (1)  & 1887.4462 (3) &               &               &               & 2619.8514 (6) &               &  & ~  \\ 
1978.6571 (3)  & 1911.4826 (4) &               &               &               & 2635.0261 (6) &               &  & ~ \\ 
1999.4092 (2)  & 1883.4114 (3) &               &               &               &               &               &  & ~  \\ 
2032.9343 (2)  & 2432.1352 (4) &               &               &               &               &               &  & ~  \\ 
2420.5842 (2)  & 2617.3825 (3) &               &               &               &               &               &  & ~  \\ 
2627.4571 (3)  & 2677.7260 (5) &               &               &               &               &               &  & ~  \\ 
2715.3350 (3)  &               &               &               &               &               &               &  & ~  \\ 
2772.1633 (3)  &               &               &               &               &               &               &  & ~  \\ 
\hline
    \end{tabular}
\end{table*}

\setcounter{table}{4}
\renewcommand{\thetable}{A\arabic{table}}
\begin{table}
\centering 
\caption{Additional cyclical variation parameters and derived values in orbital period analysis.}\label{tab:a6}
\begin{tabular}{lcccccc}
\hline
Star & $A_{cyclic}$ (d) & $P_{cyclic}$ (yr) & $e_3$ & $\omega_3$ ($^\circ$) & $f(m_3)$ ($M_\odot$) & $B$ (kG) \\
\hline
XZ Aql & 0.054  (1) & 150  (9) & 0.77  (3) & 176  (5) & 0.134 (3) & 1.63  (1) \\
Y Cam & 0.052  (1) & 61  (1)   & 0.55 (6) & 164  (4) & 0.316  (4) & 2.64  (1) \\
AB Cas & 0.008  (2) & 42 (1)  & 0.71 (5) & 125  (6) & 0.002  (2) & 3.13  (4) \\
RZ Cas & 0.007  (6) & 55  (1) & 0.77 (10) & 156 (10) & 0.001  (7) & 2.44  (1) \\
R Cma & 0.036  (2) & 93 (1) & 0.83  (5) & 275 (8) & 0.028  (4) & 2.98  (1) \\
XX Cep & 0.019  (2) & 60  (8) & 0.83 (11) & 188 (6) & 0.054 (1) & 2.11  (3) \\
WY Cet & 0.042  (1) & 48  (2) & 0.42  (29) & 287  (79) & 0.168  (1) & 4.69  (1) \\
SX Dra & 0.063  (6) & 79  (4) & 0.40  (19) & 0.89  (13) & 0.268  (1) & 1.45  (1) \\
TW Dra & 0.167  (35) & 96 (1) & 0.00  (2) & 338  $1.4\times 10^{7}$ & 2.640  (17) & 8.4  (8) \\
TZ Eri & 0.038  (2) & 51  (1) & 0.62  (11) & 9  (3) & 0.216  (7) & 3.08  (2) \\
CT Her & 0.010  (2) & 34  (1) & 0.83  (21) & 268  (41) & 0.005 (3) & 3.45  (5) \\
RX Hya & 0.017  (2) & 80  (1) & 0.83  (1) & 252 (26) & 0.004  (2) & 1.50  (1) \\
Y Leo & 0.0375  (1) & 63  (1) & 0.63  (5) & 187  (2) & 0.145  (2) & 3.92  (3) \\
\hline
\end{tabular}
\end{table}

\setcounter{table}{5}
\renewcommand{\thetable}{A\arabic{table}}
\begin{table*} 
    \scriptsize
    \centering
    \caption{Frequency list for the top 5 highest amplitude frequencies of all systems.}\label{tab:all_frequencies}
\begin{tabular}{llllllllllll}
\hline
XZ Aql & $f$ (d$^{-1}$) & A (mmag) & $\phi$ (rad) & $Q{_p}$ (d$^{-1}$)& Mode & SX Dra & $f$ (d$^{-1}$) & A (mmag) & $\phi$ (rad) & $Q{_p}$ (d$^{-1}$) & Mode \\
\hline
 & 30.634499 & 2.300125 & 0.468964 & 0.013427 & 5H  &  & 22.742214 & 7.764158 & 0.298433 & 0.014573 & 4H \\
 & 31.490861 & 1.794540 & 0.687556 & 0.013062 & 5H  &  & 19.635305 & 3.725023 & 0.870801 & 0.016879 & 3H \\
 & 32.539857 & 0.974672 & 0.588068 & 0.012641 & 5H  &  & 22.934717 & 0.819285 & 0.641045 & 0.014451 & 4H \\
 & 28.713882 & 0.897603 & 0.748284 & 0.014325 & 4H &  & 23.321961 & 0.748890 & 0.521823 & 0.014211 & 4H \\
 & 37.485167 & 0.819684 & 0.332691 & 0.010973 & 6H &  & 22.547472 & 0.775899 & 0.743522 & 0.014699 & 4H \\
\hline
Y Cam & $f$ (d$^{-1}$) & A (mmag) & $\phi$ (rad) & $Q{_p}$ (d$^{-1}$)& Mode & TW Dra & $f$ (d$^{-1}$) & A (mmag) & $\phi$ (rad) & $Q{_p}$ (d$^{-1}$)& Mode \\
\hline
 & 15.046989 & 4.685257 & 0.291013 & 0.016872 & 3H &  & 17.397582 & 1.554720 & 0.403145 & 0.019861 & 2H \\
 & 19.312900 & 1.308305 & 0.362999 & 0.013146 & 5H &  & 18.110169 & 0.812353 & 0.174390 & 0.019080 & 2H \\
 & 14.988838 & 1.361465 & 0.727384 & 0.016938 & 3H &  & 22.663752 & 0.458741 & 0.350006 & 0.015246 & 3H/4H \\
 & 14.445257 & 1.078681 & 0.007896 & 0.017575 & 2H/3H &  & 24.826994 & 0.409716 & 0.015191 & 0.013918 & 4H/5H \\
 & 18.689758 & 0.555778 & 0.883392 & 0.013584 & 4H &  & 24.949931 & 0.399313 & 0.393148 & 0.013849 & 4H/5H \\
\hline
AB Cas & $f$ (d$^{-1}$) & A (mmag) & $\phi$ (rad) & $Q{_p}$ (d$^{-1}$)& Mode & TZ Eri & $f$ (d$^{-1}$) & A (mmag) & $\phi$ (rad) & $Q{_p}$ (d$^{-1}$)& Mode \\
\hline
& 17.156371 & 14.053822 & 0.750927 & 0.032770 & F & & 18.718807 & 3.547766 & 0.625115 & 0.033512 & F \\
 & 33.003752 & 1.157385 & 0.072497 & 0.017035 & 2H/3H &  & 25.429762 & 0.427272 & 0.515919 & 0.024668 & 1H   \\
$f_2$+2$f_{orb}$ & 34.466876 & 0.715543 & 0.480929 & 0.016312 & 3H &  & 40.728396 & 0.395014 & 0.606018 & 0.015402 & 3H/4H \\
& 16.678421 & 0.589489 & 0.008270 & 0.033709 & F &  & 20.451418 & 0.293895 & 0.128707 & 0.030673 & F \\
$f_1$+3$f_{orb}$ & 19.351285 & 0.532805 & 0.518528 & 0.029053 & F/1H &  & 18.986856 & 0.257406 & 0.613651 & 0.033039 & F \\
\hline
RZ Cas & $f$ (d$^{-1}$) & A (mmag) & $\phi$ (rad) & $Q{_p}$ (d$^{-1}$)& Mode & CT Her & $f$ (d$^{-1}$)) & A (mmag) & $\phi$ (rad) & $Q{_p}$ (d$^{-1}$)& Mode \\
\hline
 & 64.195094 & 3.754291 & 0.625766 & 0.010724 & 6H &  & 49.206978 & 1.550506 & 0.924581 & 0.010197 & 6H/7H \\
$f_2$-2$f_{orb}$ & 62.407067 & 0.896527 & 0.369533 & 0.011031 & 6H &  & 52.935809 & 1.238294 & 0.820136 & 0.009479 & 7H \\
 & 64.080530 & 0.827514 & 0.745654 & 0.010743 & 6H &  & 52.115332 & 0.874471 & 0.062444 & 0.009628 & 7H \\
$f_2$-$f_{orb}$ & 63.243798 & 0.593360 & 0.524925 & 0.010885 & 6H &  & 47.828542 & 0.740779 & 0.481725 & 0.010491 & 6H/7H\\
 & 54.919447 & 0.354719 & 0.877212 & 0.012535 & 5H   &  & 45.511430 & 0.661840 & 0.830137 & 0.011025 & 6H \\
\hline
R CMa & $f$ (d$^{-1}$) & A (mmag) & $\phi$ (rad) & $Q{_p}$ (d$^{-1}$)& Mode & RX Hya & $f$ (d$^{-1}$) & A (mmag) & $\phi$ (rad) & $Q{_p}$ (d$^{-1}$)& Mode \\
\hline
 & 9.663835 & 0.040994 & 0.463342 & 0.057845 & - &  & 19.343643 & 2.528380 & 0.224804 & 0.010971 & 6H \\
 & 9.660969 & 0.039000 & 0.956387 & 0.057862 & - &  & 20.283325 & 0.790625 & 0.802924 & 0.010463 & 6H/7H \\
 & 12.310398 & 0.017304 & 0.366559 & 0.045409 & - &  & 41.988959 & 0.433117 & 0.025963 & 0.005054 & - \\
 & 6.179354 & 0.016665 & 0.496529 & 0.090463 & - &  & 32.618932 & 0.357293 & 0.402601 & 0.006506 & - \\
   &          &          &          &          &    &  & 44.355415 & 0.322092 & 0.984468 & 0.004784 & - \\
\hline
XX Cep & $f$ (d$^{-1}$) & A (mmag) & $\phi$ (rad) & $Q{_p}$ (d$^{-1}$) & Mode          &  Y Leo & $f$ (d$^{-1}$) & A (mmag) & $\phi$ (rad) & $Q{_p}$ (d$^{-1}$) & Mode \\
\hline
 & 31.768739 & 1.427969 & 0.209643 & 0.014420 & 4H &      & 43.703268 & 0.917806 & 0.896744 & 0.012683 & 5H \\
 & 30.541201 & 0.646936 & 0.701525 & 0.015000 & 4H    &  & 30.471724 & 0.566052 & 0.150525 & 0.018190 & 2H/3H \\
 & 30.507192 & 0.635518 & 0.999595 & 0.015017 & 4H    &  & 35.267880 & 0.534972 & 0.621348 & 0.015717 & 3H/4H\\
 & 37.308976 & 0.604466 & 0.717343 & 0.012279 & 5H      &  & 41.015959 & 0.489931 & 0.900594 & 0.013514 & 4H/5H\\
 & 39.422124 & 0.605575 & 0.022126 & 0.011621 & 5H/6H &  & 3.588491 & 0.477522 & 0.267368 & 0.154465 & 3H/4H\\
\hline
WY Cet & $f$ (d$^{-1}$) & A (mmag) & $\phi$ (rad) & $Q{_p}$ (d$^{-1}$)& mode & FR Ori & $f$ (d$^{-1}$) & A (mmag) & $\phi$ (rad) & $Q{_p}$ (d$^{-1}$)& mode \\
\hline
 & 13.208935 & 3.178748 & 0.566065 & 0.032650 & F&  & 44.835340 & 2.135725 & 0.332494 & 0.013045 & 4H/5H \\
 & 12.180724 & 1.991945 & 0.492841 & 0.035406 & F &  & 37.637494 & 1.980102 & 0.285332 & 0.015540 & 3H/4H \\
 & 12.392778 & 1.372572 & 0.753907 & 0.034800 & F&  & 40.398182 & 1.472366 & 0.195860 & 0.014478 & 4H \\
 & 15.768367 & 0.979712 & 0.900066 & 0.027350 & 1H&  & 44.926567 & 1.372170 & 0.759023 & 0.013019 & 4H/5H\\
 & 13.951121 & 0.915440 & 0.189426 & 0.030913 & F &  & 41.611863 & 1.366499 & 0.204584 & 0.014056 & 4H \\
\hline
UW Cyg & $f$ (d$^{-1}$) & A (mmag) & $\phi$ (rad) & $Q{_p}$ (d$^{-1}$)& Mode & AC Tau & $f$ (d$^{-1}$) & A (mmag) &$\phi$ (rad) & $Q{_p}$ (d$^{-1}$)& Mode \\
\hline
 & 32.078325 & 1.620833 & 0.701284 & 0.012613 & 5H &  & 17.532092 & 1.869619 & 0.462830 & 0.024678 & 1H \\
 & 24.167021 & 0.520474 & 0.362475 & 0.016742 & 3H &  & 17.676053 & 1.040748 & 0.725594 & 0.024477 & 1H \\
 & 23.318215 & 0.405423 & 0.197357 & 0.017352 & 3H &  & 33.663484 & 0.805995 & 0.947770 & 0.012852 & 5H \\
 & 23.840557 & 0.337344 & 0.597578 & 0.016972 & 3H &  & 18.656542 & 0.492329 & 0.864152 & 0.023191 & 1H/2H \\
 & 26.212860 & 0.335554 & 0.443395 & 0.015436 & 3H &  & 33.949460 & 0.402263 & 0.693977 & 0.012744 & 5H \\
\hline
BW Del & $f$ (d$^{-1}$) & A (mmag) & $\phi$ (rad) & $Q{_p}$ (d$^{-1}$)& Mode &            &      &      &      &      &      \\
\hline
& 5.014200 & 1.800000 & 0.322000 & 0.074917 & -        &      &      &      &      &      &      \\
 & 23.274200 & 1.020000 & 0.951000 & 0.016140 & 3H       &      &      &      &      &      &      \\
 & 23.829200 & 0.690000 & 0.802000 & 0.015764 & 3H     &      &      &      &      &      &      \\
 & 27.107800 & 0.590000 & 0.344000 & 0.013858 & 4H/5H  &      &      &      &      &      &      \\
\hline
\end{tabular}
\end{table*}

\clearpage

\setcounter{figure}{0}
\renewcommand{\thefigure}{A\arabic{figure}}
\begin{figure*}
 \begin{minipage}[b]{0.33\textwidth}
 \includegraphics[alt={Binary light curve model of XZ Aql},height=4.8cm, width=1.0\textwidth]{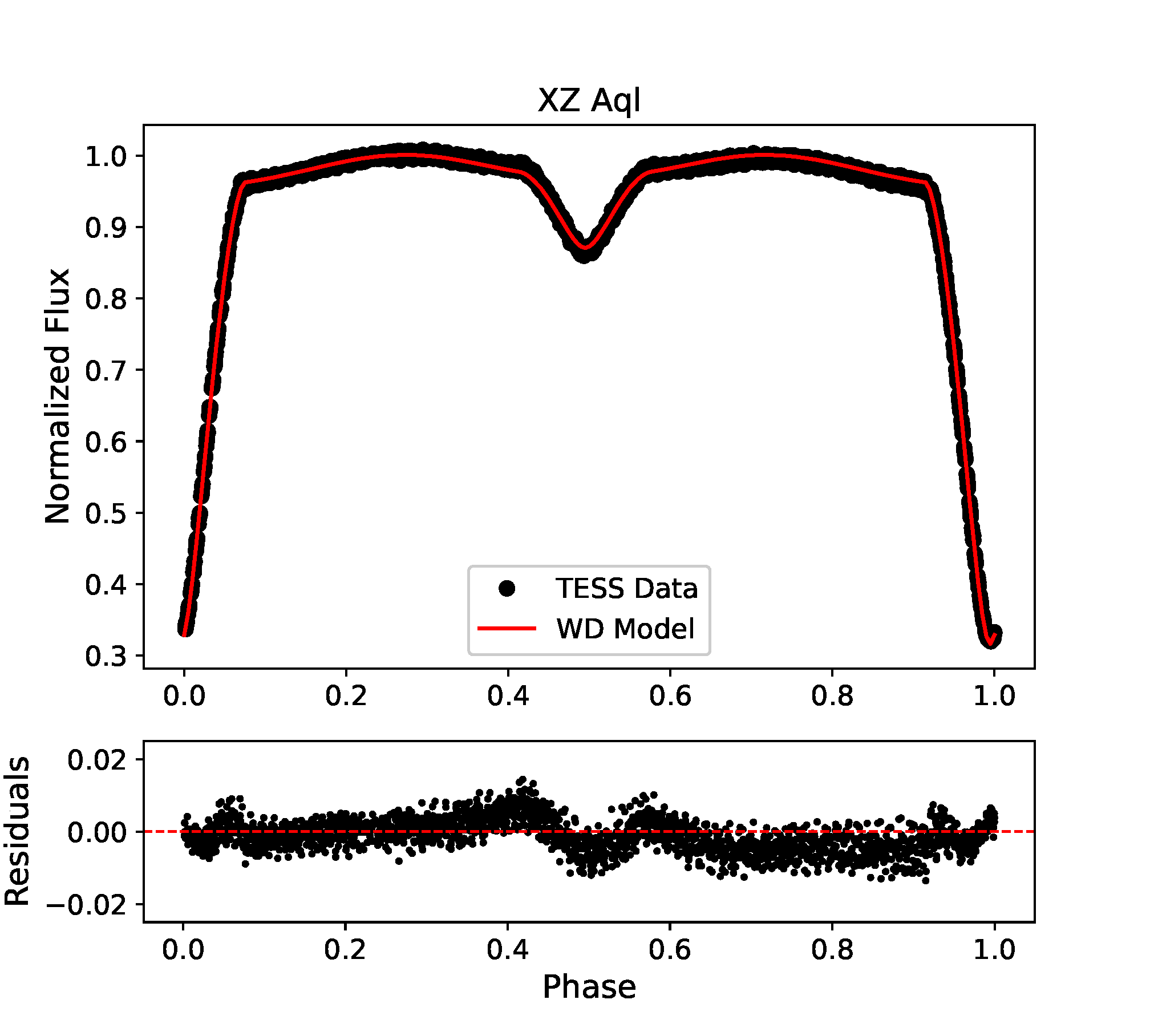}
 \end{minipage}
 \begin{minipage}[b]{0.33\textwidth}
 \includegraphics[alt={Binary light curve model of Y Cam Aql},height=4.8cm, width=1.0\textwidth]{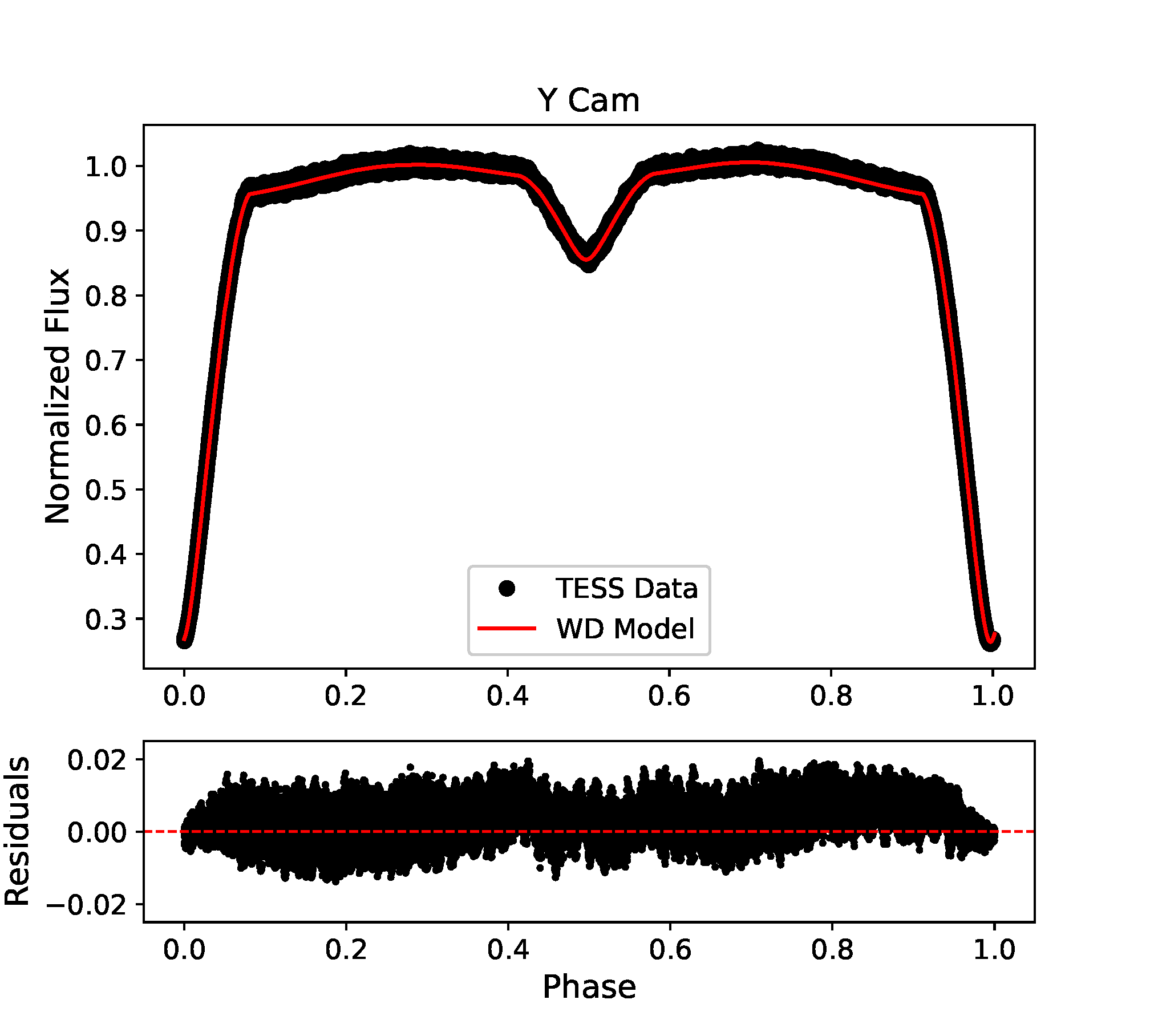}
 \end{minipage}
 \begin{minipage}[b]{0.33\textwidth}
 \includegraphics[alt={Binary light curve model of AB Cas},height=4.8cm, width=1.0\textwidth]{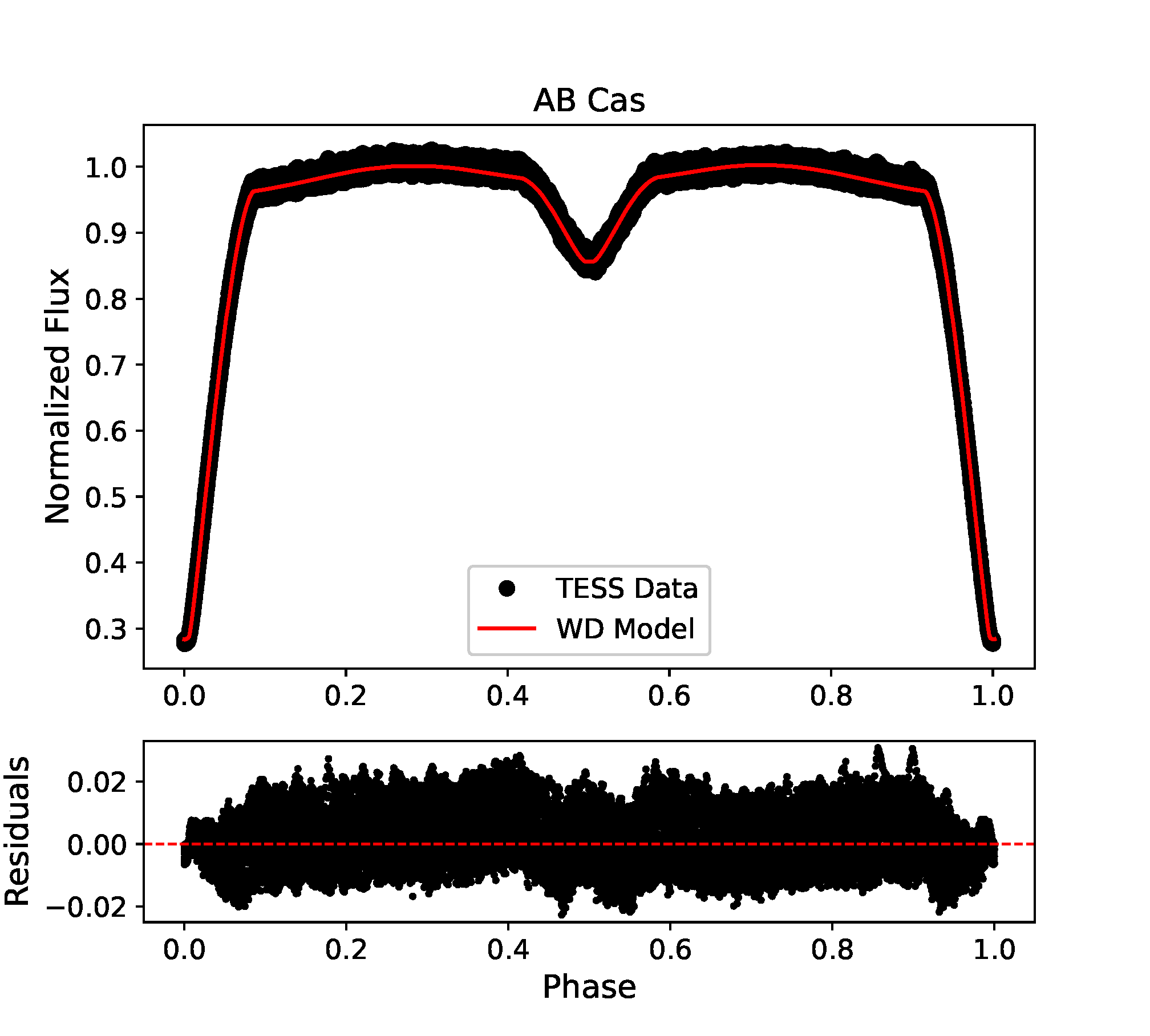}
 \end{minipage}
  \begin{minipage}[b]{0.33\textwidth}
 \includegraphics[alt={Binary light curve model of RZ Cas},height=4.8cm, width=1.0\textwidth]{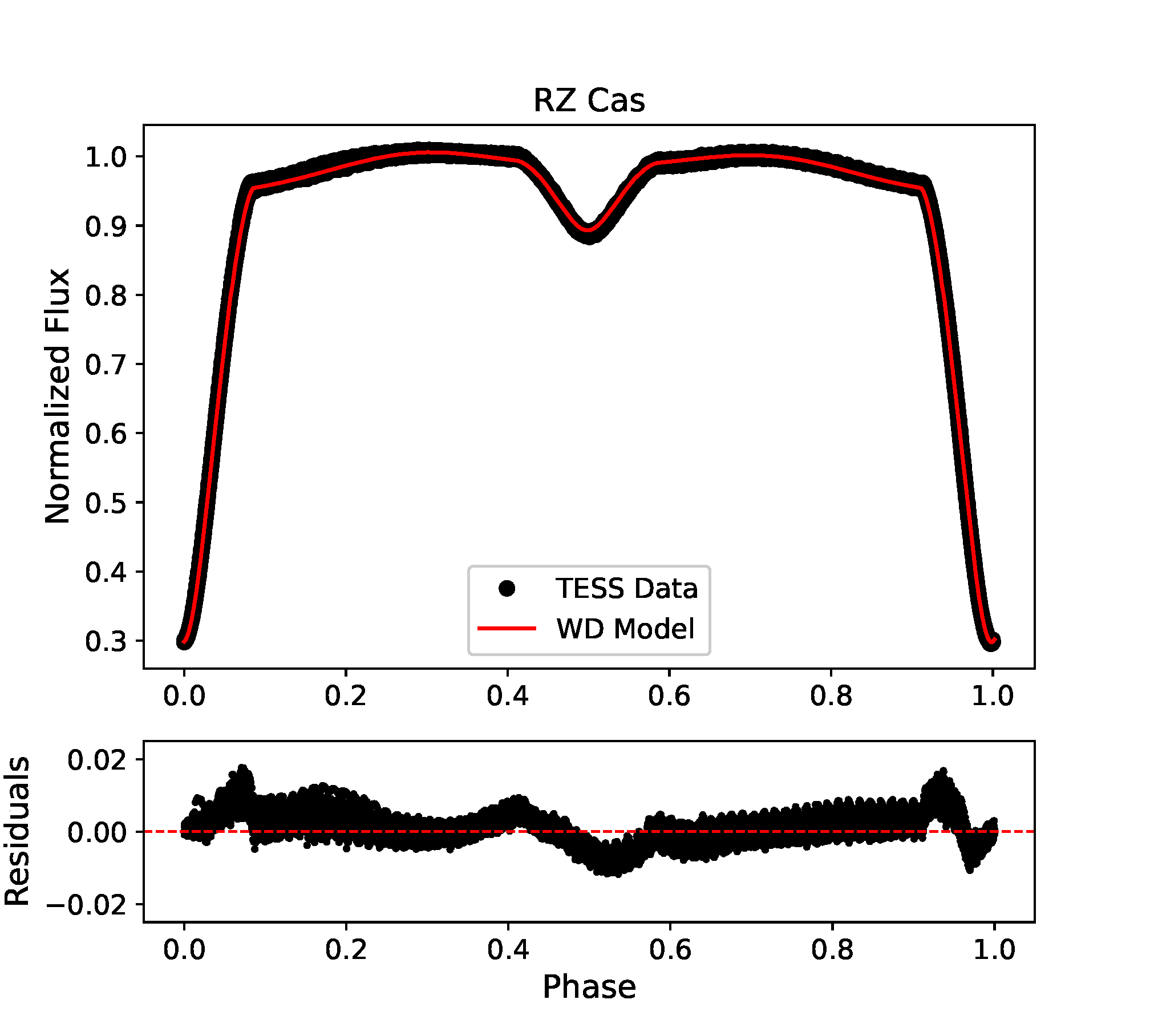}
 \end{minipage}
 \begin{minipage}[b]{0.33\textwidth}
  \includegraphics[alt={Binary light curve model of R CMa},height=4.8cm, width=1\textwidth]{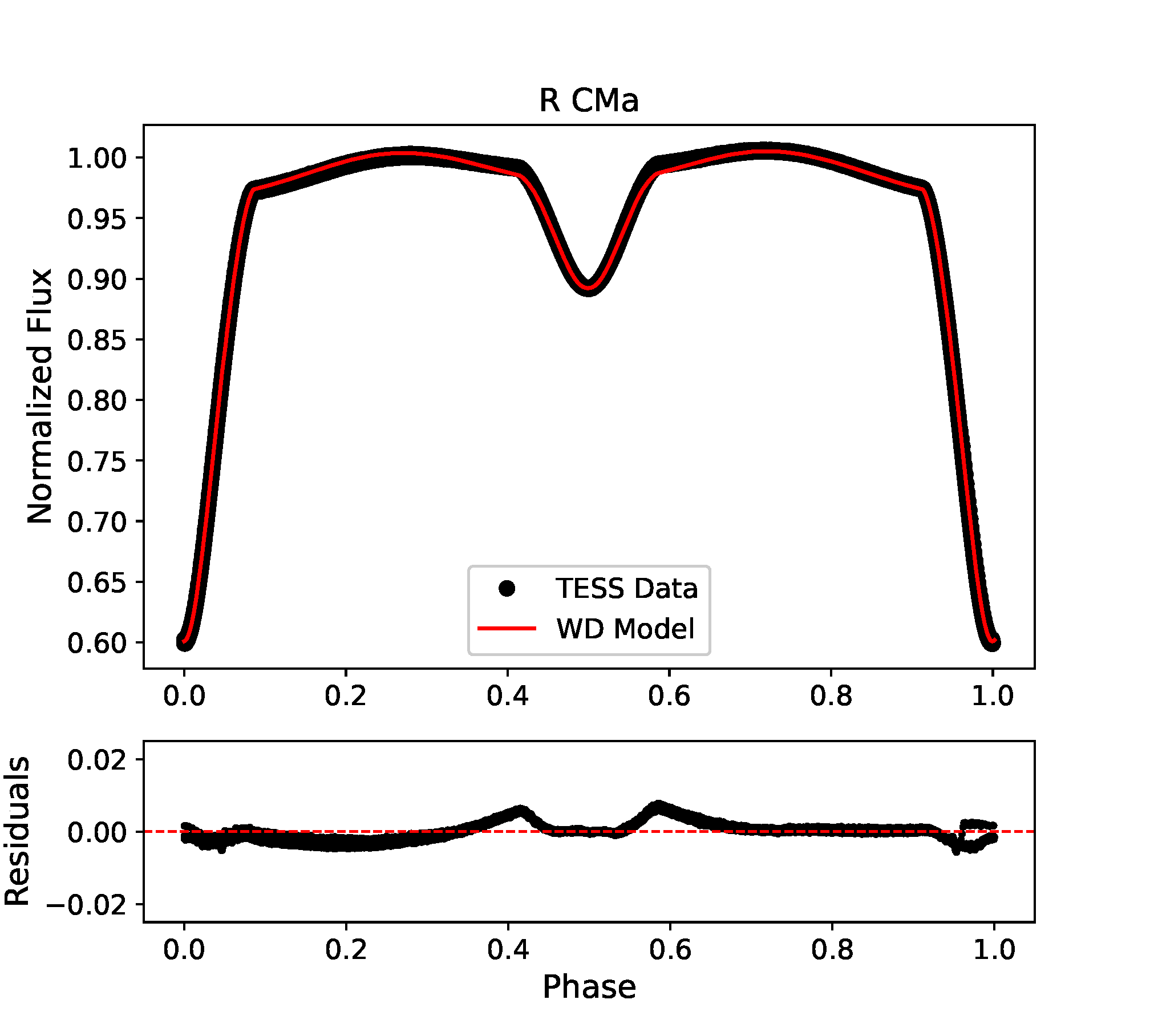}
  \end{minipage}
  \begin{minipage}[b]{0.33\textwidth}
  \includegraphics[alt={Binary light curve model of UW Cyg},height=4.8cm, width=1\textwidth]{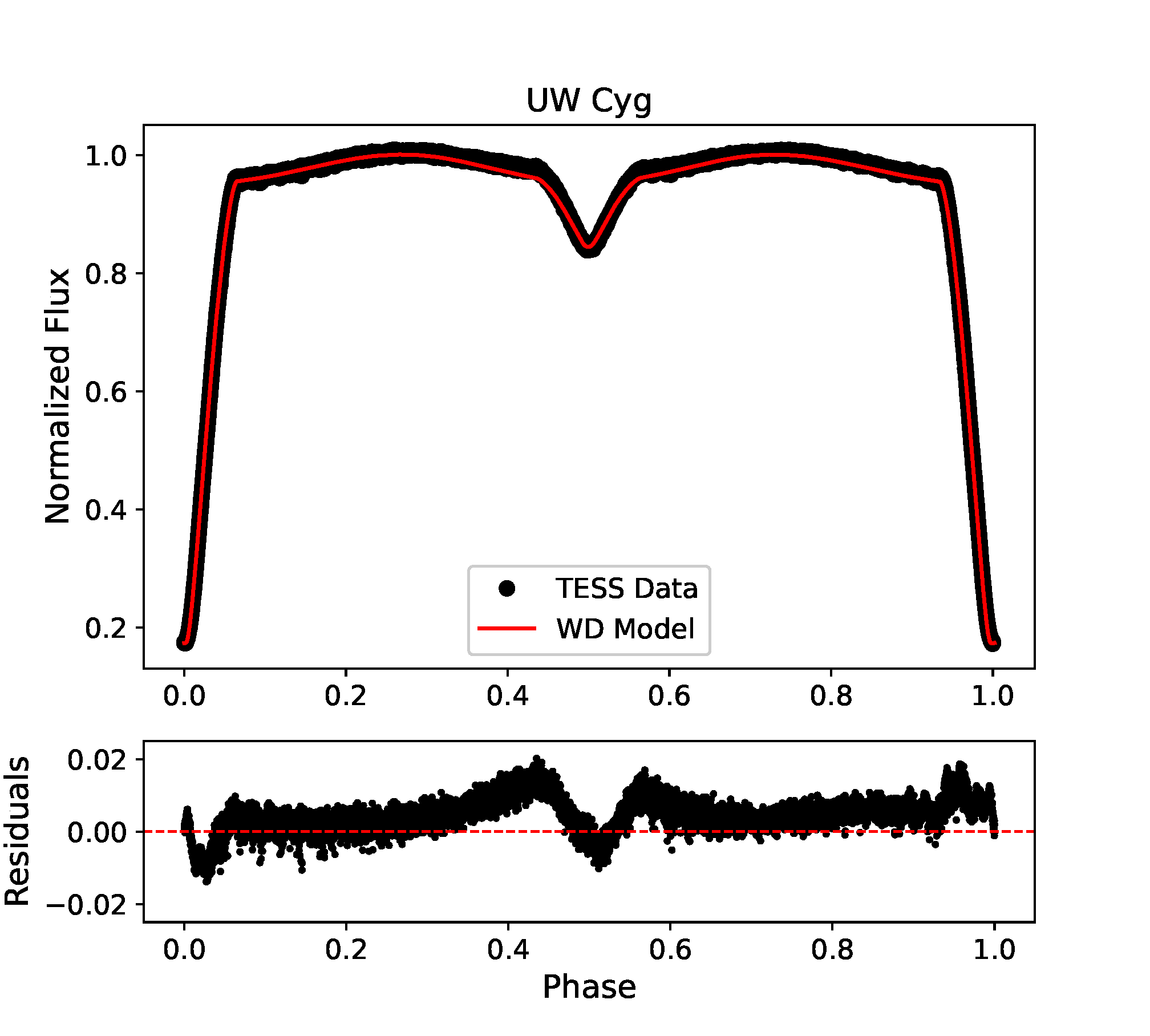}
  \end{minipage}
  \begin{minipage}[b]{0.33\textwidth}
  \includegraphics[alt={Binary light curve model of BW Del},height=4.8cm, width=1\textwidth]{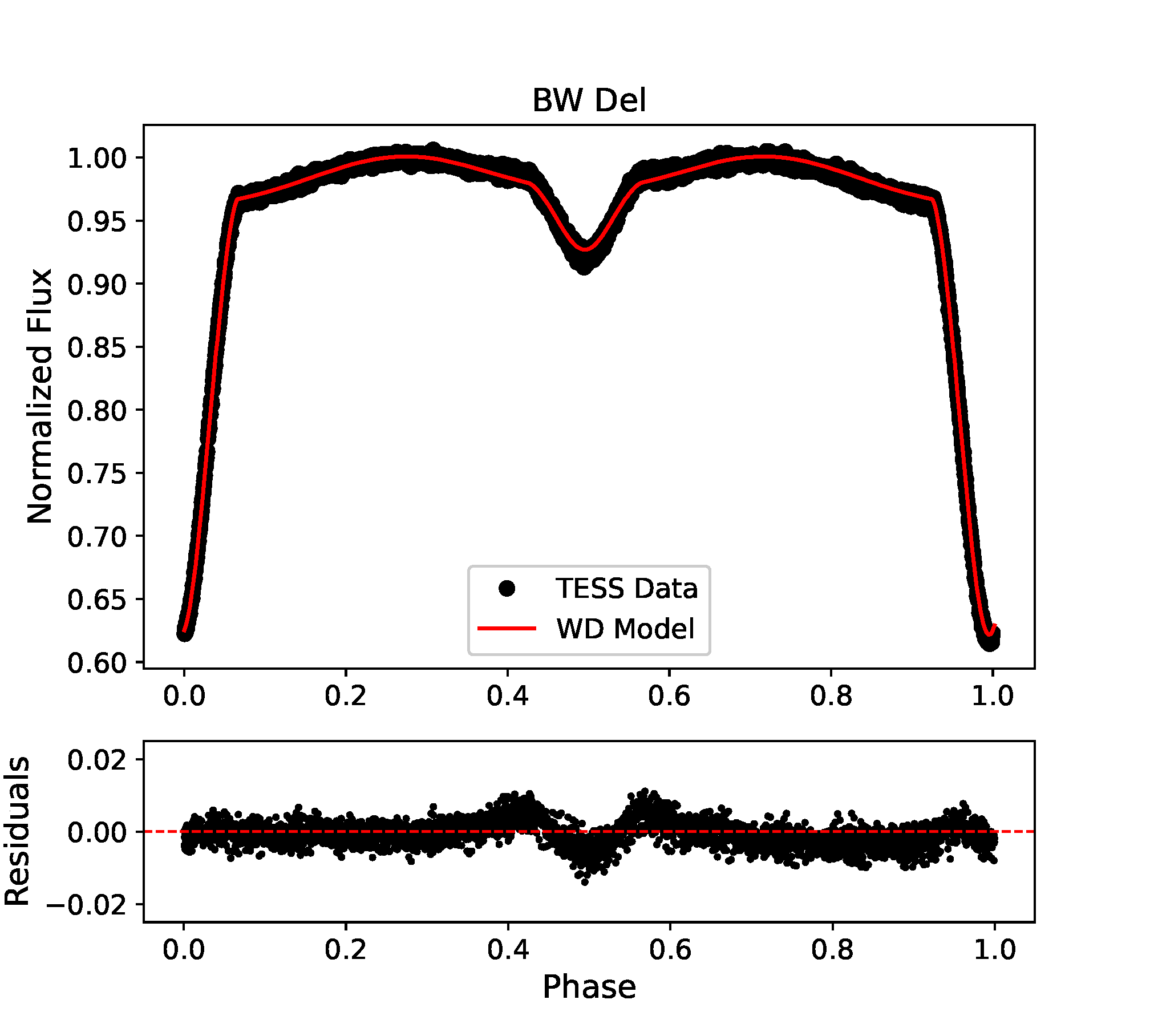}
 \end{minipage}
  \begin{minipage}[b]{0.33\textwidth}
  \includegraphics[alt={Binary light curve model of SX Dra},height=4.8cm, width=1\textwidth]{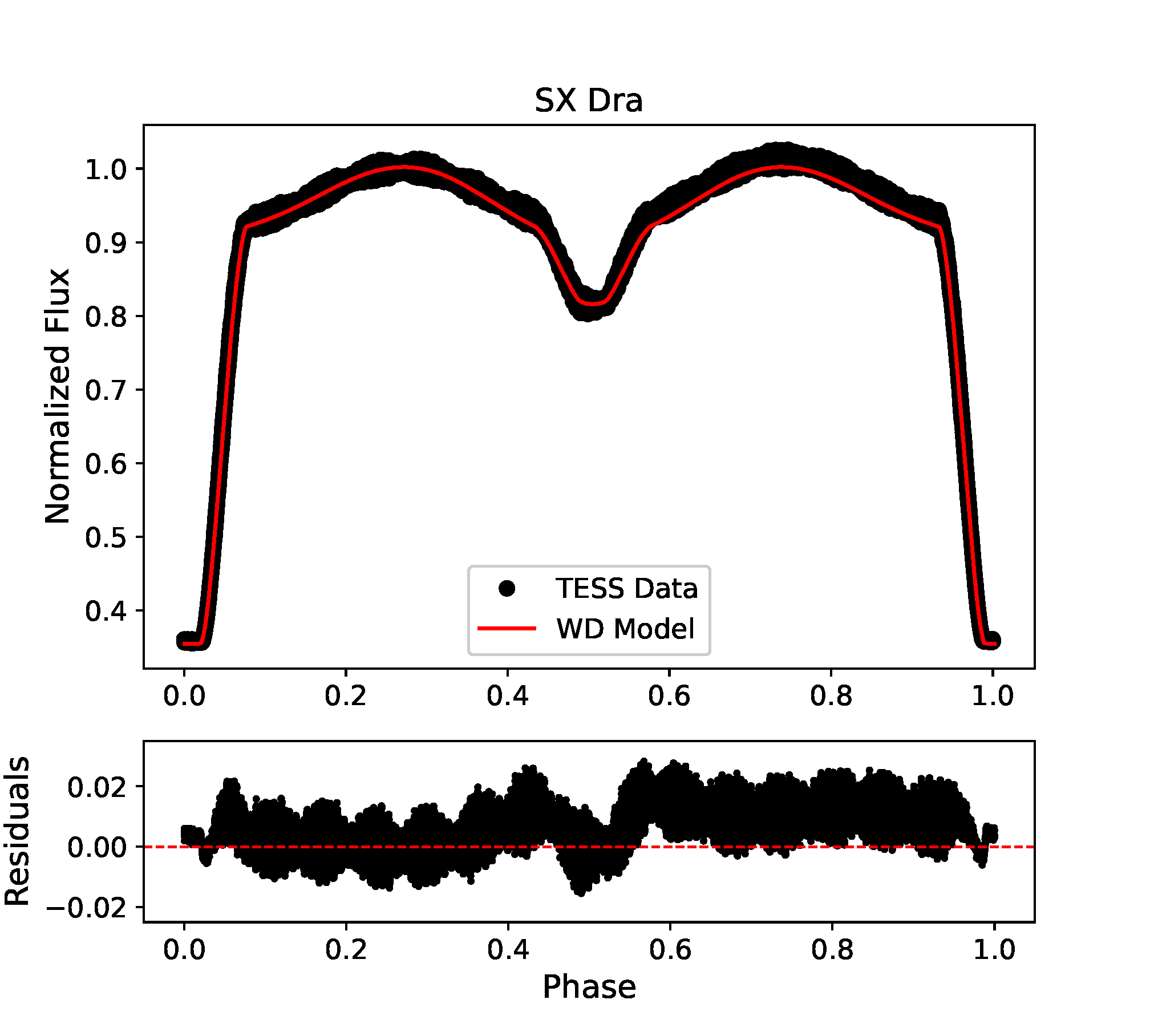}
 \end{minipage}
 \begin{minipage}[b]{0.33\textwidth}
  \includegraphics[alt={Binary light curve model of TW Dra},height=4.8cm, width=1\textwidth]{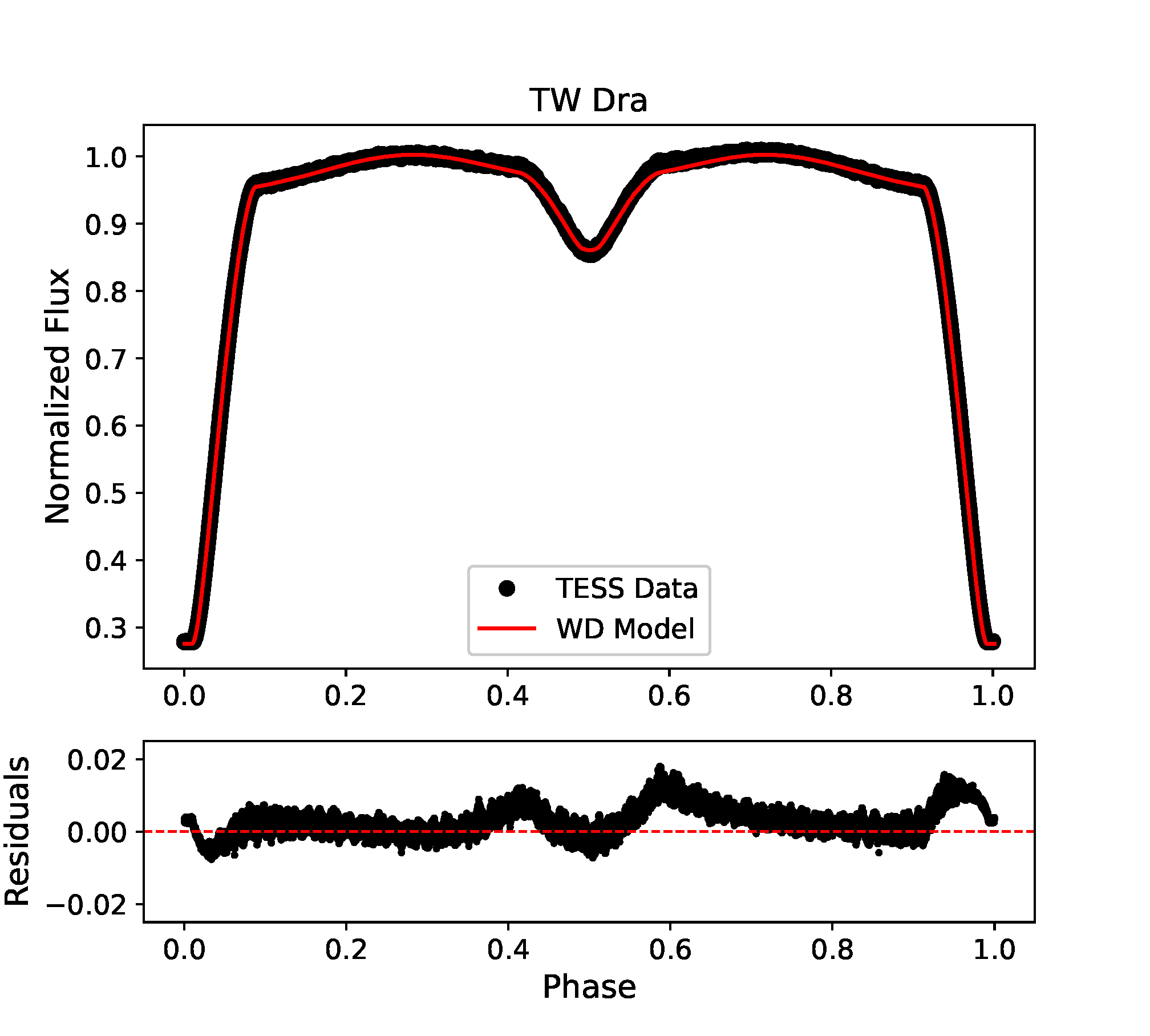}
  \end{minipage}
  \begin{minipage}[b]{0.33\textwidth}
  \includegraphics[alt={Binary light curve model of TZ ERi},height=4.8cm, width=1\textwidth]{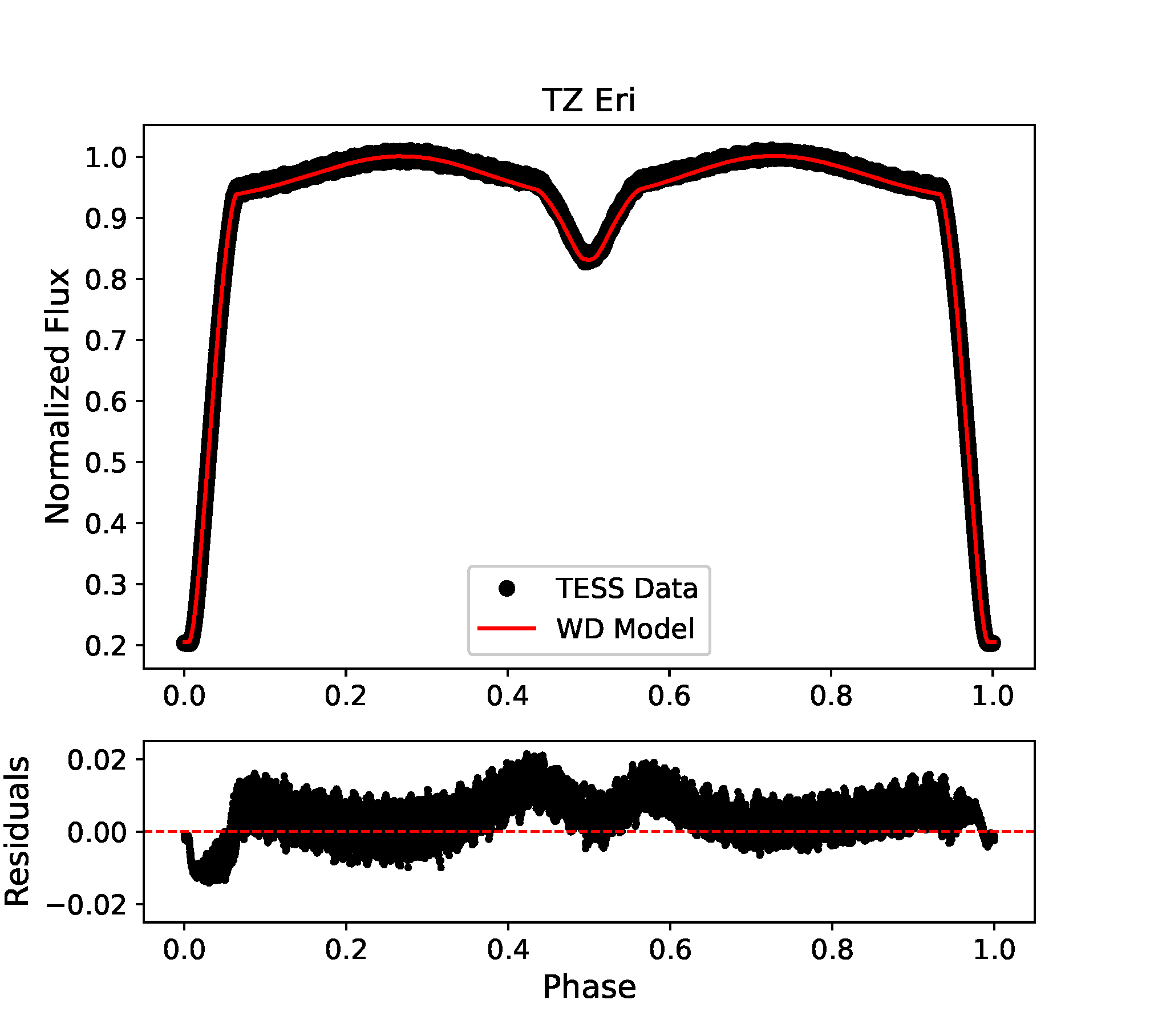}
  \end{minipage}
   \begin{minipage}[b]{0.33\textwidth}
  \includegraphics[alt={Binary light curve model of CT Her},height=4.8cm, width=1\textwidth]{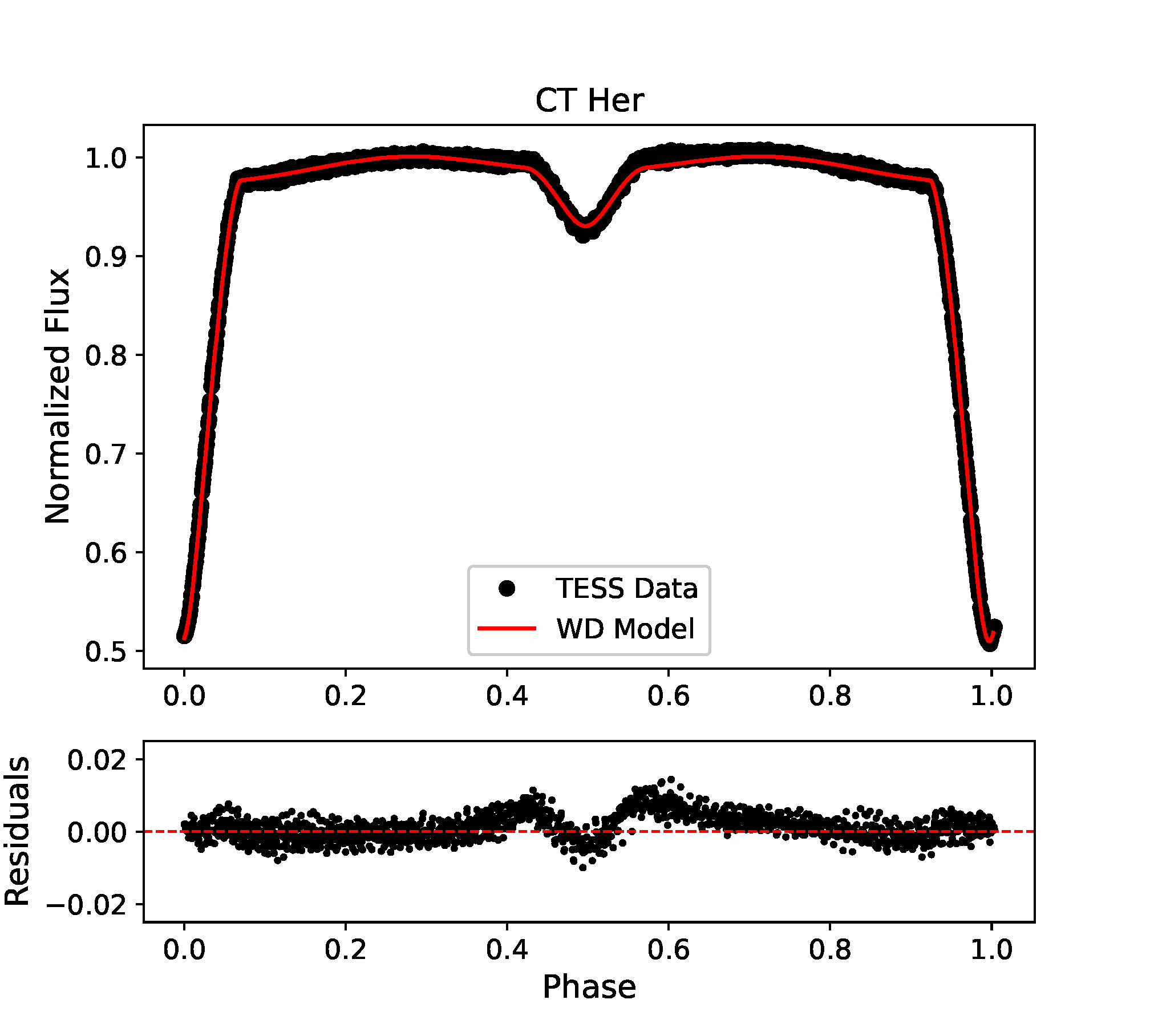}
 \end{minipage}
\begin{minipage}[b]{0.33\textwidth}
 \includegraphics[alt={Binary light curve model of RX Hya},height=4.8cm, width=1\textwidth]{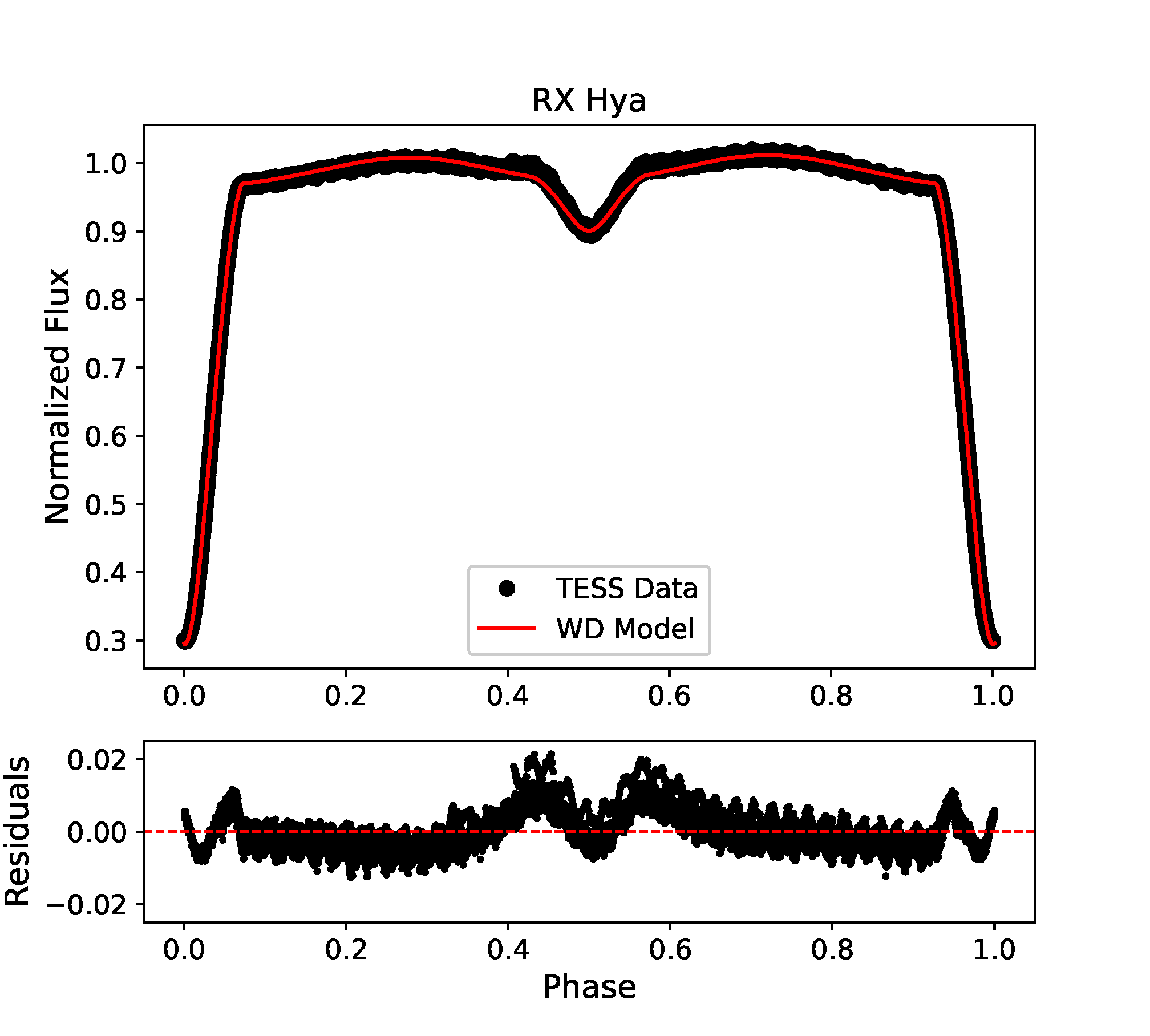}
 \end{minipage}
 \begin{minipage}[b]{0.33\textwidth}
  \includegraphics[alt={Binary light curve model of Y Leo},height=4.8cm, width=1\textwidth]{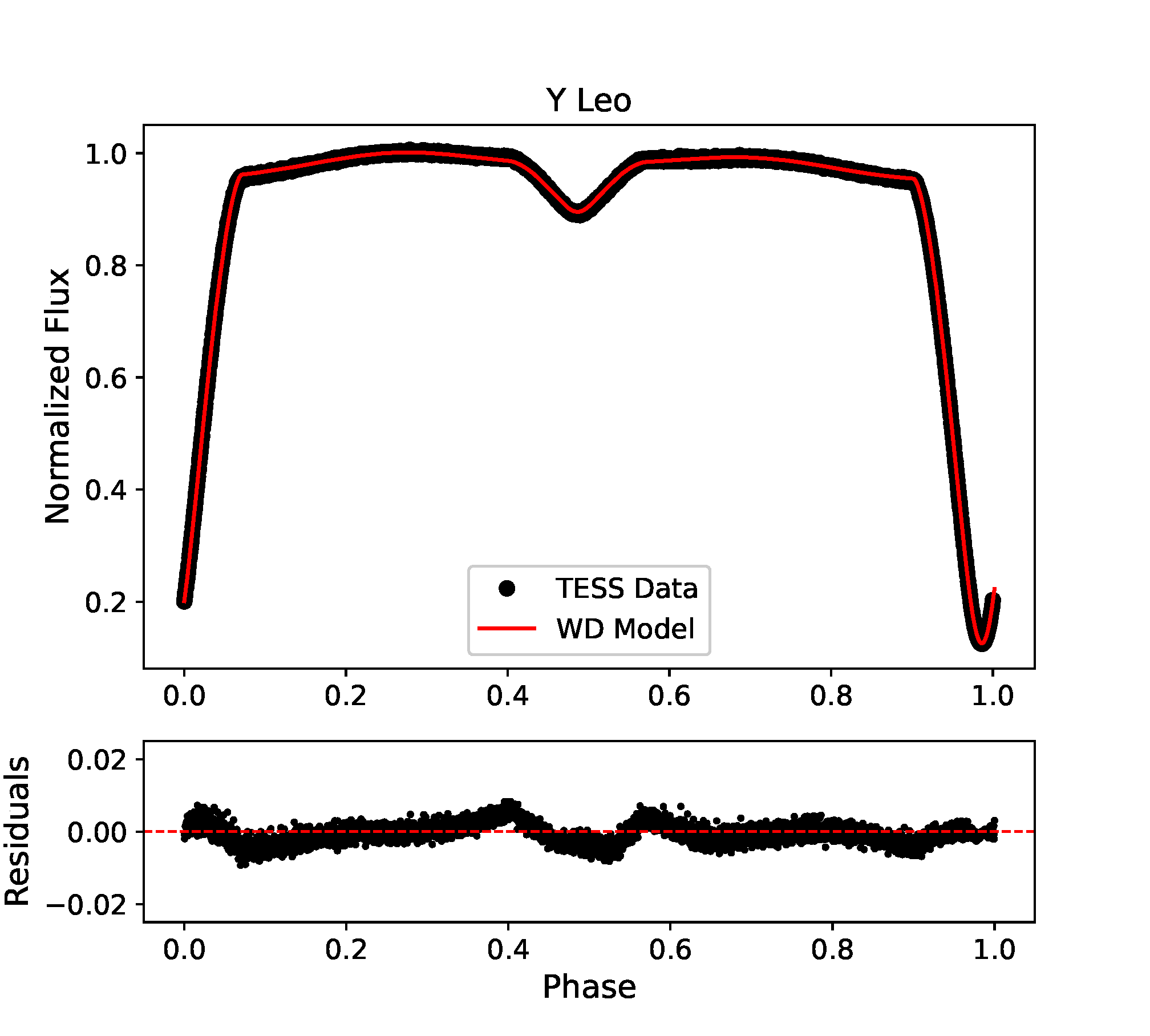}
 \end{minipage}
\begin{minipage}[b]{0.33\textwidth}
  \includegraphics[alt={Binary light curve model of FR Ori},height=4.8cm, width=1\textwidth]{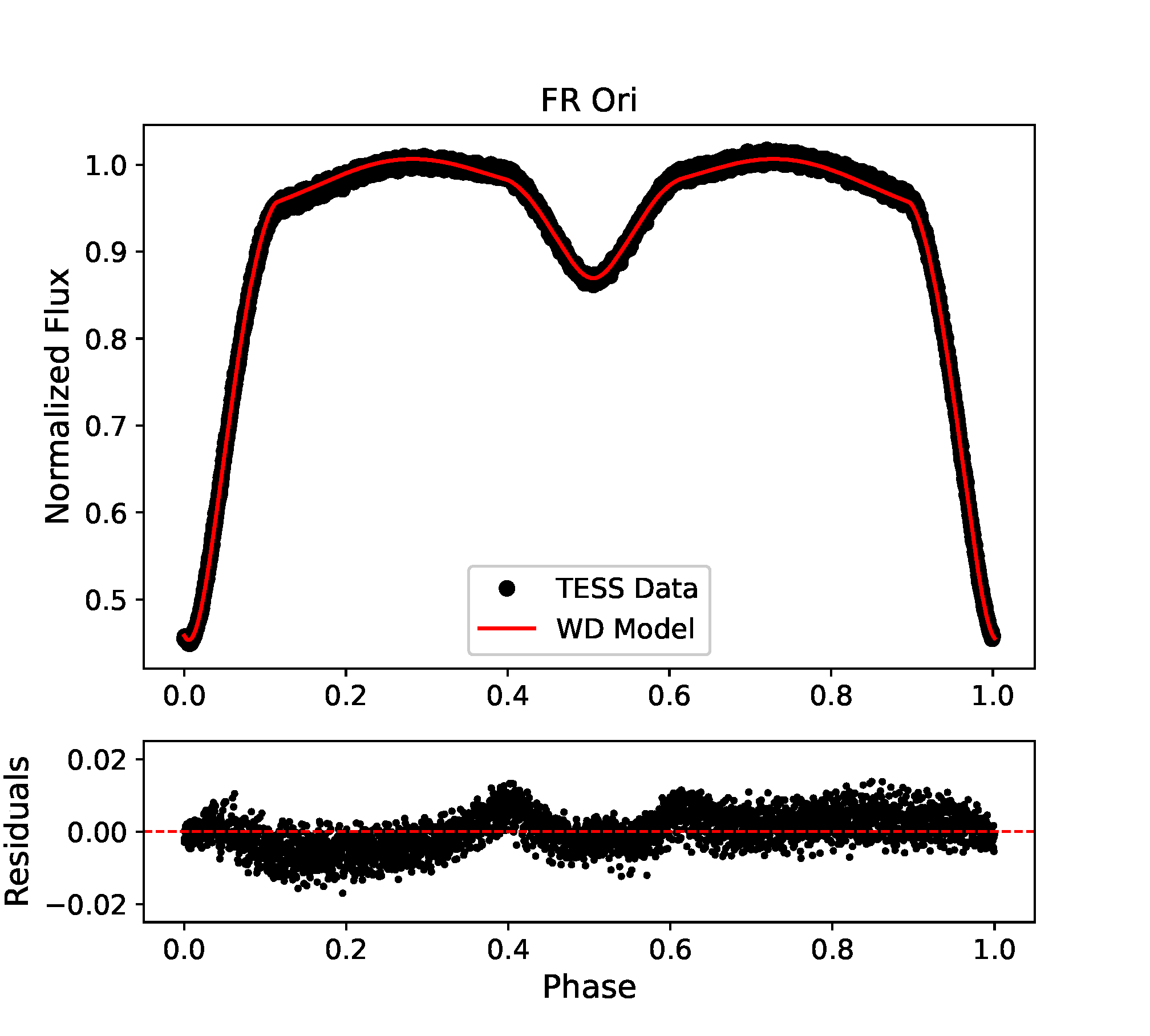}
  \end{minipage}
    \begin{minipage}[b]{0.33\textwidth}
  \includegraphics[alt={Binary light curve model of AC Tau},height=4.8cm, width=1\textwidth]{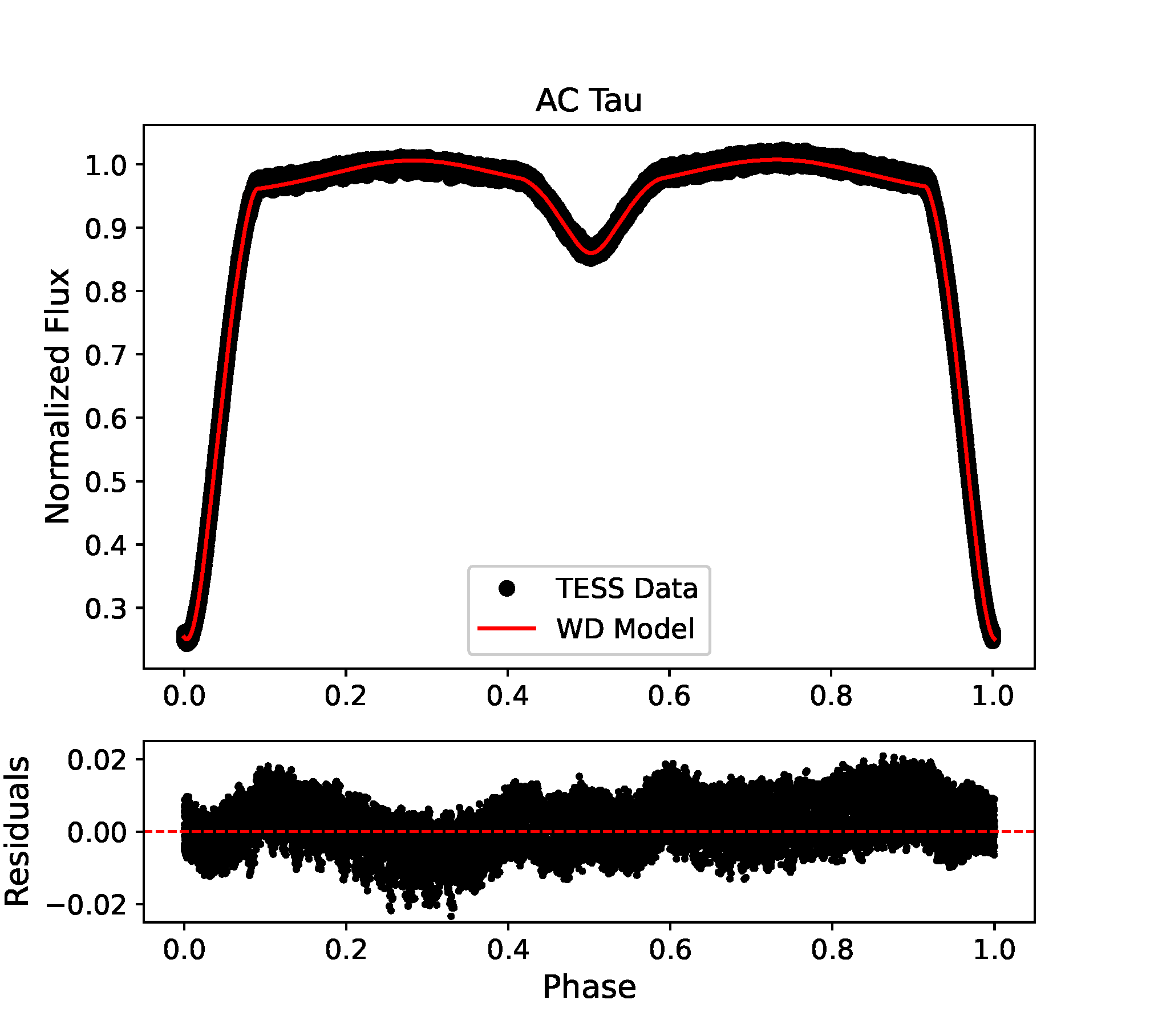}
  \end{minipage}
\caption{Binary modeling and its residuals for all systems. Black dots represent TESS observations, while red continuous lines represent WD models.}\label{fig:all_LC}
\end{figure*}

\setcounter{figure}{1}
\renewcommand{\thefigure}{A\arabic{figure}}
\begin{figure*}
 \begin{minipage}[b]{0.33\textwidth}
 \includegraphics[alt={O-C diagram of XZ Aql},height=4.3cm, width=1\textwidth]{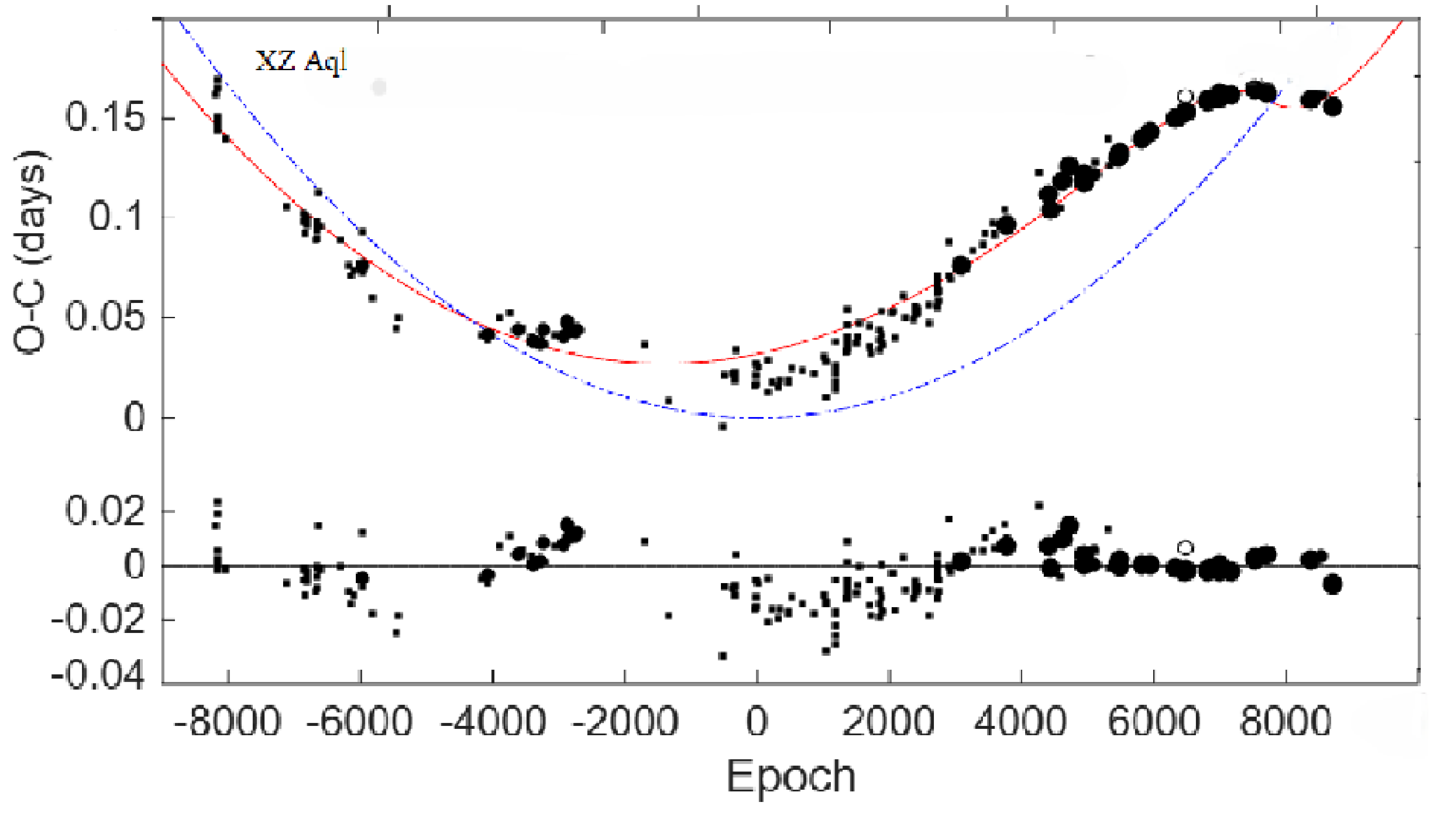}
 \end{minipage}
 \begin{minipage}[b]{0.33\textwidth}
 \includegraphics[alt={O-C diagram of Y Cam},height=4.3cm, width=1\textwidth]{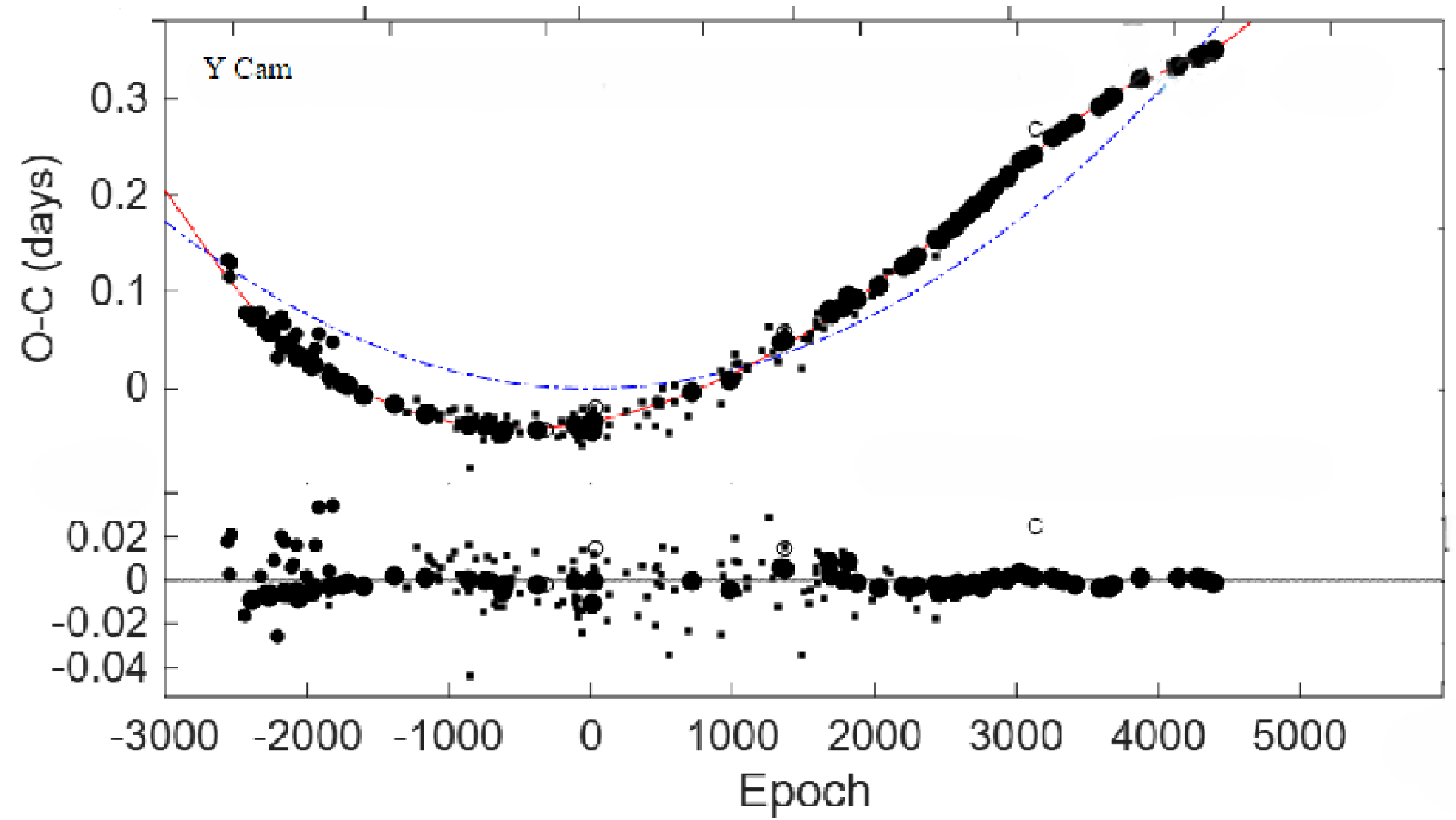}
 \end{minipage}
 \begin{minipage}[b]{0.33\textwidth}
 \includegraphics[alt={O-C diagram of AB Cas},height=4.3cm, width=1.0\textwidth]{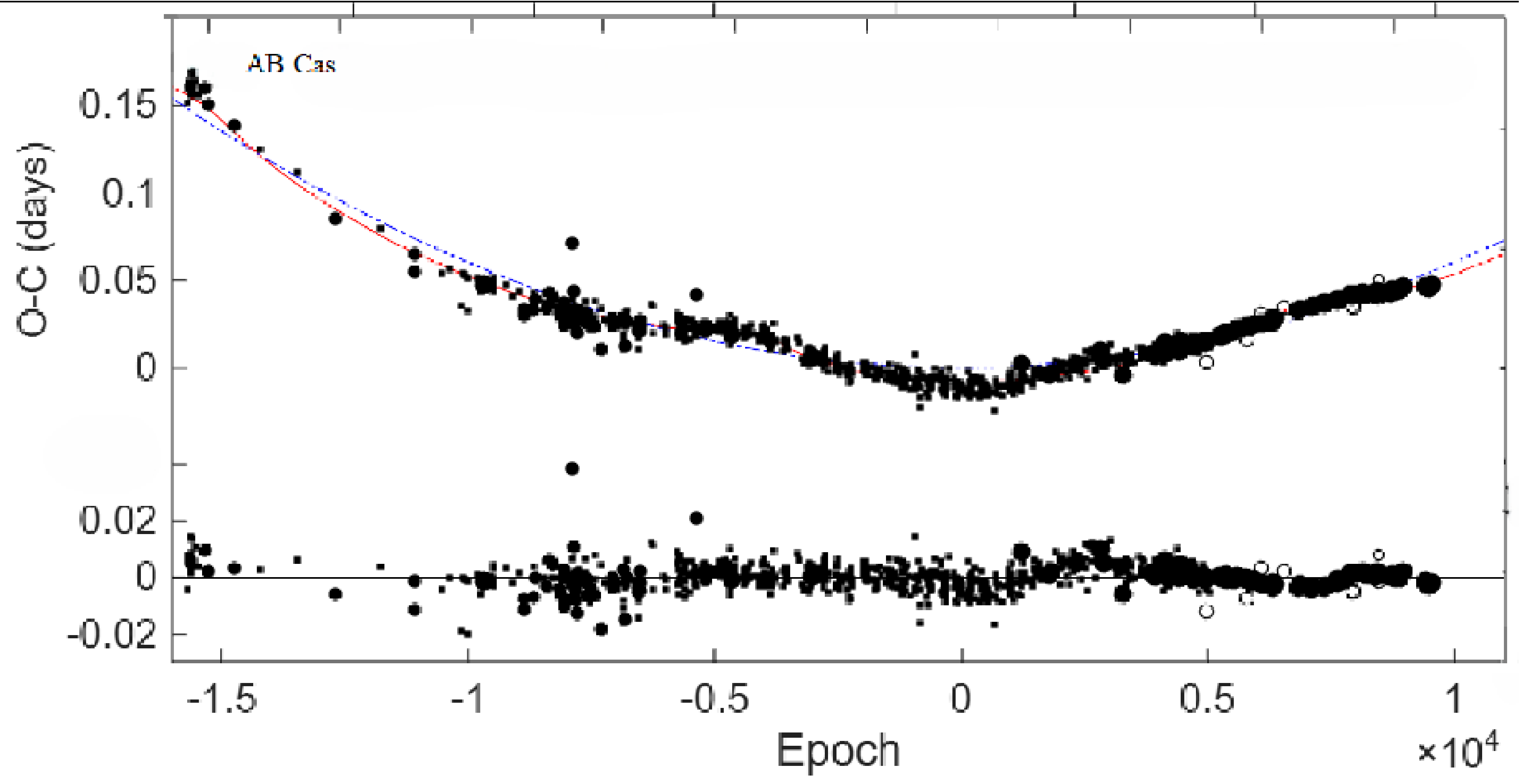}
 \end{minipage}
 \begin{minipage}[b]{0.33\textwidth}
 \includegraphics[alt={O-C diagram of RZ Cas},height=4.3cm, width=1\textwidth]{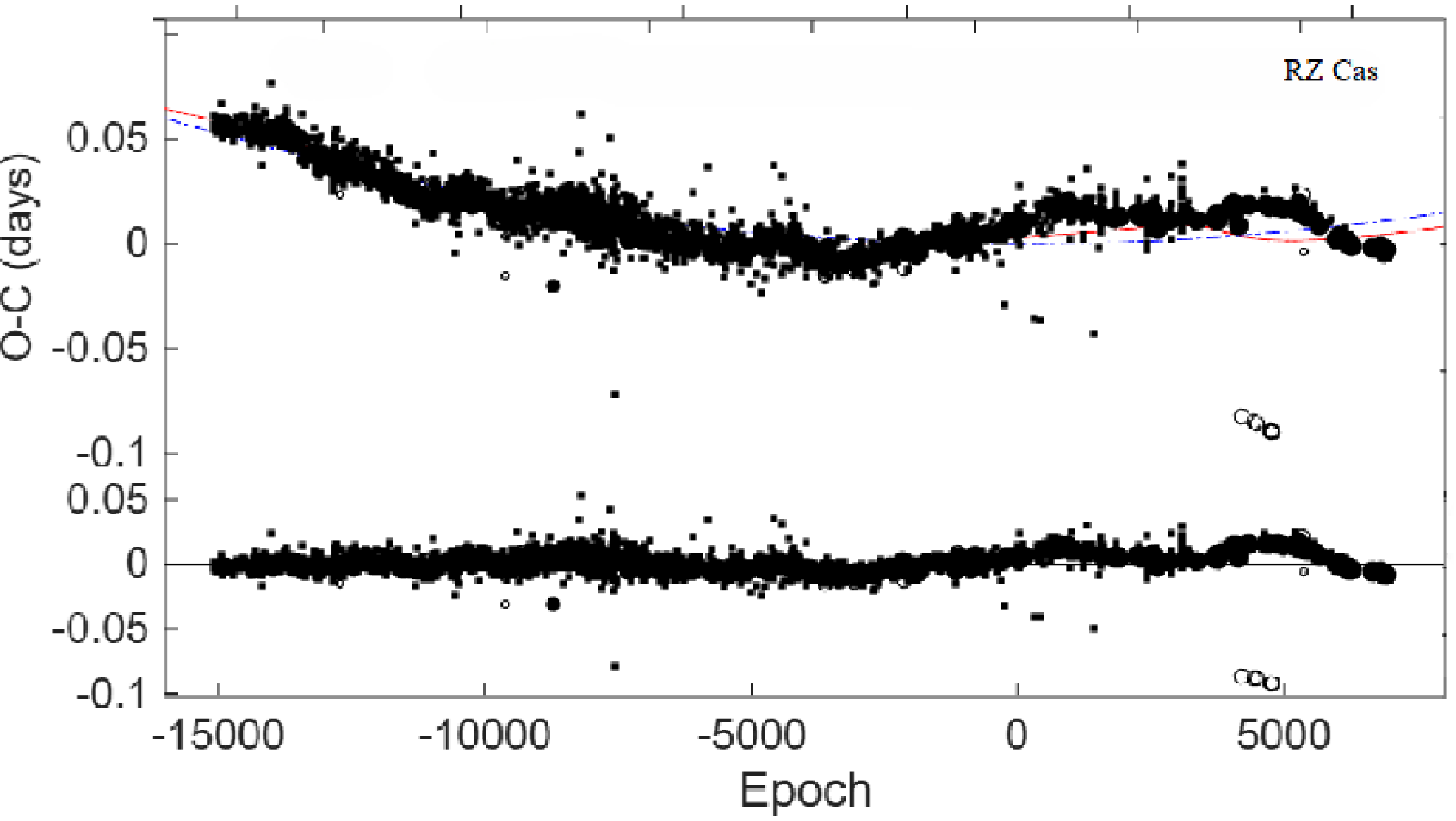}
 \end{minipage}
 \begin{minipage}[b]{0.33\textwidth}
 \includegraphics[alt={O-C diagram of R CMa},height=4.3cm, width=1\textwidth]{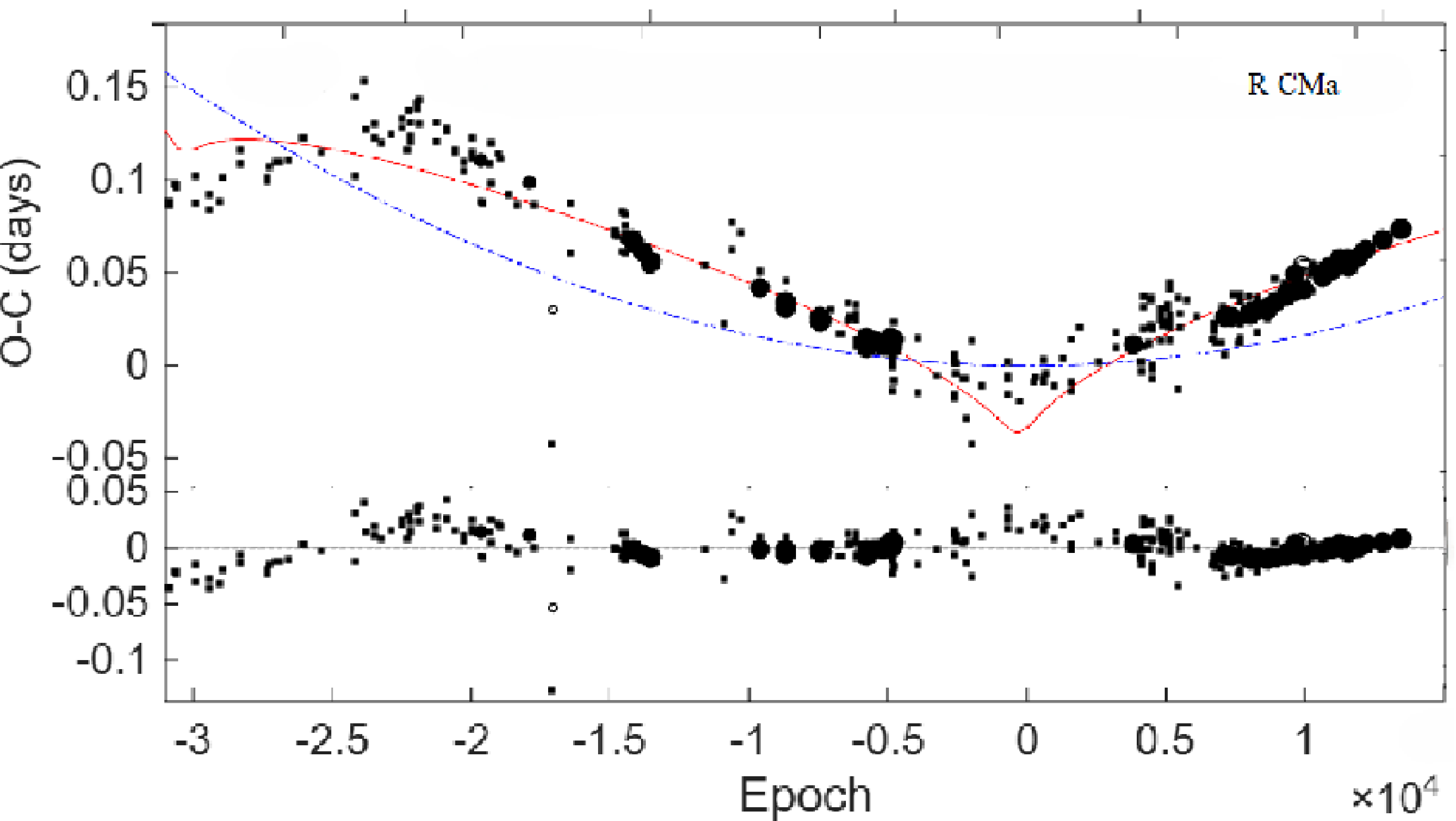}
 \end{minipage}
 \begin{minipage}[b]{0.33\textwidth}
 \includegraphics[alt={O-C diagram of XX Cep},height=4.3cm, width=1\textwidth]{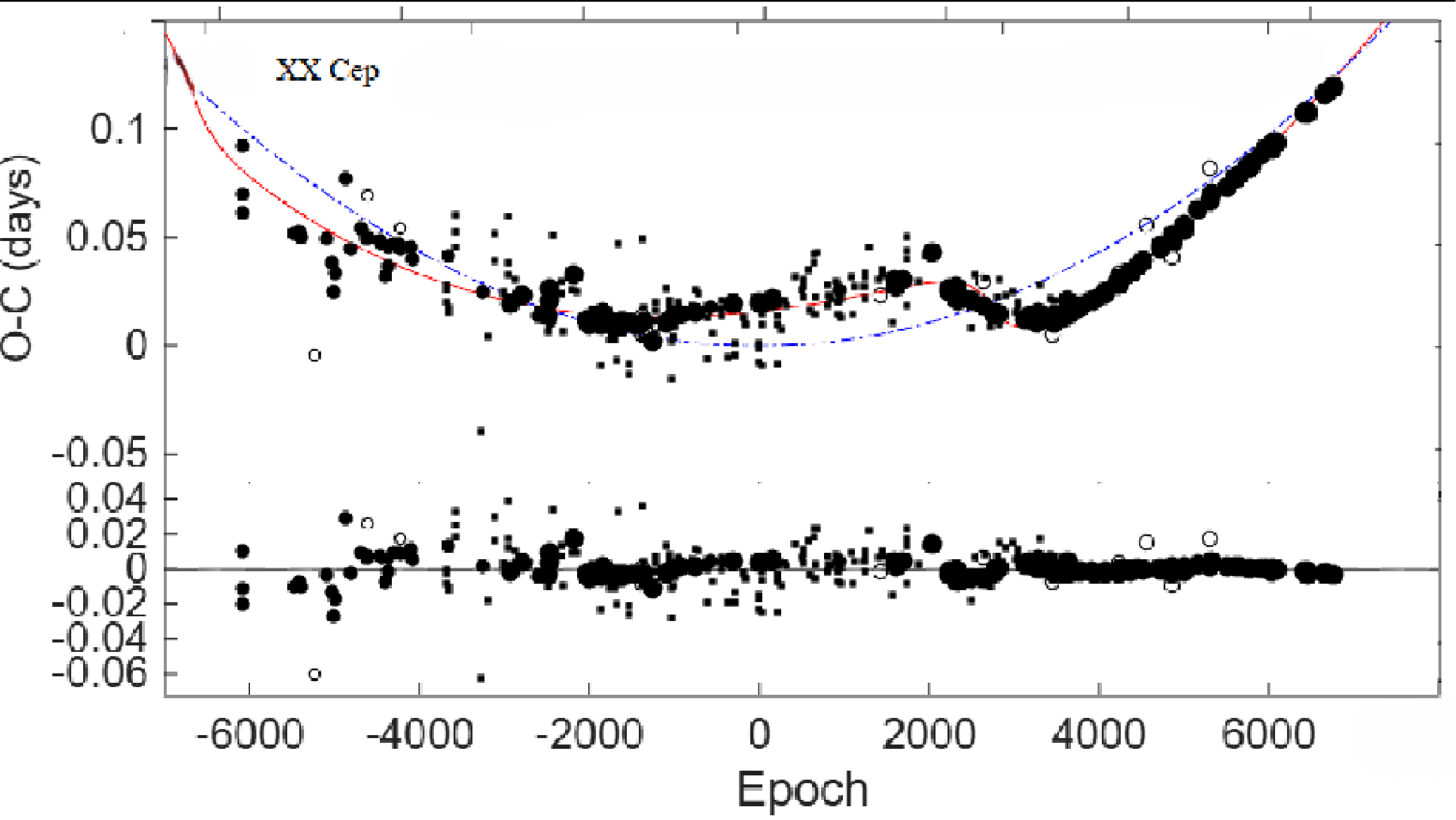}
 \end{minipage}
 \begin{minipage}[b]{0.33\textwidth}
 \includegraphics[alt={O-C diagram of UW Cyg},height=4.3cm, width=1\textwidth]{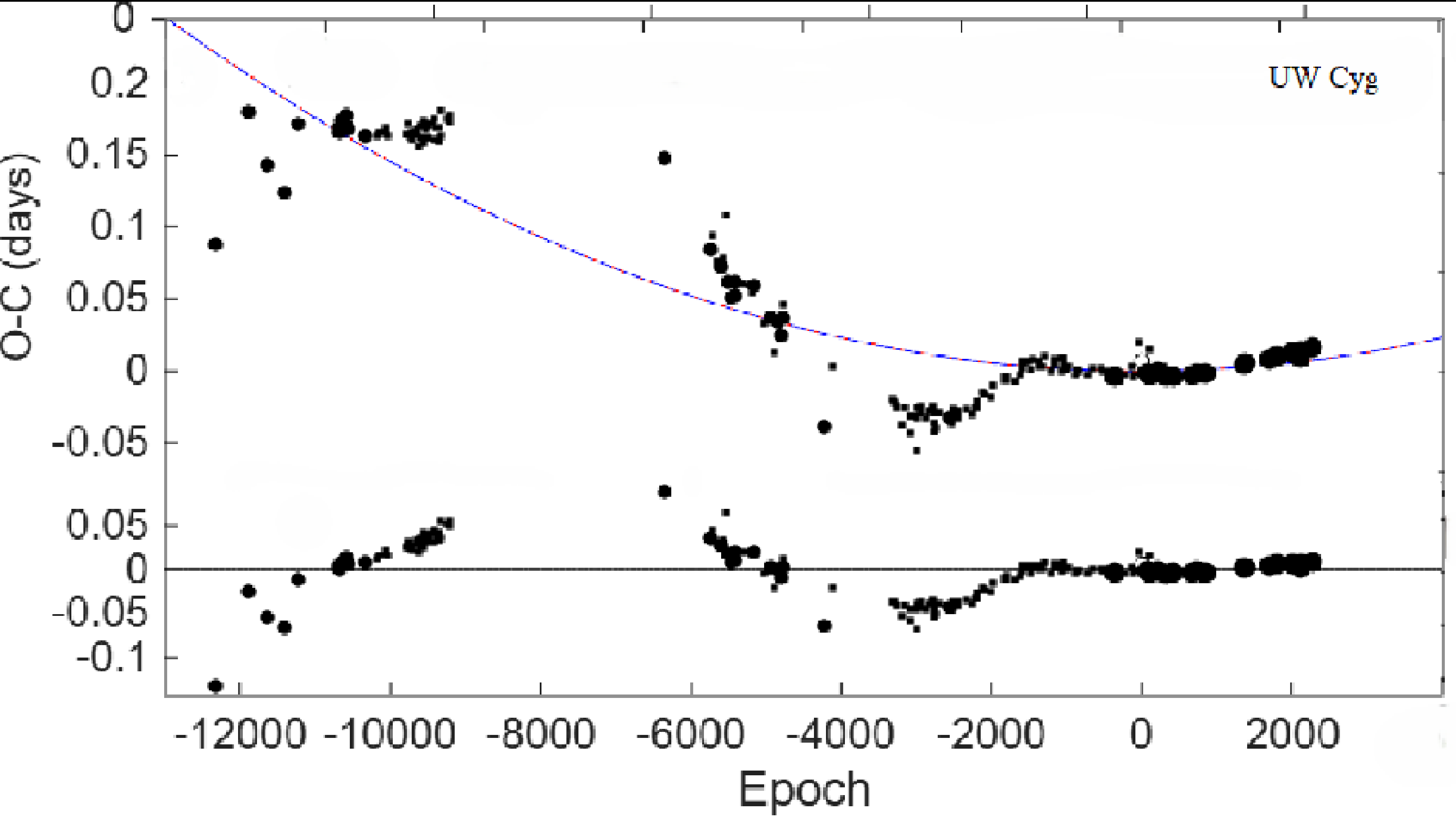}
 \end{minipage}
 \begin{minipage}[b]{0.33\textwidth}
 \includegraphics[alt={O-C diagram of SX Dra},height=4.3cm, width=1\textwidth]{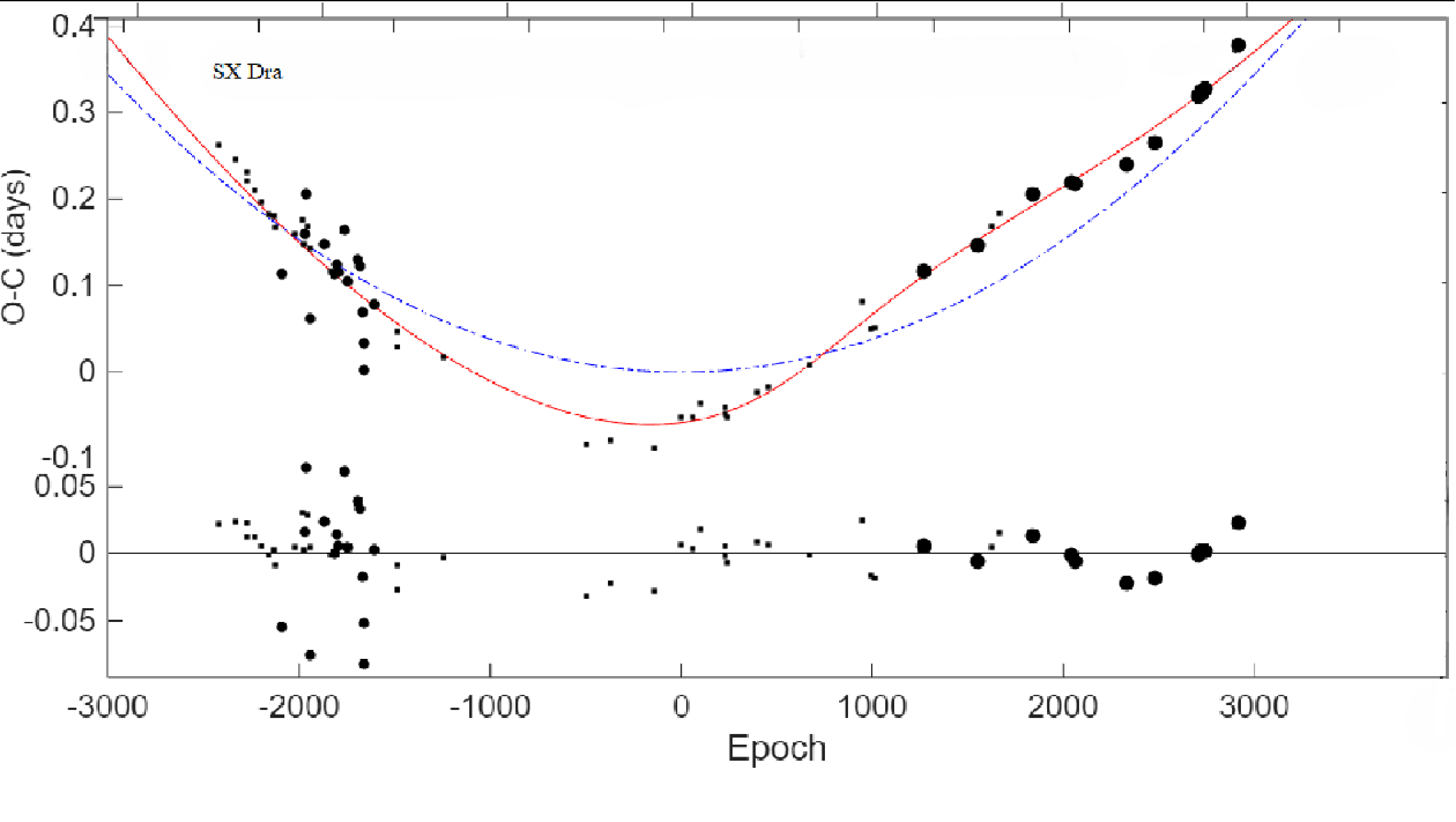}
 \end{minipage}
 \begin{minipage}[b]{0.33\textwidth}
 \includegraphics[alt={O-C diagram of TW Dra},height=4.3cm, width=1\textwidth]{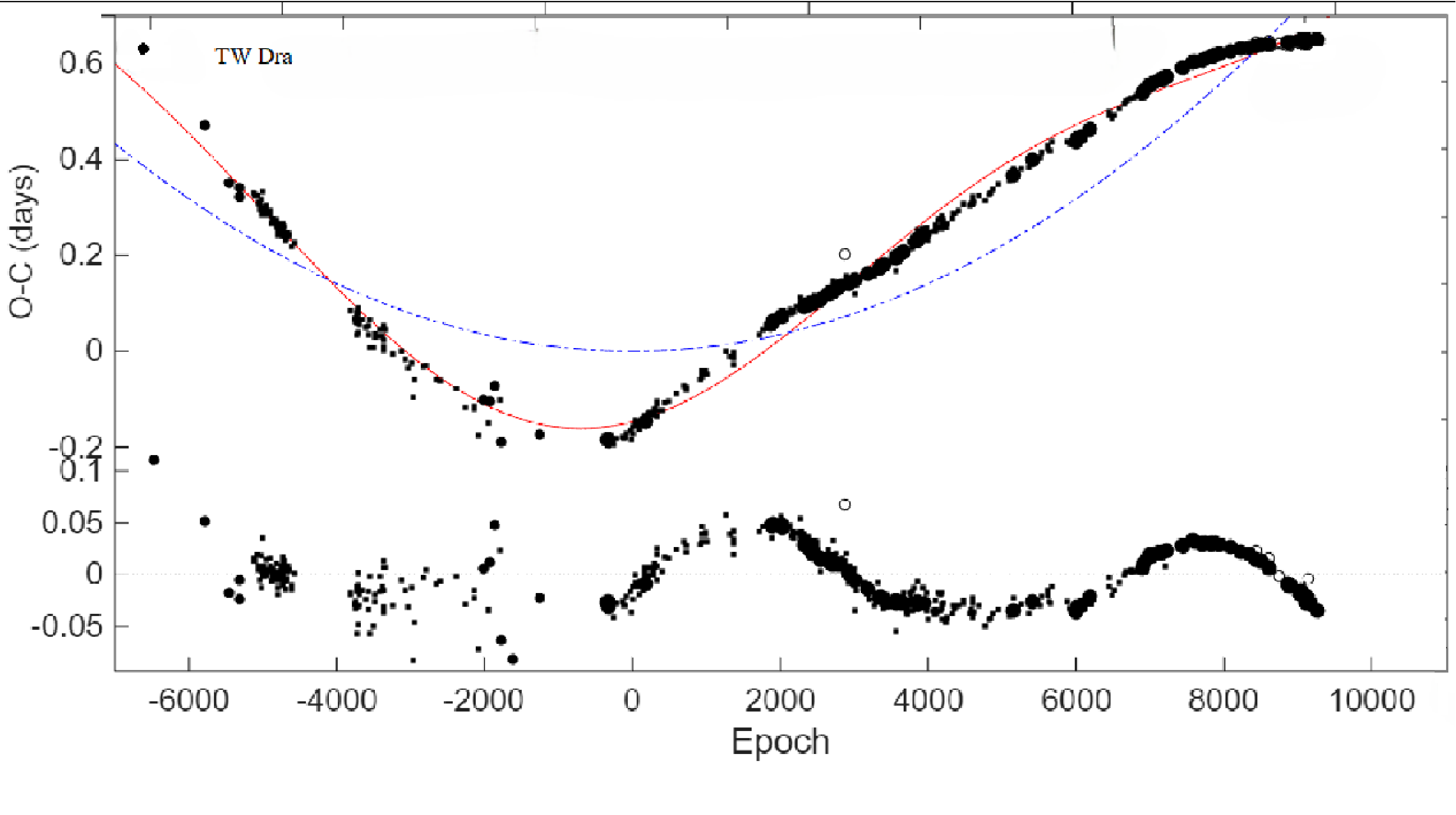}
 \end{minipage}
 \begin{minipage}[b]{0.33\textwidth}
 \includegraphics[alt={O-C diagram of TZ Eri},height=4.3cm, width=1\textwidth]{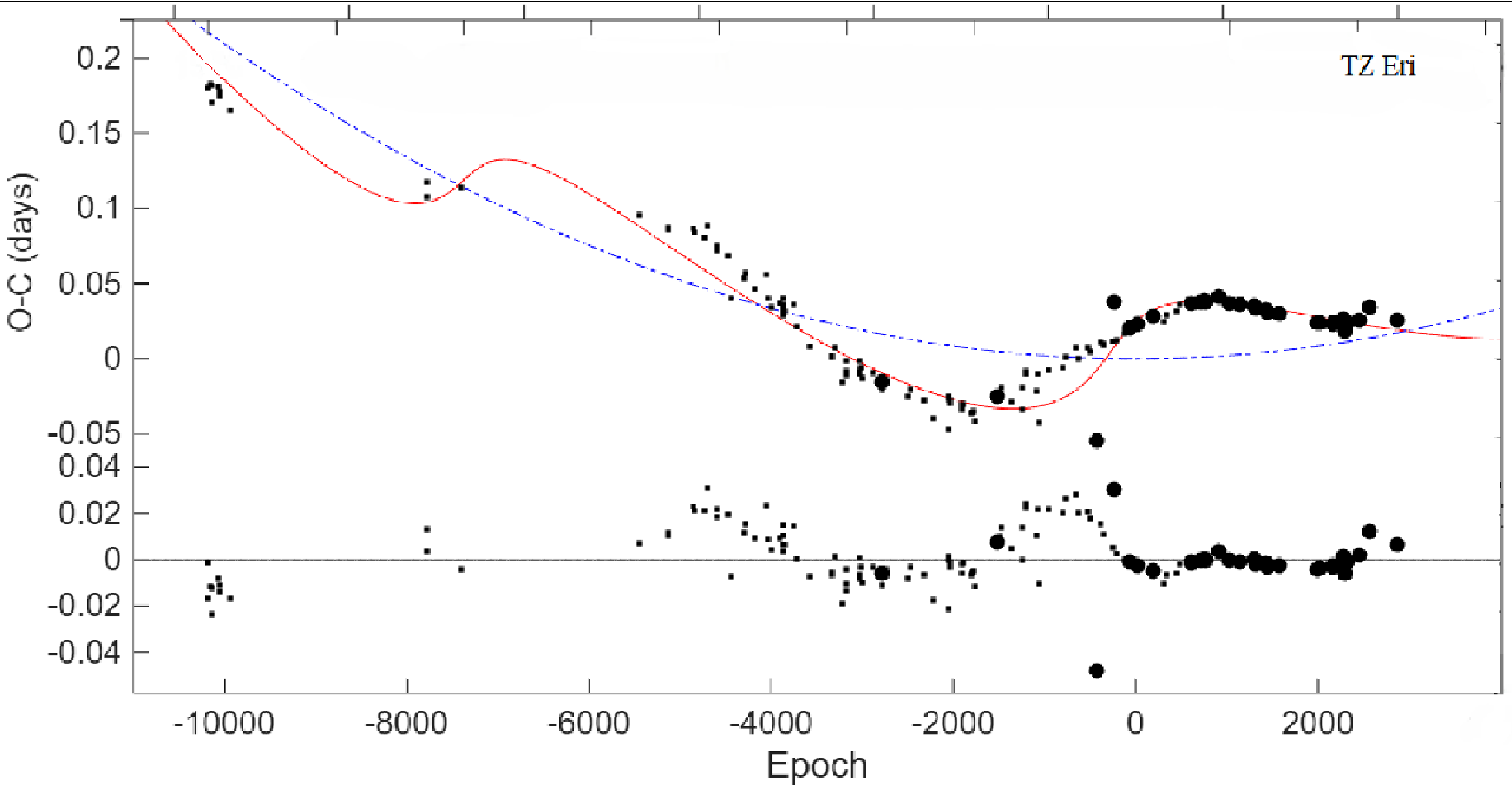}
 \end{minipage}
 \begin{minipage}[b]{0.33\textwidth}
 \includegraphics[alt={O-C diagram of CT Her},height=4.3cm, width=1.0\textwidth]{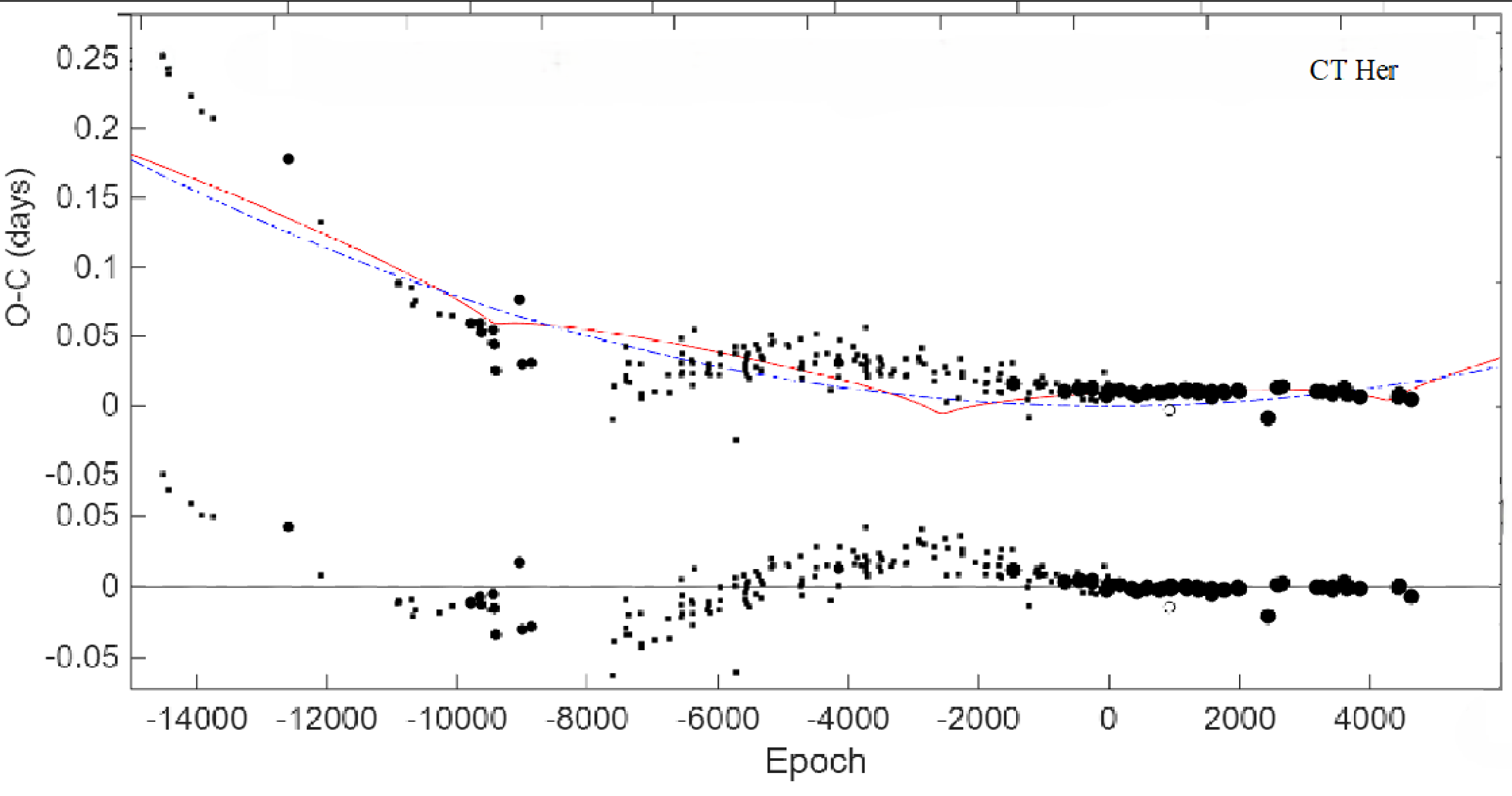}
 \end{minipage}
 \begin{minipage}[b]{0.33\textwidth}
 \includegraphics[alt={O-C diagram of RX Hya},height=4.3cm, width=1\textwidth]{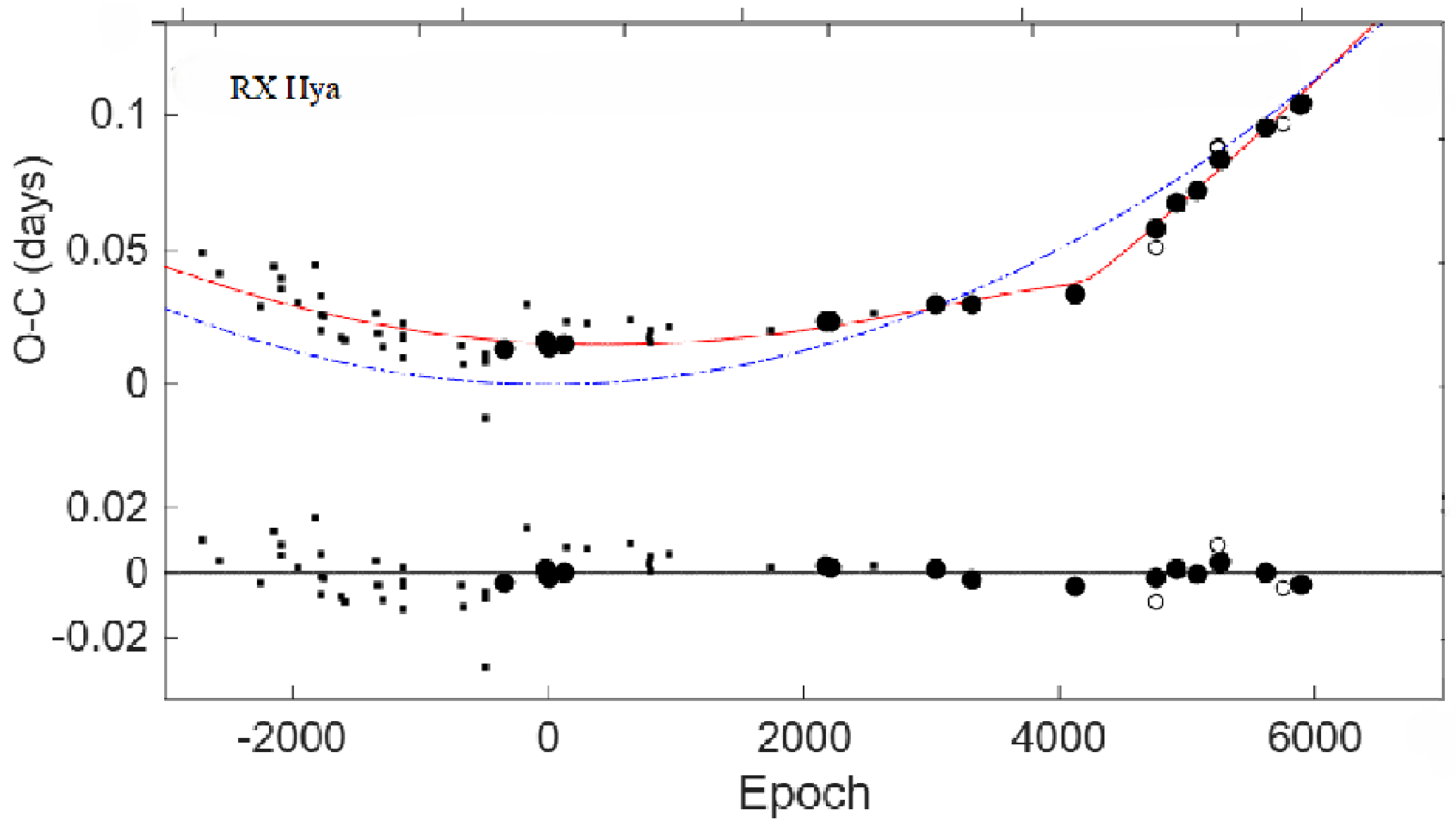}
 \end{minipage}
 \begin{minipage}[b]{0.33\textwidth}
 \includegraphics[alt={O-C diagram of Y Leo},height=4.3cm, width=1\textwidth]{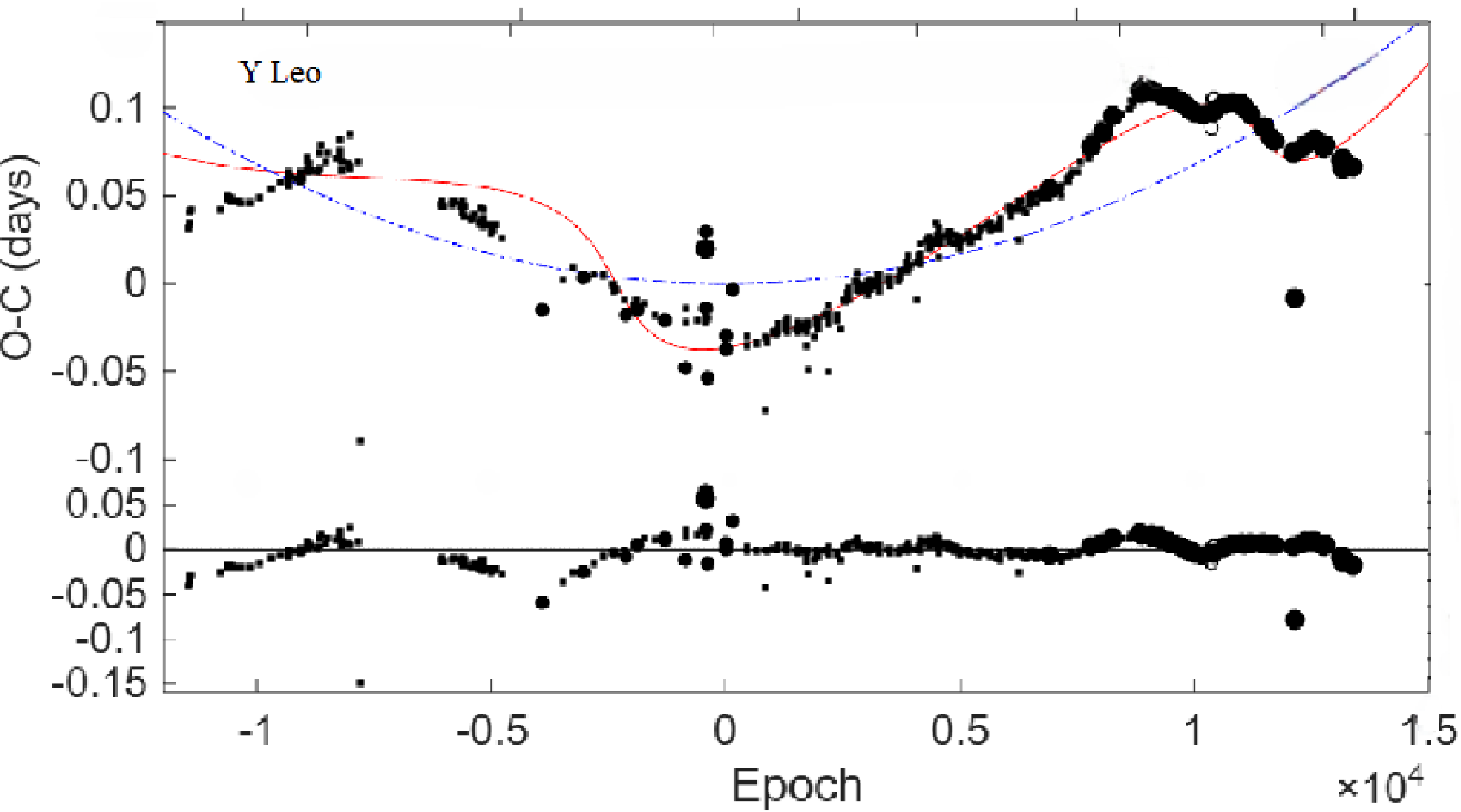}
 \end{minipage}
 \begin{minipage}[b]{0.33\textwidth}
 \includegraphics[alt={O-C diagram of FR Ori},height=4.3cm, width=1.0\textwidth]{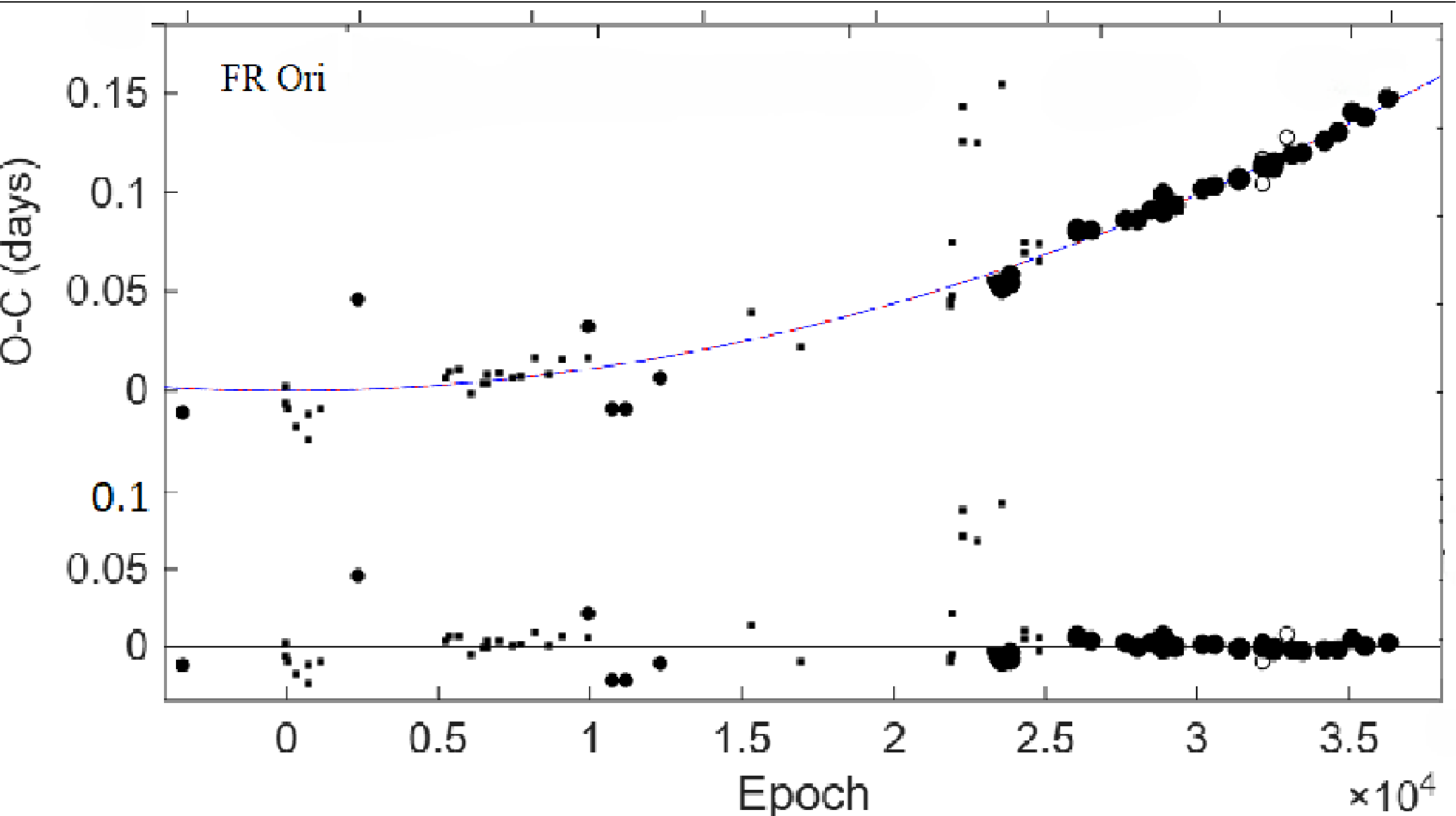}
 \end{minipage}
 \begin{minipage}[b]{0.33\textwidth}
 \includegraphics[alt={O-C diagram of AC Tau},height=4.3cm, width=1.0\textwidth]{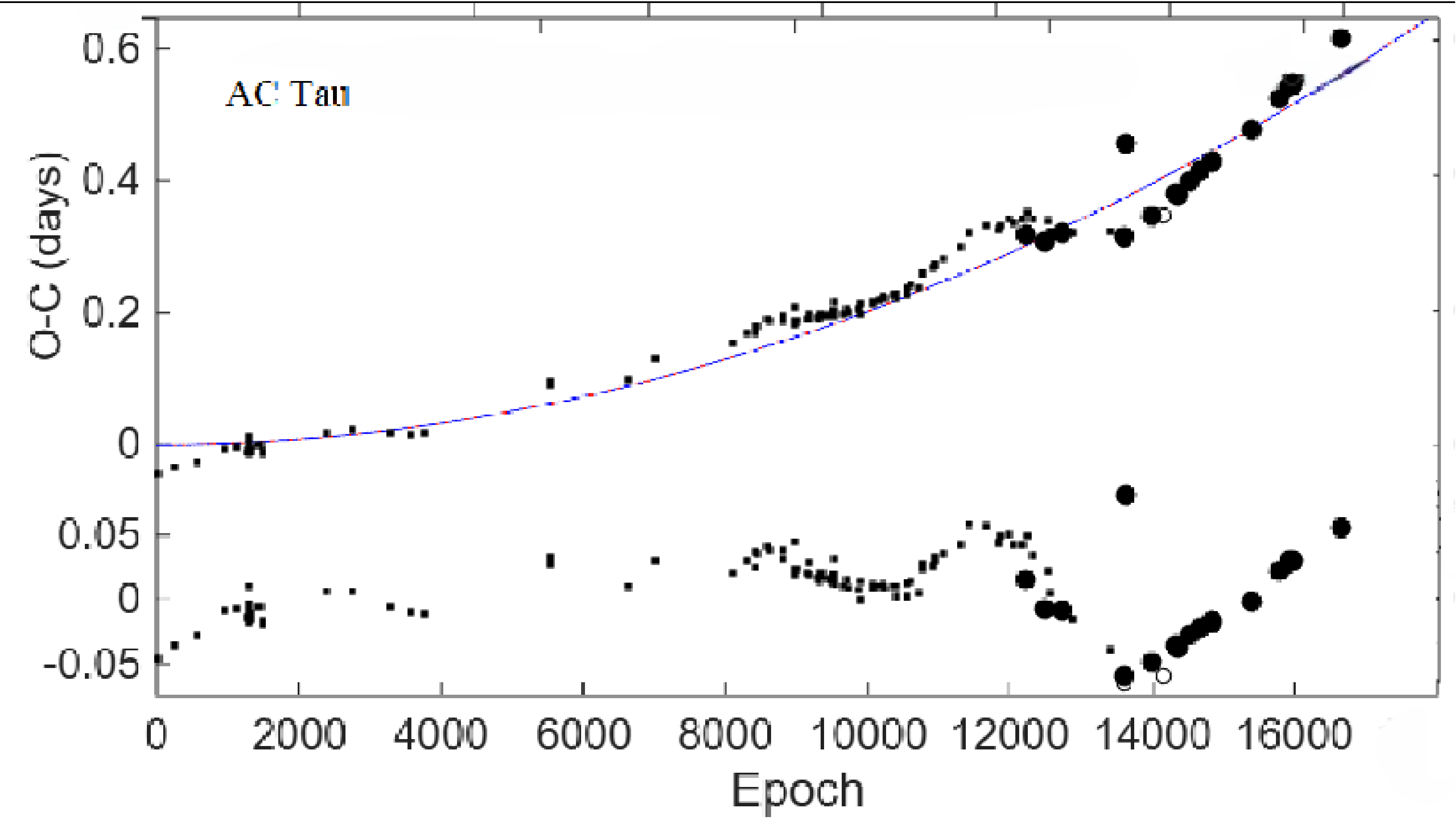}
 \end{minipage}
\caption{O–C diagram showing cyclic period variations for the targets. The blue dashed lines represent the upward parabolic variation, while the red lines represent the combined cyclic variation models, such as sinusoidal and parabolic.}\label{fig:ap}
\end{figure*}

\setcounter{figure}{2}
\renewcommand{\thefigure}{A\arabic{figure}}
\begin{figure*}
 \begin{minipage}[b]{0.33\textwidth}
 \includegraphics[alt={Scatter plot showing the correlation between $\log Q$ and $\log P_{\text{puls}}$ for SB2 systems, with a linear fit and $1\sigma$ error band.},height=4.8cm, width=1.0\textwidth]{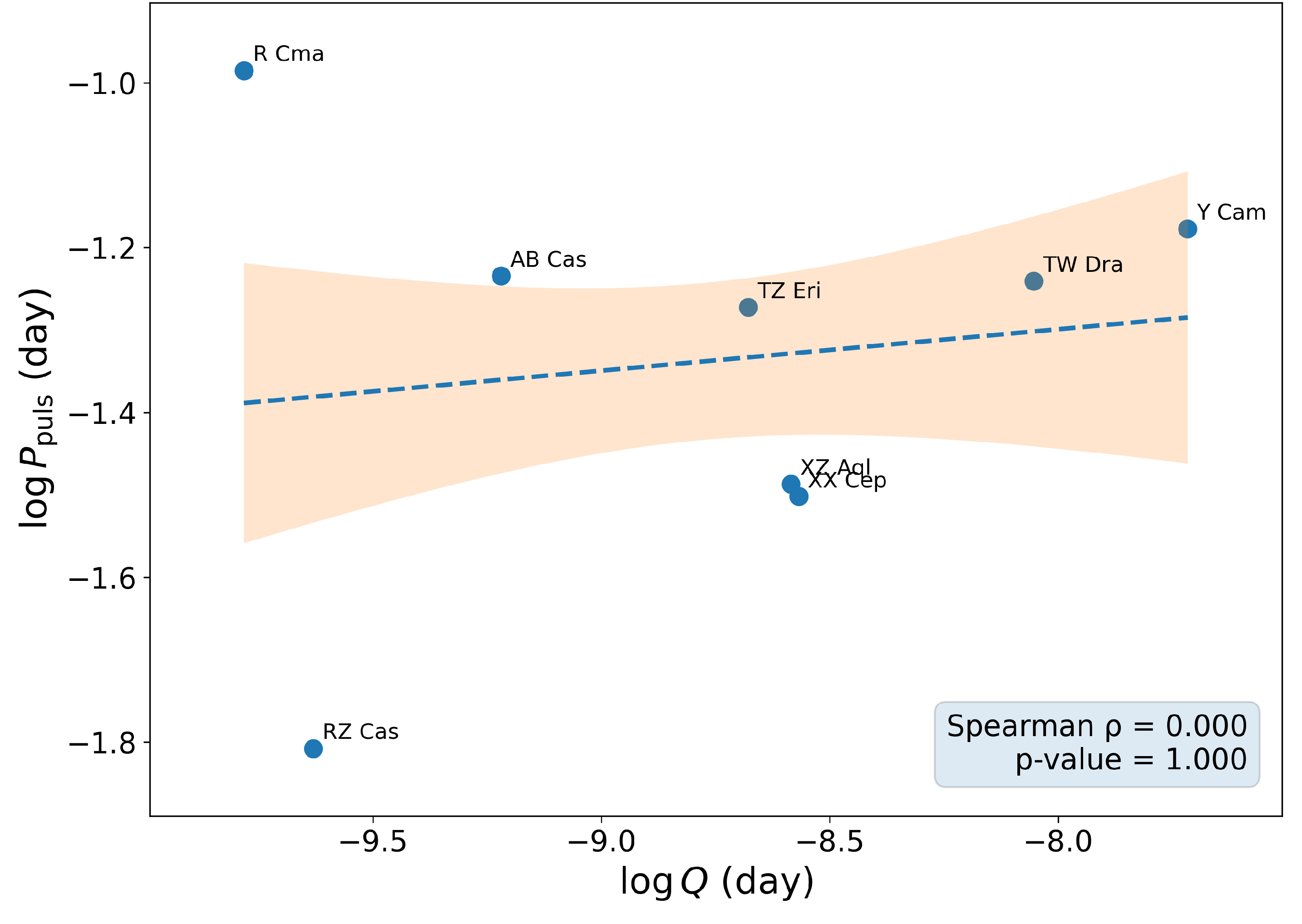}
 \end{minipage}
 \begin{minipage}[b]{0.33\textwidth}
 \includegraphics[alt={Scatter plot showing the correlation between $\log Q$ and $\log \text{Amp}$ for SB2 systems, with a linear fit and $1\sigma$ error band.},height=4.8cm, width=1.0\textwidth]{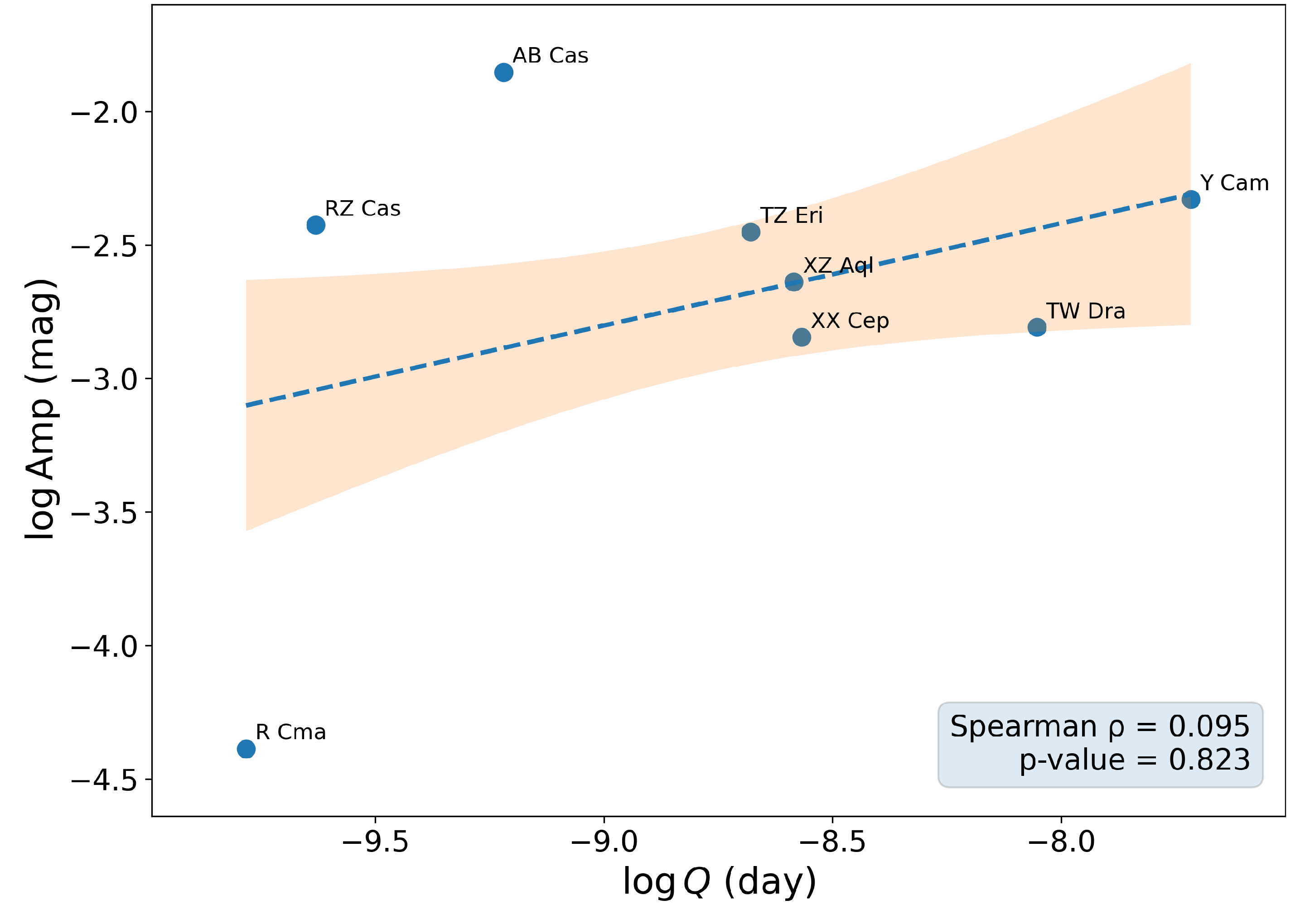}
 \end{minipage}
 \begin{minipage}[b]{0.33\textwidth}
 \includegraphics[alt={Scatter plot showing the correlation between $\log dP/dt$ and $\log P_{\text{puls}}$ for SB2 systems, with a linear fit and $1\sigma$ error band.},height=4.8cm, width=1.0\textwidth]{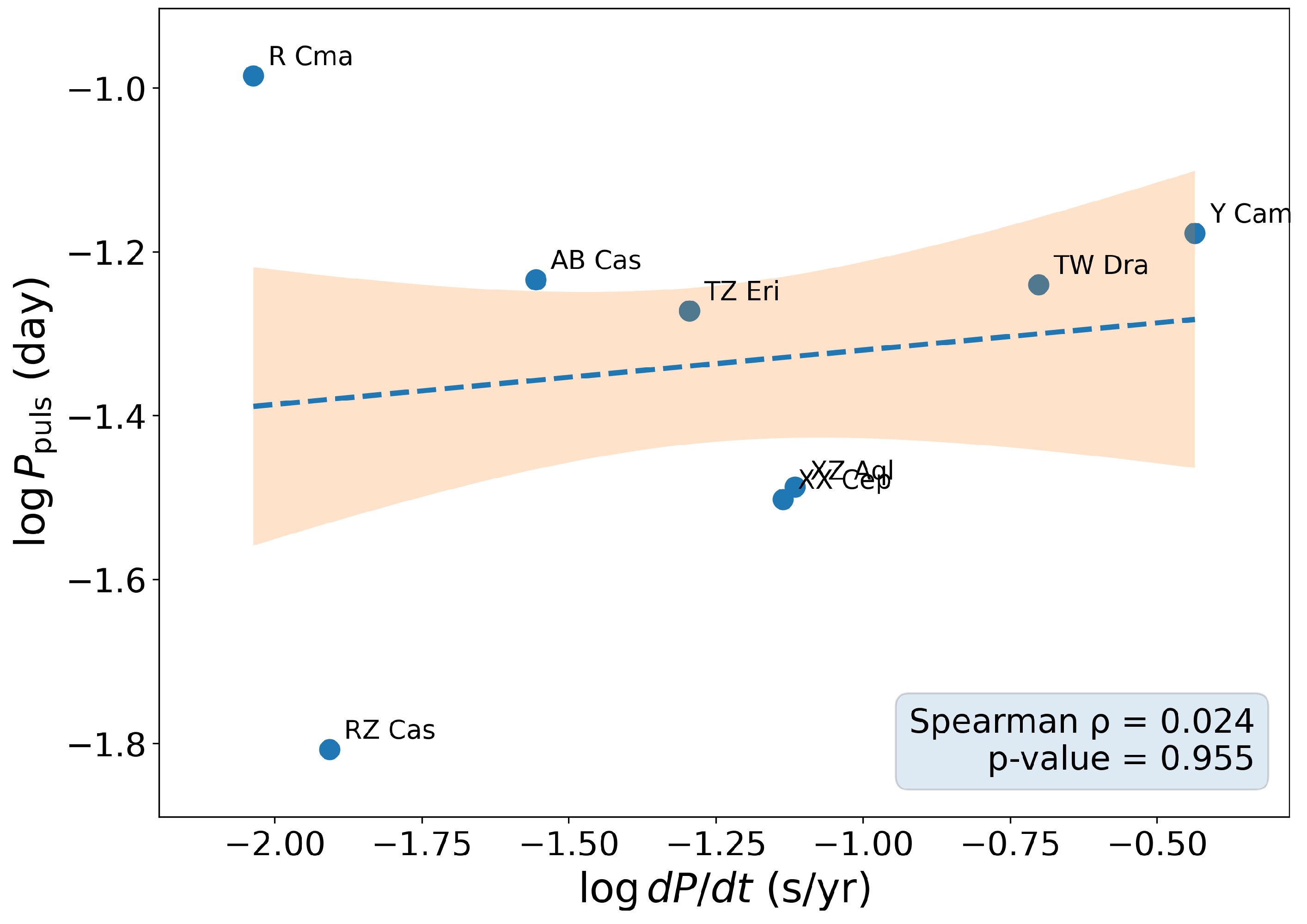}
 \end{minipage}
  \begin{minipage}[b]{0.33\textwidth}
 \includegraphics[alt={catter plot showing the correlation between $\log dP/dt$ and $\log \text{Amp}$ for SB2 systems, with a linear fit and $1\sigma$ error band.},height=4.8cm, width=1.0\textwidth]{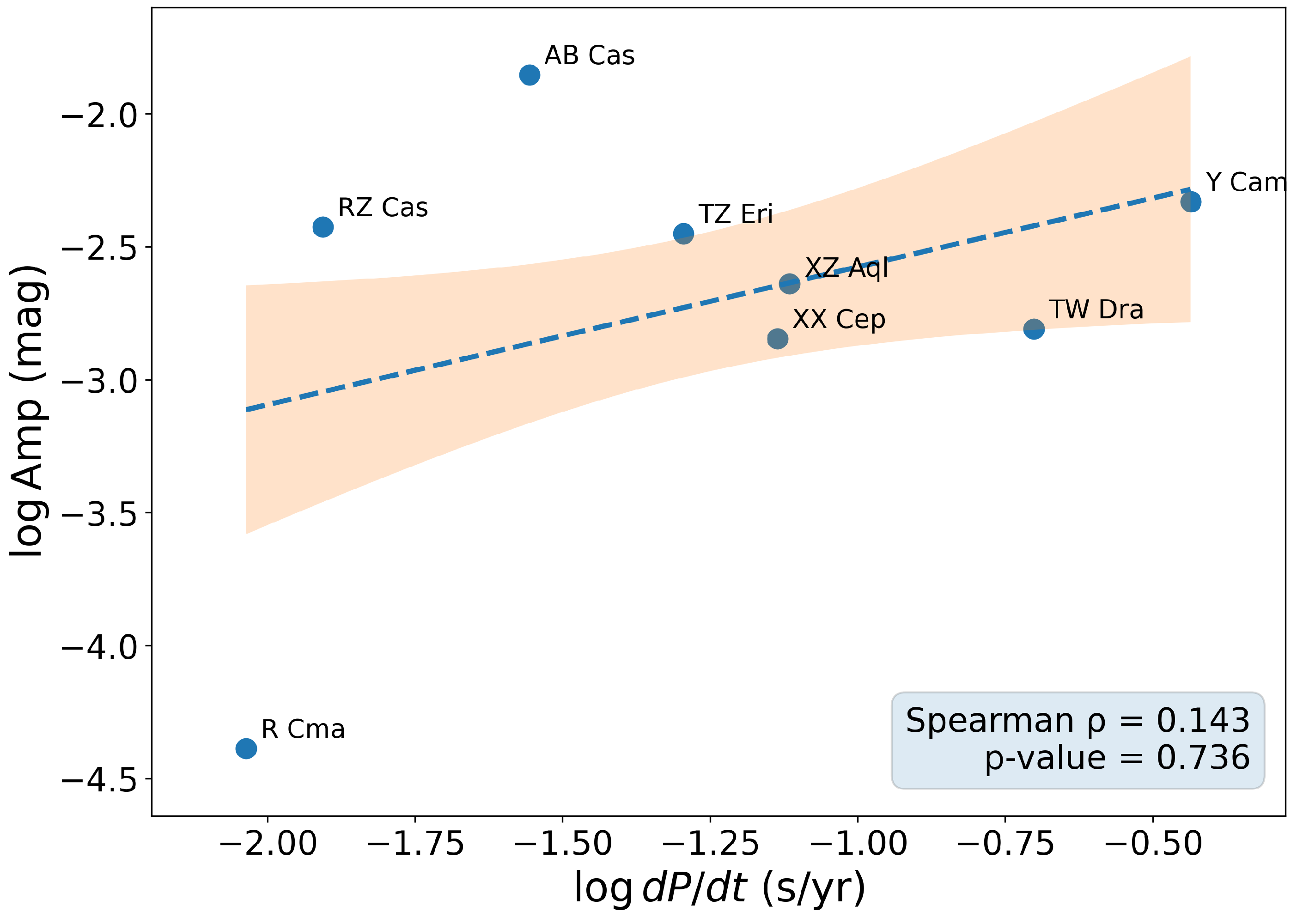}
 \end{minipage}
 \begin{minipage}[b]{0.33\textwidth}
  \includegraphics[alt={Scatter plot showing the correlation between conservative $\log dM/dt$ and $\log P_{\text{puls}}$ for SB2 systems, with a linear fit },height=4.8cm, width=1\textwidth]{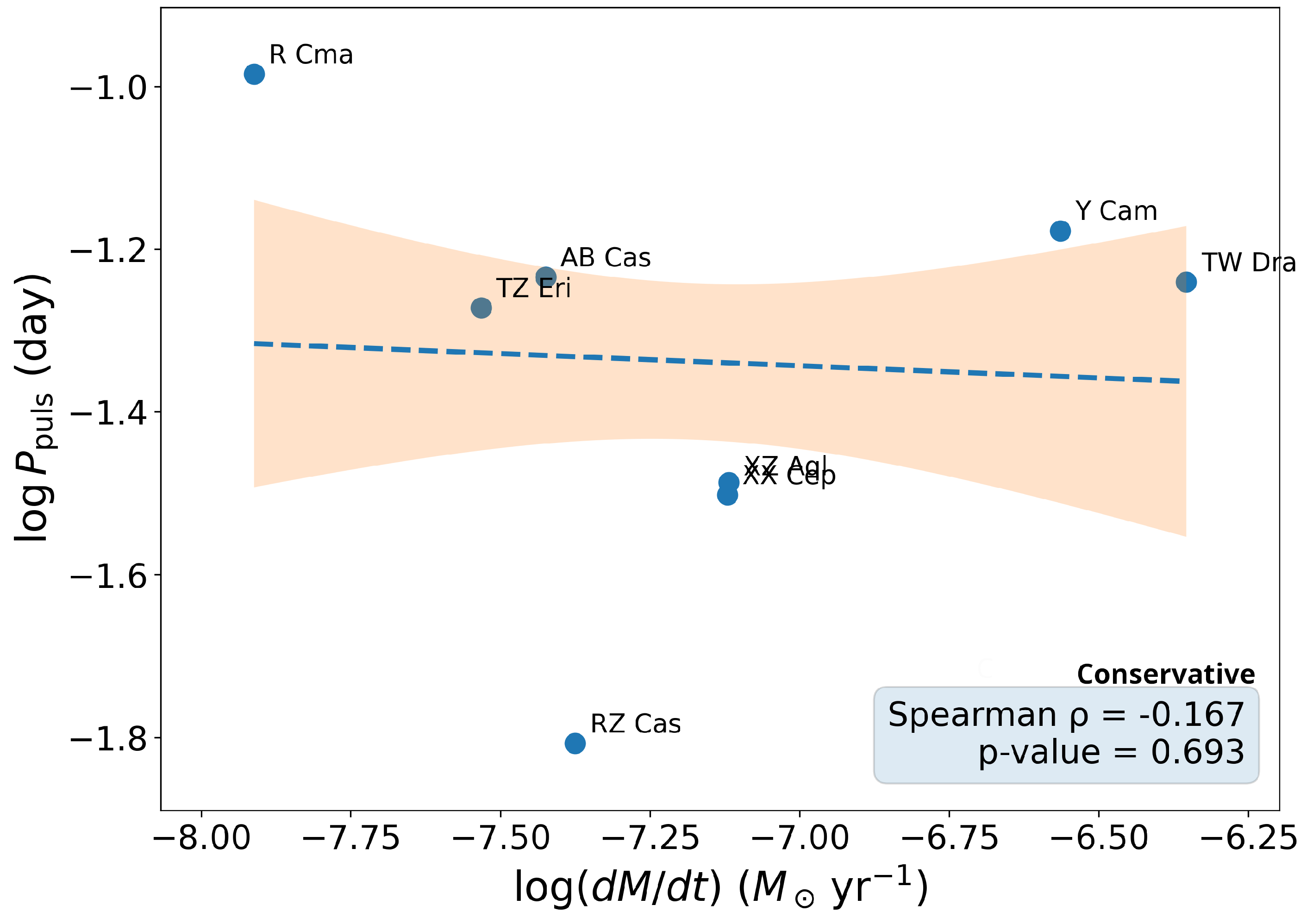}
  \end{minipage}
  \begin{minipage}[b]{0.33\textwidth}
  \includegraphics[alt={catter plot showing the correlation between conservative $\log dM/dt$ and $\log \text{Amp}$ for SB2 systems, with a linear fit and $1\sigma$ error band.},height=4.8cm, width=1\textwidth]{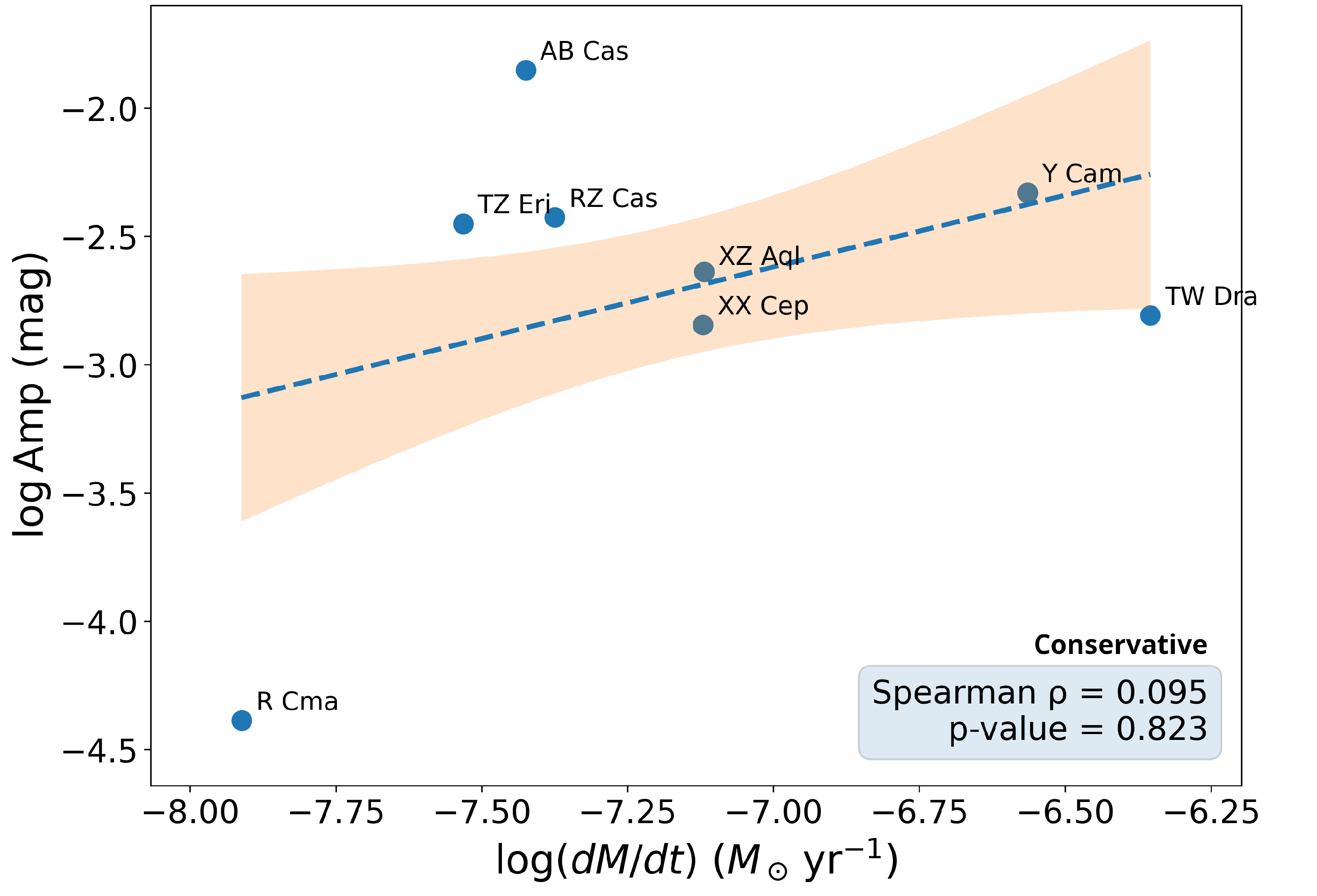}
  \end{minipage}
  \begin{minipage}[b]{0.33\textwidth}
  \includegraphics[alt={Scatter plot showing the correlation between non-conservative $\log dM/dt$ and $\log P_{\text{puls}}$ for SB2 systems, with a linear fit and $1\sigma$ error band.},height=4.8cm, width=1\textwidth]{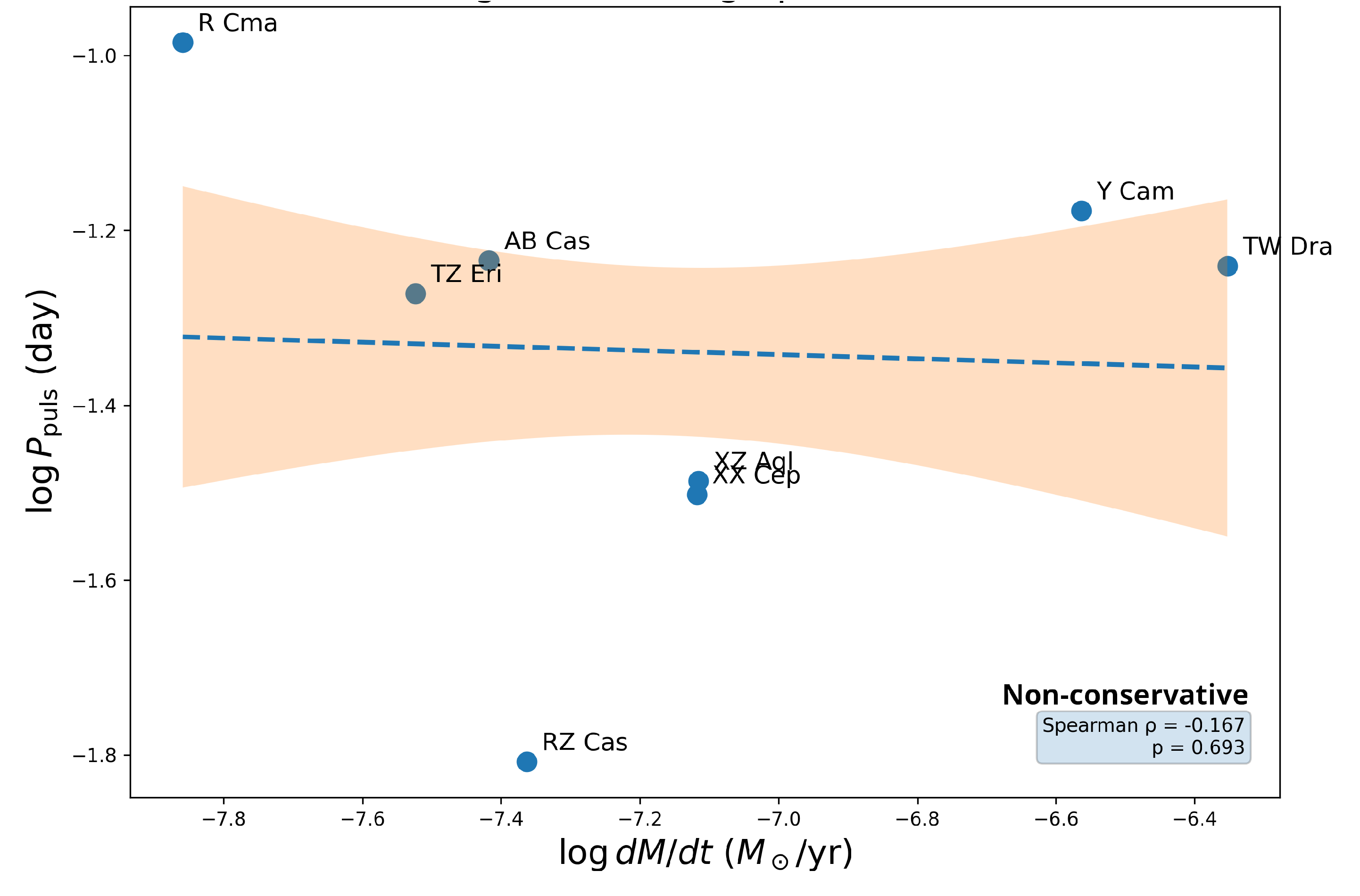}
 \end{minipage}
  \begin{minipage}[b]{0.33\textwidth}
  \includegraphics[alt={Scatter plot showing the correlation between non-conservative $\log dM/dt$ and $\log \text{Amp}$ for SB2 systems, with a linear fit and $1\sigma$ error band.},height=4.8cm, width=1\textwidth]{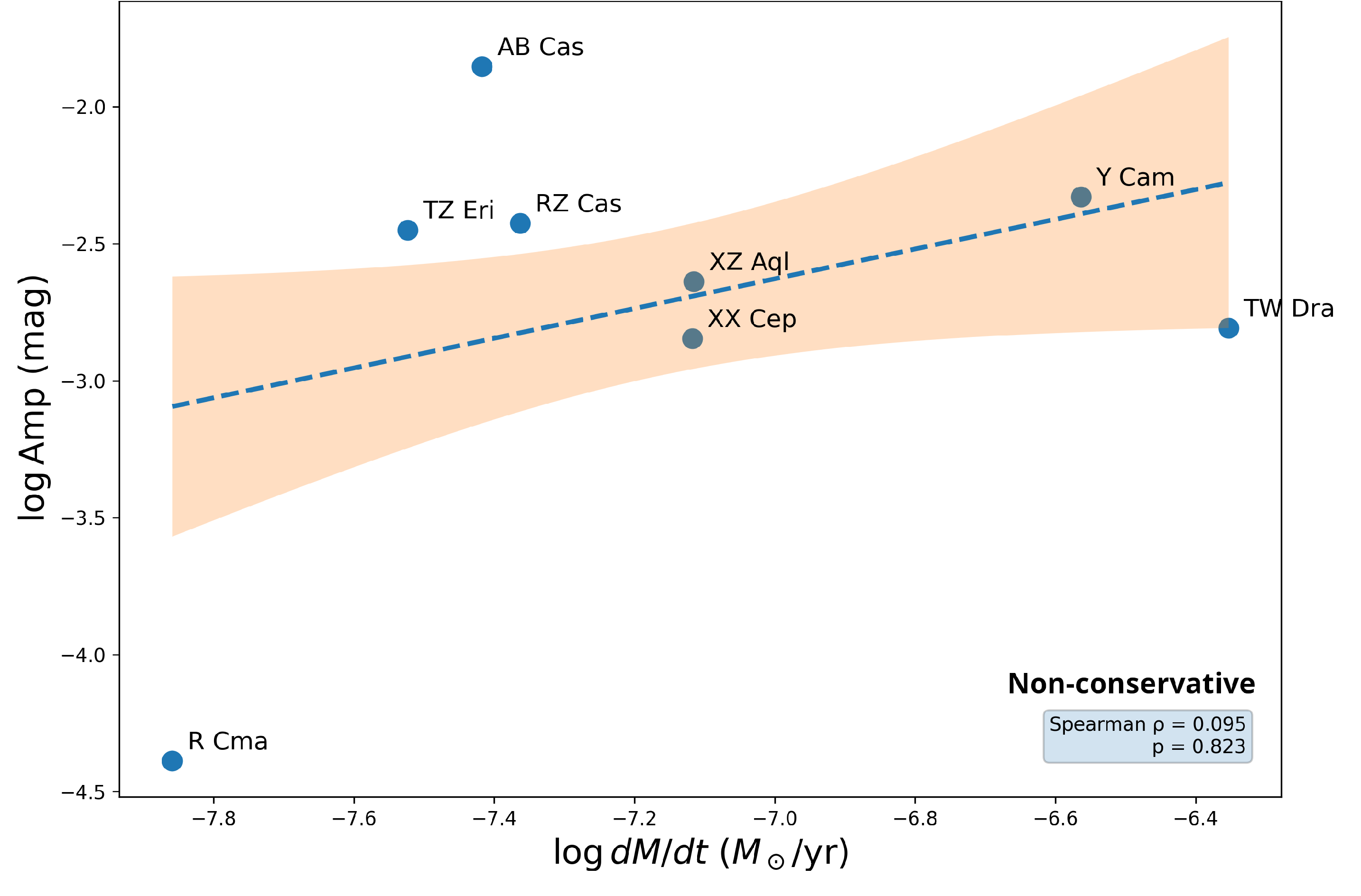}
 \end{minipage}
\caption{Same as Figs. 5, 6, 7, and 8, respectively, but only for the SB2 systems.}\label{fig:relationship_app}
\end{figure*}

%
%

\clearpage
\bibliography{PASJ-2026-0124}{}
\bibliographystyle{aasjournalv}

\end{document}